%% file: 0_main.tex
\documentclass[5p,twocolumn,10pt,times]{elsarticle}
\usepackage{amsmath}
\usepackage{hyperref} % added [draft] to avoid compilation issues that happen if a link is split and appears in two pages
	\newif\ifcolorrevise
	
	\colorrevisefalse

	\makeatletter
	\providecommand{\doi}[1]{%
	  \begingroup
	    \let\bibinfo\@secondoftwo
	    \urlstyle{rm}%
	    \href{http://dx.doi.org/#1}{%
	      doi:\discretionary{}{}{}%
	      \nolinkurl{#1}%
	    }%
	  \endgroup
	}
	\makeatother

	\usepackage[labelformat=parens]{subcaption} % subfigures
	\usepackage[export]{adjustbox} % [right] alignment for includegraphics
	
	\usepackage{rotating} % turn env for rotating text in figures

	\usepackage{wrapfig} % inline figures

	\usepackage{multirow} % multicolumn, multirow
	\usepackage{colortbl} % \cellcolor{<color>}
	\newcolumntype{C}[1]{>{\centering\arraybackslash}m{#1}}   %% centered
	\newcolumntype{R}[1]{>{\raggedleft\arraybackslash}m{#1}}  %% right aligned

	\usepackage[capitalise]{cleveref} % automatically add `Fig.'  etc before a reference.

        \usepackage{ amssymb } % \therefore
	
	\newcommand{\degree}{^\circ}
	
	\usepackage[binary-units]{siunitx} % mm and stuff
	\DeclareSIUnit\pixel{px}

	\usepackage{units} % \nicefrac{3}{8}

	\DeclareMathOperator*{\argmax}{arg\,max}
	\DeclareMathOperator*{\argmin}{arg\,min}
	
	\usepackage{amsthm} % \begin{proof}
	
	\theoremstyle{definition}
	\newtheorem{definition}{Definition}[section]

	\usepackage[inline]{enumitem} % inline enumerate*

	\usepackage[toc,page]{appendix} % appendicces
	
	\usepackage{pgfplots}
	\usepackage{pgfplotstable} % tikzpicture table plots
	\pgfplotsset{compat=1.14}
	\usetikzlibrary{backgrounds}

	\usepackage[noend]{algpseudocode} % algorithmic
	\usepackage{algorithm} % wrapper for pseudocode to give a caption and label

	\usepackage{upgreek} % \uplambda

	\usepackage{listings} % for listing C++ code instead of pseudocode
    \usepackage{color,soul} % caps, highlight (\hl)

	\newcommand{\comment}[1]{}

	\usepackage[normalem]{ulem}
	\newcommand{\stkout}[1]{\ifmmode\text{\sout{\ensuremath{#1}}}\else\sout{#1}\fi}
	
	\newcommand{\revise}[2]{#2}
	\ifcolorrevise
	\renewcommand{\revise}[2]{{\color{blue}{#2}}}
	\fi
	\setulcolor{red}

	\usepackage[normalem]{ulem} % squigly underline

	\newlength{\figwidth}
	\newlength{\figwidthTwo}
	\newlength{\figwidthTree}
	\newlength{\figheight}
	\newlength{\figheightTwo}
	\newlength{\tempheight}
	\newlength{\tempheightTwo}

\begin{document}
\baselineskip11pt 

\begin{frontmatter} 

\title{A framework for adaptive width control of dense contour-parallel toolpaths in \revise{additive manufacturing}{fused deposition modeling}}

%\author{Paper ID: xxx}

\author[um,tud]{Tim Kuipers}
\author[tud]{Eugeni L. Doubrovski}
\author[tud]{Jun Wu\corref{cor1}}
\ead{j.wu-1@tudelft.nl}
\cortext[cor1]{Corresponding author}
\author[cuhk]{Charlie C. L. Wang}
% \ead{cwang@mae.cuhk.edu.hk}
\address[um]{Ultimaker, Utrecht, The Netherlands}
\address[tud]{Department of Sustainable Design Engineering, Delft University of Technology, The Netherlands}
\address[cuhk]{Department of Mechanical and Automation Engineering, The Chinese University of Hong Kong, Hong Kong SAR, China}

\begin{abstract}
3D printing techniques such as Fused Deposition Modeling (FDM) have enabled the fabrication of complex geometry quickly and cheaply. 
\revise{By densely filling consecutive 2D layers with contour-parallel extrusion toolpaths, FDM can produce parts with high stiffness and strength.}
{High stiffness parts are produced by filling the 2D polygons of consecutive layers with contour-parallel extrusion toolpaths.}
\revise{Toolpath with uniform }{Uniform width toolpaths consisting of} inward offsets from the outline polygons produce over- and underfill regions in the center of the shape, which are especially \revise{problematic for }{detrimental to the mechanical performance of} thin parts\revise{ such as casings and microstructures}{}.
\revise{}{In order to fill shapes with arbitrary diameter densely the toolpaths require adaptive width.}
Existing approaches for generating toolpaths with adaptive width result in a large variation in widths, which for some hardware systems is difficult to realize accurately\revise{, if not beyond their capabilities.}{.}
In this paper we present a framework which supports multiple schemes to generate toolpaths with adaptive width, by \revise{using}{employing} a function to decide the number of beads and their widths\revise{ which are applied from the center outward.}{.}
Furthermore, we propose a novel scheme\revise{ for FDM printing}{} which \revise{avoids}{reduces} extreme bead width\revise{ deviation from the nozzle size}{s}, \revise{and limits}{while limiting} the number of altered toolpaths.
We \revise{}{statistically }validate the effectiveness of our framework and this novel scheme on a data set of \revise{300 slices.}{representative 3D models, }%sgd
\revise{}{and physically validate it by developing a technique, called \emph{back pressure compensation}, for off-the-shelf FDM systems to effectively realize adaptive width.}
\end{abstract}

%
% The code below should be generated by the tool at
% http://dl.acm.org/ccs.cfm
% Please copy and paste the code instead of the example below.
%
%\begin{CCSXML}
%\end{CCSXML}

%\ccsdesc[500]{Computer systems organization~Embedded systems}
%\ccsdesc[300]{Computer systems organization~Redundancy}
%\ccsdesc{Computer systems organization~Robotics}
%\ccsdesc[100]{Networks~Network reliability}


\end{frontmatter}

%  \temp{%Table of contents just for reviewing purposes
%  \tableofcontents
%  }

\input{1_intro}

\input{2_related_work}

\input{4_method}

\input{5_generalization}

\input{6_2_printing_results}

\input{6_validation}

\input{8_discussion}

\input{12_conclusions_future_work}

\section*{References}
\interlinepenalty=100000 % prevents pdfendlink ended up accross pages error. see https://tex.stackexchange.com/a/449633/129190
\bibliography{99_mybib}

\begin{appendices}

\input{19_edge_discretization}
\input{20_dataset}
\input{25_accuracy}

\end{appendices}
\end{document}

%% file: 1_intro.tex
\section{Introduction}
3D printing enables the fabrication of complex geometry under few design constraints compared to conventional fabrication techniques.
Recent developments have seen a rapid growth in both the use and capabilities of desktop 3D printing systems.
%The rapid spread of 3D printing through different industries and types of application calls for the possibility to manufacture a wide range of geometries while guaranteeing mechanical properties of the resulting parts.
Fused Deposition Modeling (FDM) is one of the most common 3D printing techniques.
It is widely used because of the versatility in the types of plastic which can be used and the relatively low running costs.
FDM printers are used, for example, in showcasing scale models of buildings, casings for electronics, prototypes for blow molded parts, jigs and fixtures.
Recent research \revise{developments have investigated}{adressed} manufacturing complex volumetric structures such as microstructures~\cite{bates2018compressive,Al-Ketan2018,Maskery2018} and topology optimized structures~\cite{Zegard2016SMO,Wu2019a,Cheng2019}.
Many of these applications involve 3D models with detailed features within the order of magnitude of the \revise{printing resolution}{nozzle size}, which restrains the field of the process planning algorithms.

FDM printers extrude semi-continuous beads of molten plastic through a nozzle, which moves along a planned toolpath within each layer of a 3D object.
\revise{
A common strategy to accurately manufacture a given 3D model is to extrude along a contour-following path,
because the position and shape of the toolpath can be controlled relatively accurately.
Filling up a shape using parallel straight lines would expose defects of the size of the hole in the nozzle, which is generally an order of magnitude larger than the resolution of the positioning system.
Contour-parallel extrusion therefore leads to a less bumpy outline shape than direction-parallel extrusion does.
}{%sgd
A common strategy is to extrude along a number of parallel toolpaths which follow the shape of the contour of the layer and fill up the remaining area using parallel straight toolpaths.
Contour-parallel toolpaths fit to the layer outlines more accurately, because the resolution of the positioning system is an order of magnitude smaller than the size of the hole in the nozzle.
This paper is concerned with the generation of such contour-parallel toolpaths and addresses several issues which commonly occur in 3D models with narrow geometry.
%Because contour-parallel extrusion leads to a more accurate outline shape it is common practice to print either the whole layer or only a limited number of outer perimeters that way.
%This paper improves on those contour-parallel toolpaths and addresses several issues which commonly occur in 3D models with narrow geometry.
}

\begin{figure}\centering
\setlength{\figwidth}{.9\columnwidth}
\setlength{\figwidthTwo}{.05\columnwidth}
\begin{subfigure}{\figwidth}\centering
\parbox[b]{\figwidthTwo}{\subcaption{}\label{intro_wedge_uniform}}\includegraphics[width=\figwidth]{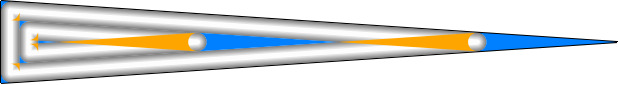}
\end{subfigure}
\begin{subfigure}{\figwidth}\centering
\parbox[b]{\figwidthTwo}{\subcaption{}\label{intro_wedge_centered}}\includegraphics[width=\figwidth]{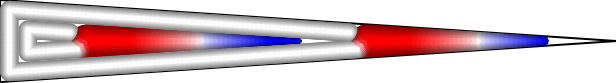}
\end{subfigure}
\begin{subfigure}{\figwidth}\centering
\parbox[b]{\figwidthTwo}{\subcaption{}\label{intro_wedge_distributed}}\includegraphics[width=\figwidth]{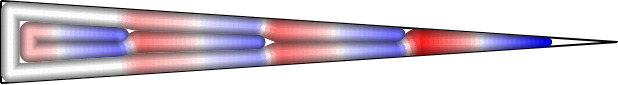}
\end{subfigure}
\caption{
Illustration of different toolpath\revise{}{s} for a \revise{wedge }{}shape \revise{}{showcasing a range of shape radii }\revise{}{(black)}.
\revise{}{These results can be read as a graph with feature size on the horizontal axis and its corresponding beading along the vertical axis.}
\subref{intro_wedge_uniform} Toolpath\revise{}{s} using uniform offsetting results in large overfill (orange) and underfill (azure).
\subref{intro_wedge_centered} Toolpath\revise{}{s} with adaptive width~\cite{Jin2017JMS} where beads that are wider or narrower than the nozzle size are \revise{indicatd}{indicated} in red and blue, respectively.
\subref{intro_wedge_distributed} Our approach minimizes over- and underfill with \revise{beads close to the nozzle size}{less extreme widths}.
}
\label{intro_wedge}
\end{figure}

The simple technique for generating the dense contour-parallel toolpaths of a layer consists \revise{in}{of} performing uniform inward offsets with the size of the nozzle from the outline shape.
However, for geometrical features which are not an exact multiple of the nozzle size this method produces areas where an extrusion bead is placed twice: \emph{overfill} areas; and areas which are not filled at all: \emph{underfill} areas.
See \cref{intro_wedge_uniform}.
Overfills cause a buildup of pressure in the mechanical extrusion system, which can result in bulges or even a full print failure.
Underfills\revise{}{,} on the other hand, can cause a drastic decrease in the part stiffness or even for small features not to be printed at all.
These problems are exacerbated for models with layer outlines with small features, because the over- and underfill areas are relatively large compared to the \revise{whole part}{those features}.

One promising direction to avoid over- and underfills is to employ toolpath\revise{}{s} with adaptive width.
\citeauthor{Ding2016a} developed a toolpath strategy for wire and arc additive manufacturing which produces a width variation typically lower than a factor of $3$, but is \revise{sometimes}{} far greater\revise{}{ for some parts}~\cite{Ding2016a,Xiong2019}.
However, the range of \revise{}{bead }widths manufacturable by FDM systems is limited.
A nozzle of \revise{\SI{0.4}{\milli\meter}}{$w=\SI{0.4}{\milli\meter}$} will typically start to cause fluttered extrusion around lines narrower than \SI{0.3}{\milli\meter} and lines will start to bulge upward if they are wider than the flat part of the nozzle, which is typically \SI{1.0}{\milli\meter}.
%Therefore, a limited range of widths is required by the hardware system.

The current \revise{state-of-the-art}{state of the art of contour-parallel toolpath generation} \revise{for FDM printing }{}developed by \citeauthor{Jin2017JMS}
employs a strategy which alters the widths of the centermost beads \revise{at most by a factor of $2$}{within a range of widths $[0.25w,1.8w]$}~\cite{Jin2017JMS},
which is similar to the strategy employed by the open source industry standard software package Ultimaker Cura~\cite{cura}.
See \cref{intro_wedge_centered}.
Still, controlling the extrusion width through movement speed changes or through volumetric flow control (e.g. linear advance) yields diminishing accuracy for deposition widths \revise{farther away}{deviating more} from the nozzle size,
since process parameters such as nozzle temperature are optimized for beads with the nozzle size.
Moreover, reducing the variation in width is beneficial for limiting the variation in mechanical properties of the resulting product, meaning it conforms better to a simulation which employs a homogeneity assumption.
\revise{
Therefore, a narrower range of widths is desirable.
}
{We therefore reduce the bead width range by distributing the workload from the centermost bead over neighboring beads.}

\revise{
In this paper we propose a framework for planning toolpaths with control over the adaptive width for minimizing over- and underfill
We show that this framework supports various control schemes for determining the bead spacing and extrusion widths. 
For FDM printing in particular we propose a novel scheme which reduces the amount of over- and underfill while ensuring the extrusion beads deviate little from the nozzle size.
See \cref{intro_wedge_distributed}. 
}{}

Our contributions are as follows:
\begin{itemize}
\item A geometric framework \revise{for generating densely filling contour-parallel toolpaths employing adaptive width, according to any beading scheme which decides on the bead spacing and widths.}{allowing various adaptive bead width control schemes used to generate contour-parallel toolpaths which minimize under- and overfill.}
\item A specific beading scheme\revise{ for FDM printing}{,} which
reduces the \revise{amount of deviation}{variation} in the extrusion widths \revise{compared to existing literature, }{to within $[0.75w,1.5w]$.}
\revise{and which promotes smooth toolpaths that are equal to the preferred width toward the outline of the shape.}{}
\revise{}{
\item A back pressure compensation approach to accurately realize adaptive bead width on Bowden style hardware systems.
}
\end{itemize}

%This work is patent pending, but the source code is available open source.

%% file: 2_related_work.tex
\section{Related Work}

% This paragraph explains what toolpath generation is, and the context where toolpath generation is positioned in
%Toolpath generation for deposition based additive manufacturing consists in generating geometric paths which the nozzle follow during printing.
%Extrusion toolpaths are generated within the planar contour of a layer at the intersection of a horizontal plane and a 3D solid object.
Toolpath generation consists in generating a path in the a planar contour, representing the intersection of a plane and a 3D solid object.
The nozzle is then instructed to move along the path while extruding material.
Sites along the toolpath\revise{}{s} are assigned several properties such as movement speed, but for this paper we will focus on the assigned width of the extruded bead.
Toolpath generation is an integral part of process planning for 3D printing.
For an overview of the processing pipeline, we refer to the survey by \citeauthor{Livesu2017CGF}~\cite{Livesu2017CGF}.
For reducing printing time and material cost, sparse infill structures such as triangular and hexagonal patterns have been used to approximate the interior of 3D shapes.
In this paper, we focus on \revise{generating toolpath that seamlessly fills the entire 2D contour.}{the generation of toolpaths for dense regions, such as the boundary shell of 3D shapes.}
This is sometimes called `dense infill'~\cite{Livesu2017CGF}.

% criteria for toolpath generation
The toolpath has a direct influence on the printing time, material cost, and mechanical properties of the printed object~\cite{N.Turner2014,ahn2002anisotropic}.
FDM calls for toolpaths with several desirable properties.
First, the extrusion path should be as continuous as possible.
A discontinuous path requires to stop and restart material extrusion.
For certain materials\revise{}{,} such as TPU, this could lead to printing defects or even print failure~\cite{KUIPERS2019CAD}.
Second, the toolpath is preferred to be smooth.
Sharp turns require to reduce the movement speed of the nozzle, and so this prolongs the printing process.
Third, the extruded path should cover \revise{a }{}the region of the contour without \revise{gaps}{underfilling}.
Such underfill negatively influences the mechanical performance of the parts.
Fourth, the extrusion paths should not overlap with one another.
Such overfill causes a pressure build up in the mechanical system, which leads to overextrusion in \revise{further}{later} paths and in extreme cases cause print failure~\cite{KUIPERS2019CAD}.
An analysis of under- and overfill from a vertical cross-section was presented in~\cite{Han2002JMSE}. 
Our method is primarily concerned with minimizing under- and overfill within horizontal cross-sections.

% basic toolpath strategies and their variations
Two basic strategies for dense toolpath generation are \revise{}{the }direction-parallel\revise{}{ strategy} and \revise{}{the }contour-parallel\revise{}{ strategy}.
Direction-parallel (or zig-zag) toolpaths fill an arbitrarily shaped contour with a set of parallel, equally spaced line segments.
These parallel segments are linked together at one of their extremities, to avoid discontinuous extrusion.
Contour-parallel toolpaths typically consists of a set of equally spaced offsets from the slice boundary outline polygons.
\citeauthor{steuben2016implicit} presented a method for generating sparse infill toolpaths based on the isocontours of surface plots of some variable generated on each 2D contour~\cite{steuben2016implicit}.
In order to increase the continuity of contour-parallel toolpaths, a strategy to connect dense toolpaths into spirals was introduced by \citeauthor{Zhao2016}~\cite{Zhao2016} and later extended to also connect a mixture of dense and sparse toolpaths together~\cite{KUIPERS2019CAD}.
\citeauthor{Jin2017RCIM} discusses several approaches for connecting direction-parallel and contour-parallel toolpaths into continuous paths~\cite{Jin2017RCIM}.
Spiral toolpaths have also been applied to (CNC) machining~\cite{Held2009,Huang2017}.
One of the problems with contour-parallel toolpath\revise{}{s} is that it tends to leave gaps between the toolpaths (see \cref{intro_wedge_uniform}).
This is due to the fact that the diameter of the part is not exact multiple of the (constant) deposition width in those regions.
To avoid problems with such gaps, hybrid approaches that combine direction and contour-parallel are often used~\cite{Mcmains2000DETC,Jin2013adaptive}.
Close to the slice boundary, there are several contour-parallel curves, while \revise{in }{}the interior is filled using a zig-zag pattern.
For complex shapes, the entire cross-section could be decomposed into a set of patches, and for each of them the basic strategies can be applied~\cite{Ding2014,Jin2017RCIM}.
Alternative toolpath patterns, seen also in CNC machining, include space-filling curves~\cite{Cox1994CAD,Griffiths1994,Shaikh2016}.

% the idea, and (some of) the most related work
Reducing under- and over-filling can be accurately achieved by making use \revise{over}{of} adaptive \revise{depisition}{deposition} width.
Adaptive width can be used to locally match the nonuniform space between adjacent paths, and thus to ensure a better filling of the area.
%Material extrusion with a varying width can be realized by changing the extrusion rate or the nozzle travel velocity~\cite{Ertay2018,Kuipers2018}.
\citeauthor{kao1998optimal} propose smooth adaptive toolpaths based on the medial axis skeleton of the boundary contour~\cite{kao1998optimal}.
Their approach handles simple geometry where there are no branches in the medial axis.
An extension was proposed by \citeauthor{Ding2016a} to handle complex shapes~\cite{Ding2016a}.
However, this extension inherits a problem in the original method:
from any point in the skeleton to the boundary, the number of toolpaths is constant.
\revise{%sdg
Since the distance may vary considerably for shapes with both large and small features, this strategy leads to a toolpath with widths that may differ by a factor beyond $3$, which is impossible for some FDM systems.
}{%f
Depending on the size of small and large features in the layer outlines, this strategy can require a range of bead widths beyond the capabilities of the manufacturing system.
}%sagd
\citeauthor{Jin2017JMS} proposed a strategy of adding \revise{a }{}toolpath\revise{}{s} with varying width along the center edges of the skeleton, while leaving other paths unchanged~\cite{Jin2017JMS}.
The resulting \revise{variation of width in the center is still up to a factor of $2$}{beads have widths within the wide range of $[0.25w,1.8w]$} (see \cref{intro_wedge_centered}).
In this paper we propose a novel scheme to distribute the width alterations throughout a region around the center, and thus limit the occurrence of extreme \revise{deviations}{variation} in width (see \cref{intro_wedge_distributed}).

% the context where our method can be positioned, and some related problems in that context
Under- and over-filling issues have a large proportional impact for thin geometric features.
\citeauthor{Jin2017a} proposed a sparse wavy path pattern for thin-walled parts~\cite{Jin2017a}.
Besides under- and over-filling, there are a few other robustness issues in toolpath generation for thin geometric features.
\citeauthor{Moesen2011} proposed a method to \revise{make beam compensation more reliable for}{reliably manufacture} thin geometric features\revise{}{ using laser-based additive manufacturing techniques}~\cite{Moesen2011}.
>> achange!!
\citeauthor{Behandish2019a} presented a method to characterize local- topological discrepancies due to material under- and over-deposition, and used this information to modify the as-manufactured outcomes~\cite{Behandish2019a}.

%% file: 4_method.tex
\begin{figure*}
\centering
\setlength{\figheight}{.15\textwidth}
\setlength{\figwidth}{.15\textwidth}
\setlength{\figwidthTwo}{.22\textwidth}
\begin{subfigure}{\figwidth}\centering
\includegraphics[width=\figheight,rotate=-90]{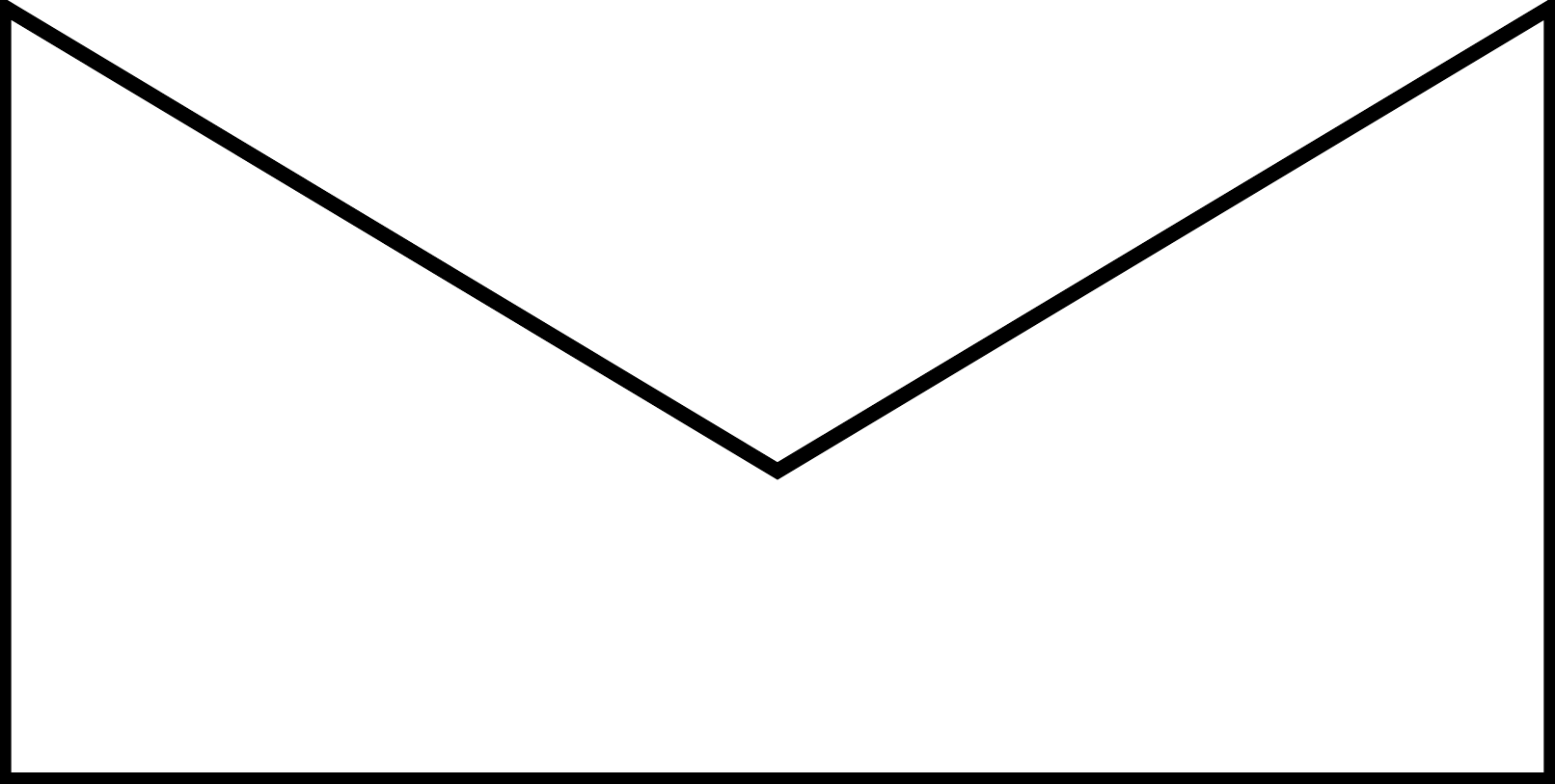}
\caption{Layer outline}\label{overview_outline}
\end{subfigure}
\begin{subfigure}{\figwidth}\centering
\includegraphics[width=\figheight,rotate=-90]{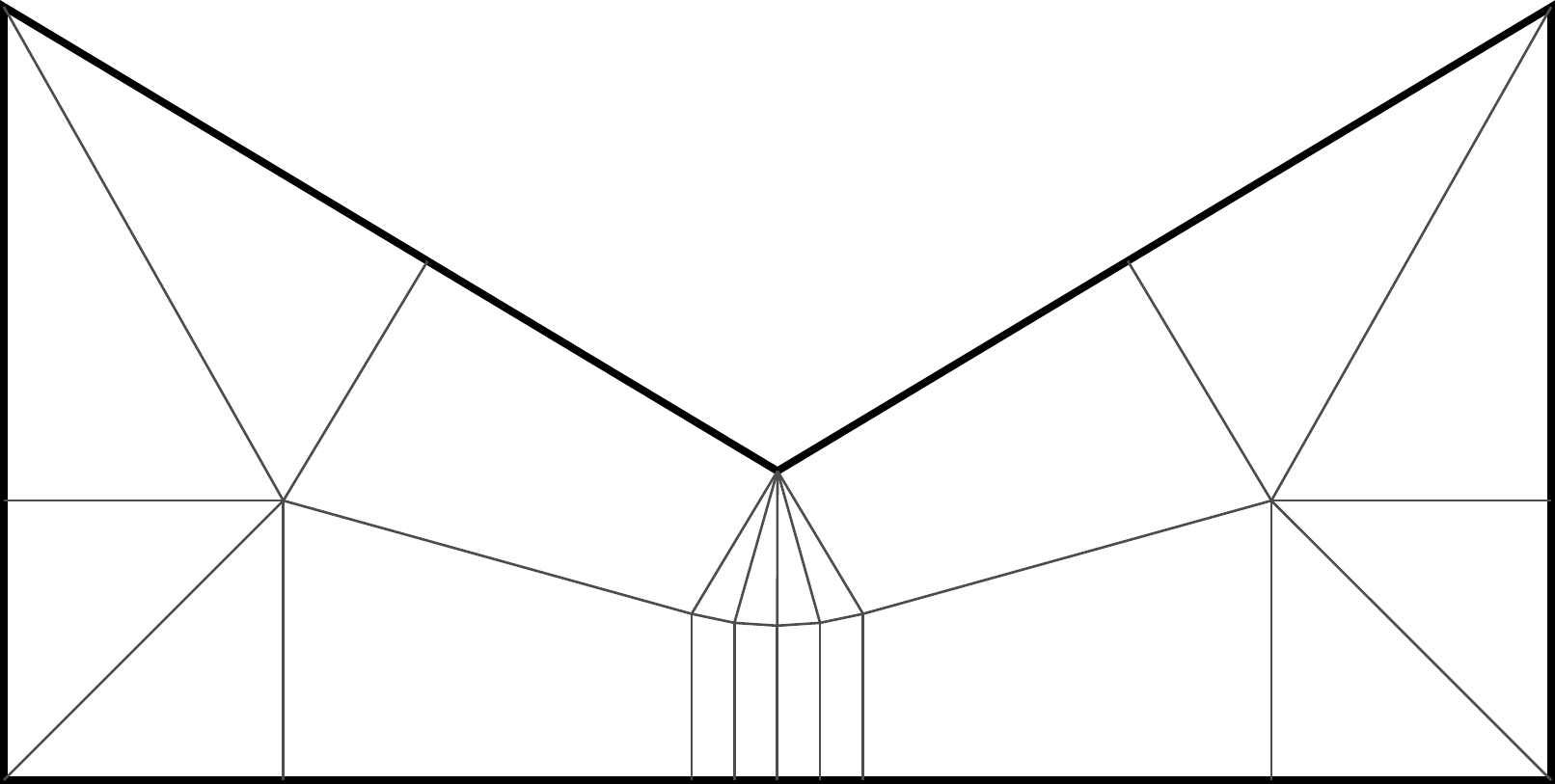}
\caption{Skeleton}\label{overview_skeleton}
\end{subfigure}
\begin{subfigure}{\figwidthTwo}\centering
\hspace*{\tempheightTwo}
\includegraphics[height=\figheight]{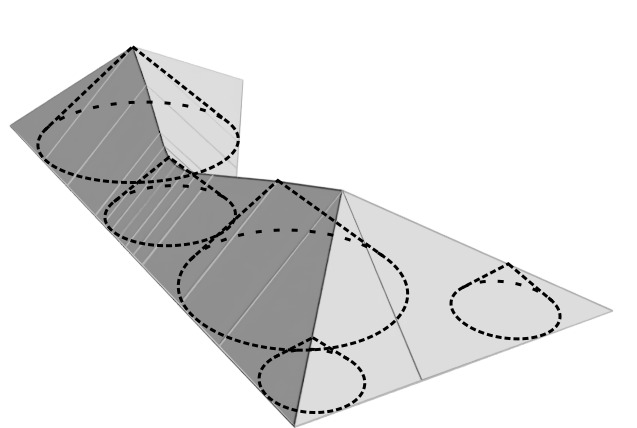}
\caption{Union of cones}\label{overview_uoc}
\end{subfigure}
\begin{subfigure}{\figwidthTwo}\centering
\hspace*{\tempheightTwo}
\includegraphics[height=\figheight]{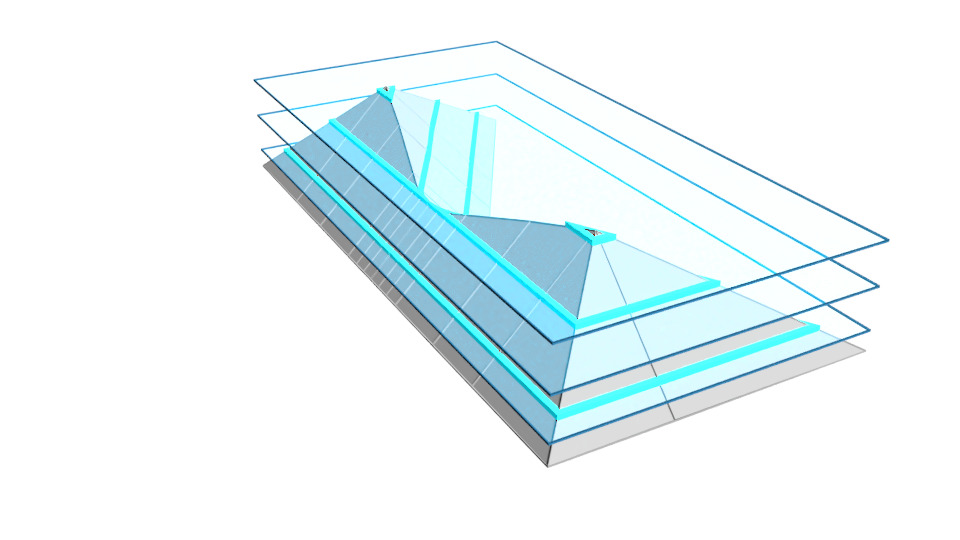}
\caption{Slicing}\label{overview_uniform_sliced}
\end{subfigure}
\begin{subfigure}{\figwidth}\centering
\includegraphics[width=\figheight,rotate=-90]{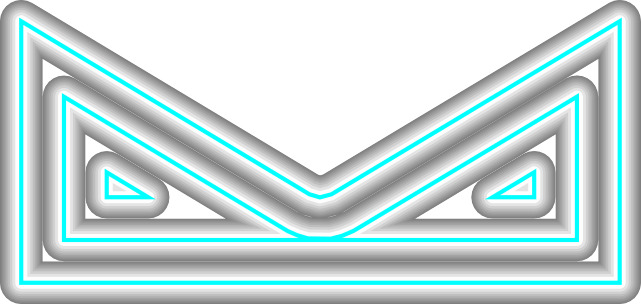}
\caption{Uniform paths}\label{overview_uniform_paths}
\end{subfigure}

\setlength{\figwidth}{.22\textwidth}
\setlength{\figwidthTwo}{.17\textwidth}
\setlength{\figwidthTree}{.22\textwidth}
\setlength{\tempheight}{-0.3cm}
\setlength{\tempheightTwo}{-0.5cm}
\begin{subfigure}{\figwidth}\centering
\hspace*{\tempheightTwo}
\includegraphics[height=\figheight]{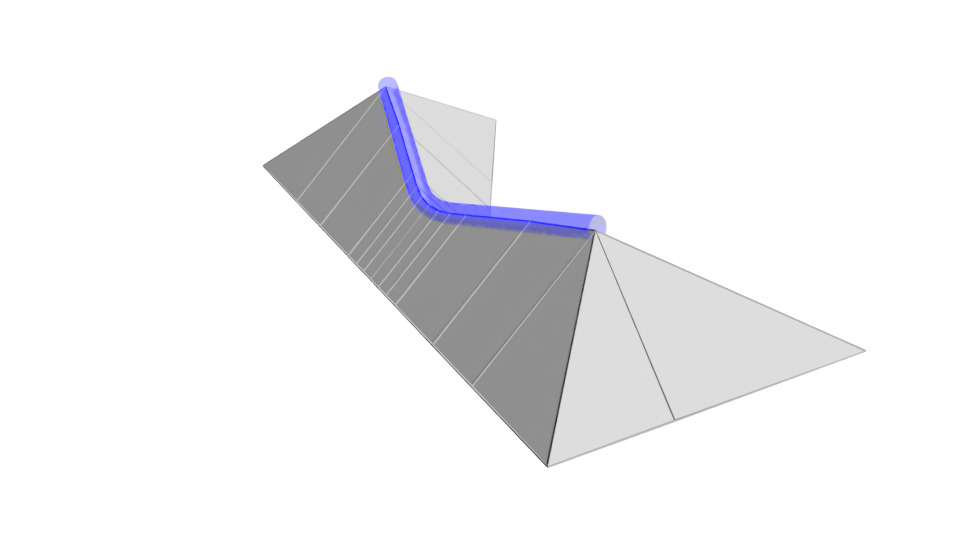}

\vspace{\tempheight}

\includegraphics[width=\figheight]{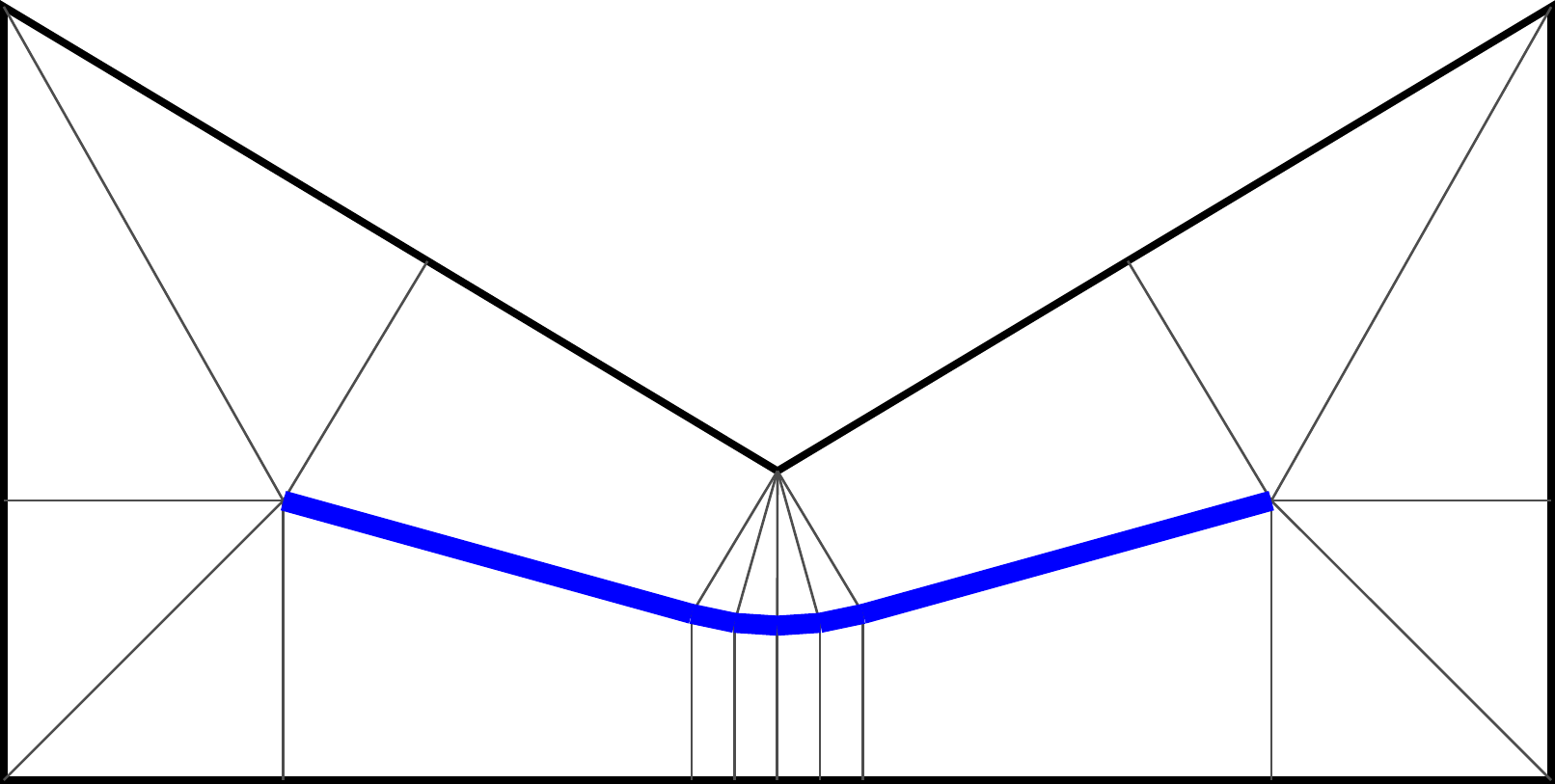}
\caption{Center identification}\label{3d_surface_overview_center}
\end{subfigure}
\begin{subfigure}{\figwidth}\centering
\hspace*{\tempheightTwo}
\includegraphics[height=\figheight]{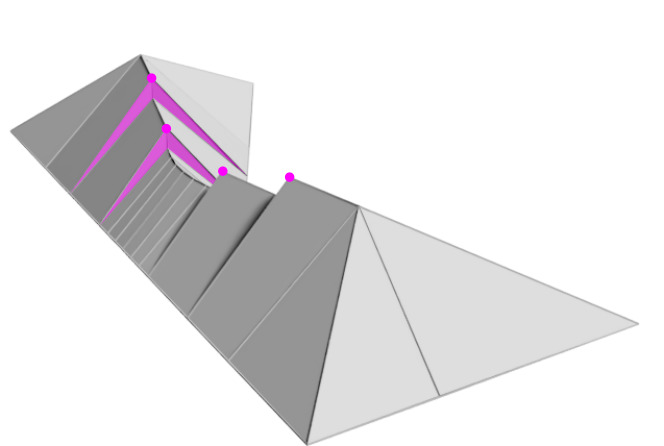}

\vspace{\tempheight}

\includegraphics[width=\figheight]{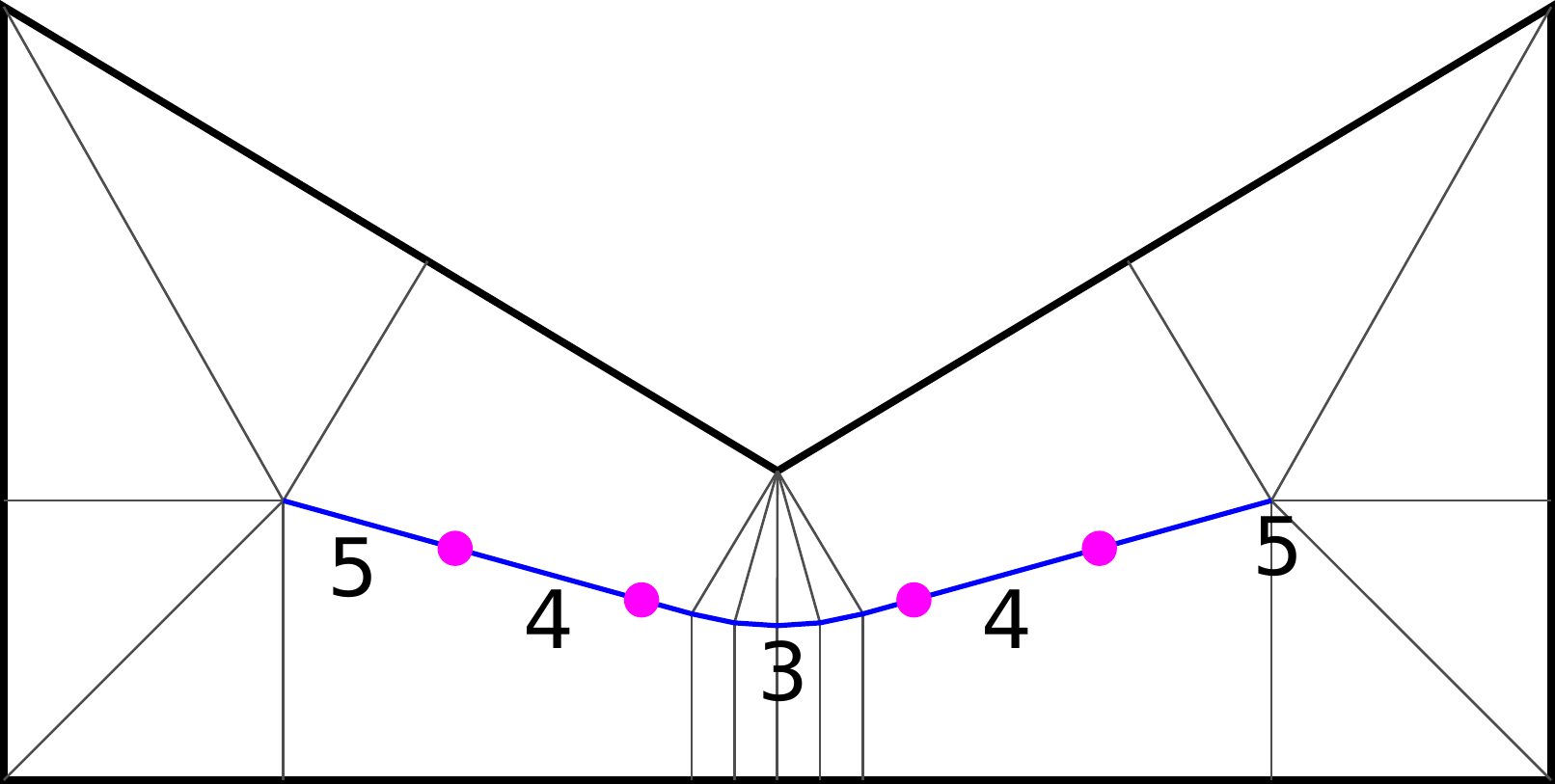}
\caption{Quantization}\label{3d_surface_overview_rounded}
\end{subfigure}
\begin{subfigure}{\figwidth}\centering
\hspace*{\tempheightTwo}
\includegraphics[height=\figheight]{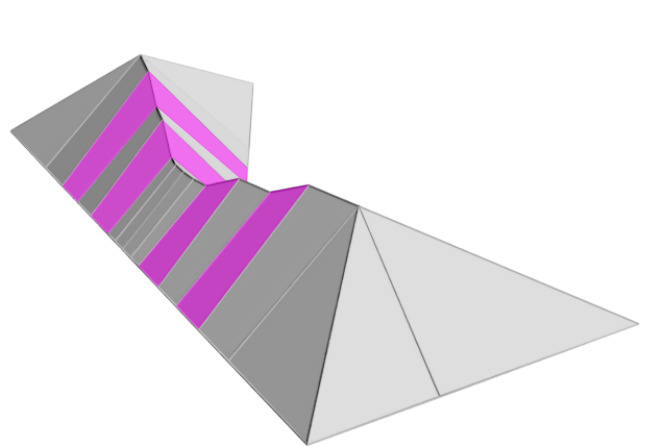}

\vspace{\tempheight}

\includegraphics[width=\figheight]{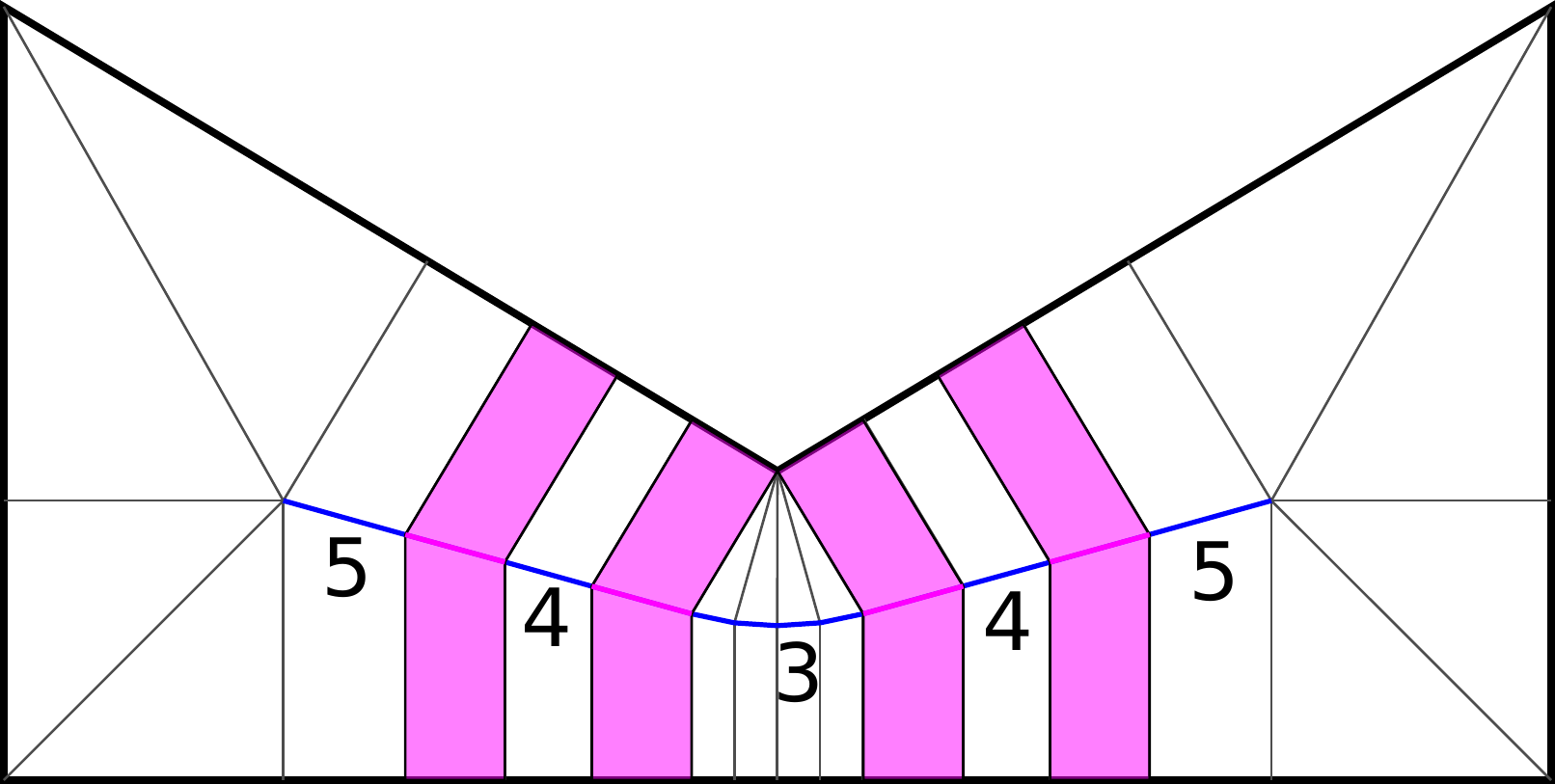}
\caption{Smoothing}\label{3d_surface_overview_smoothed}
\end{subfigure}
\begin{subfigure}{\figwidth}\centering
\hspace*{\tempheightTwo}
\includegraphics[height=\figheight]{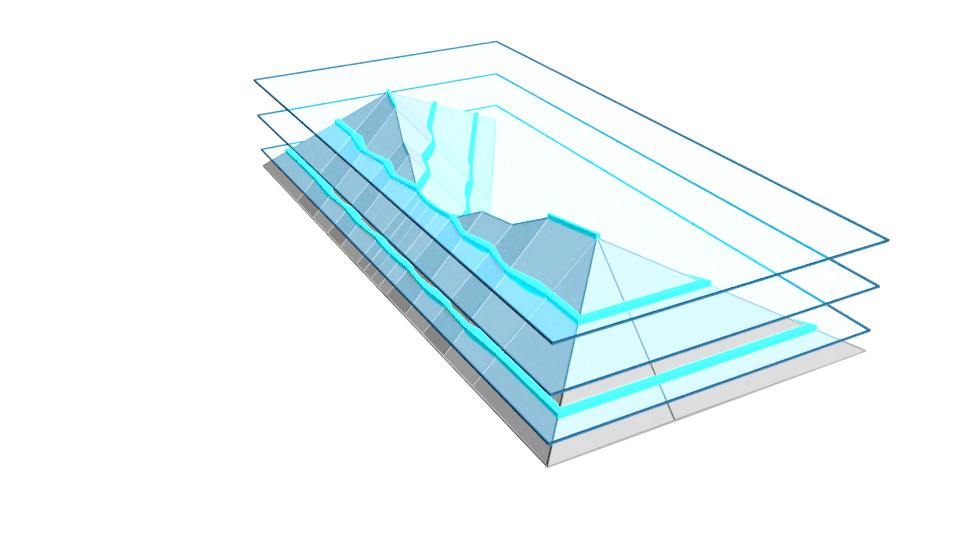}

\vspace{\tempheight}

\includegraphics[width=\figheight]{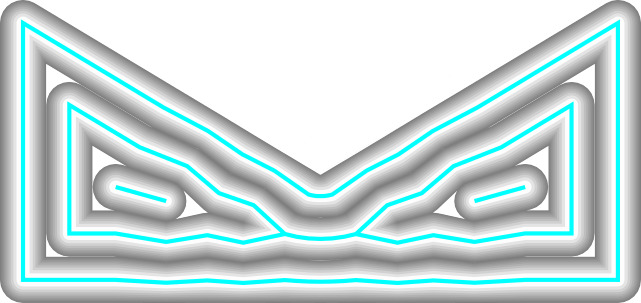}
\caption{Adaptive width paths}\label{3d_surface_overview_sliced}
\end{subfigure}
\caption{
The first row illustrates the generation of uniform paths \subref{overview_uniform_paths}
by interpreting the path as the intersection between horizontal planes and the union of cones \subref{overview_uoc}, which is an alternative visualization of the skeleton \subref{overview_skeleton}. 
The second row depicts the stages with both 2D and 3D visualizations for generating paths with adaptive width \subref{3d_surface_overview_sliced}.
Central elements in the skeleton are first identified (blue in \subref{3d_surface_overview_center}).
The heights are then quantized in terms of number of beads (the integer values in \subref{3d_surface_overview_rounded}), and smoothed \subref{3d_surface_overview_smoothed}.
}
\label{3d_surface_overview}
\end{figure*}

For ease of reference we have included a legend showing the terms employed in this manuscript in \cref{legend}.
These terms will be further explained as they first appear throughout this paper.

\section{Method}
\subsection{Overview}

Given arbitrarily shaped polygons which represent the 2D outline of a layer of a 3D model, our method generates extrusion toolpaths with varying width, i.e. a set of polylines where each site consists of a location and an extrusion width;
in between the sites we linearly interpolate the position and extrusion width.

Our method starts with computing \revise{the skeleton of the input polygon}{a graph which represents the topology of the input polygon: its \emph{skeleton}. 
Our skeleton is} based on the medial axis transform (MAT), a strategy that has been commonly used for generating contour-parallel toolpaths~\cite{eiamsa2003toward}.
We visualize the skeleton as the union of cones (UoC) (\cref{sec_surface_construction}), by raising each point in the domain to a height that equals the shortest distance from the point to the polygon boundary (\cref{overview_uoc}).
Contour-parallel toolpath\revise{}{s} with uniform width can be interpreted as the intersection of the union of cones with a set of \revise{planar}{horizontal} planes at equally spaced heights (\cref{overview_outline,overview_skeleton,overview_uoc,overview_uniform_sliced,overview_uniform_paths}).

As depicted in \cref{3d_surface_overview_center}, our method first identifies edges and nodes of the skeleton in the center of the polygon, which correspond to ridges and peaks in the UoC  (\cref{sec_center_classification}).
The heights $\tilde{b}$ at these elements are then quantized to an integer number of beads $\bar{b}$.
To ensure a smooth toolpath between regions with quantized integer heights that differ, we add new nodes in the skeleton with quantized heights and interpolate the heights $\hat{b}$ in between (\cref{sec_central_height_adjustment}).
The union of cones corresponding to the smoothed skeleton is then sliced at regular intervals to obtain toolpaths with varying spacing, which translates into varying width (\cref{sec_toolpath_extraction}).
\revise{}{%afs
The video in the supplementary material provides an example animation of this approach.
}

% We describe how to construct the mesh of the UoC in \cref{sec_surface_construction} and
% explain how to identify central regions in \cref{sec_center_classification}.
% Quantizing and smoothing the heights in the central mesh regions are handled in \cref{sec_central_height_adjustment}.
% We then describe how to deal with heights in the periphery in \cref{sec_peripheral_height_adjustment}
% and describe the toolpath extraction algorithm in \cref{sec_toolpath_extraction}.

This section explains how we generate toolpaths using our framework with uniform bead widths and evenly distributed locations between the center of the polygon and the outline.
In \cref{sec_generalization} we describe how to apply the framework to different beading schemes and we show several such beading schemes.

\subsection{Union of cones}\label{sec_surface_construction}

The union of cones (UoC) is derived from a common skeletonization of the polygonal outline shape: the medial axis.
By assigning each node in the skeleton a height equal to its shortest distance to the outline we obtain the shape of the UoC.
Starting from the medial axis we further decompose the shape into simple fragments, so that the domain contains only quads and triangles.
This decomposition constitutes an approximation of the UoC.

\begin{figure}\centering
\setlength{\figwidth}{0.24\columnwidth}
\setlength{\figwidthTwo}{0.3\columnwidth}
\begin{subfigure}[t]{\figwidth}\centering
\includegraphics[height=\figwidthTwo]{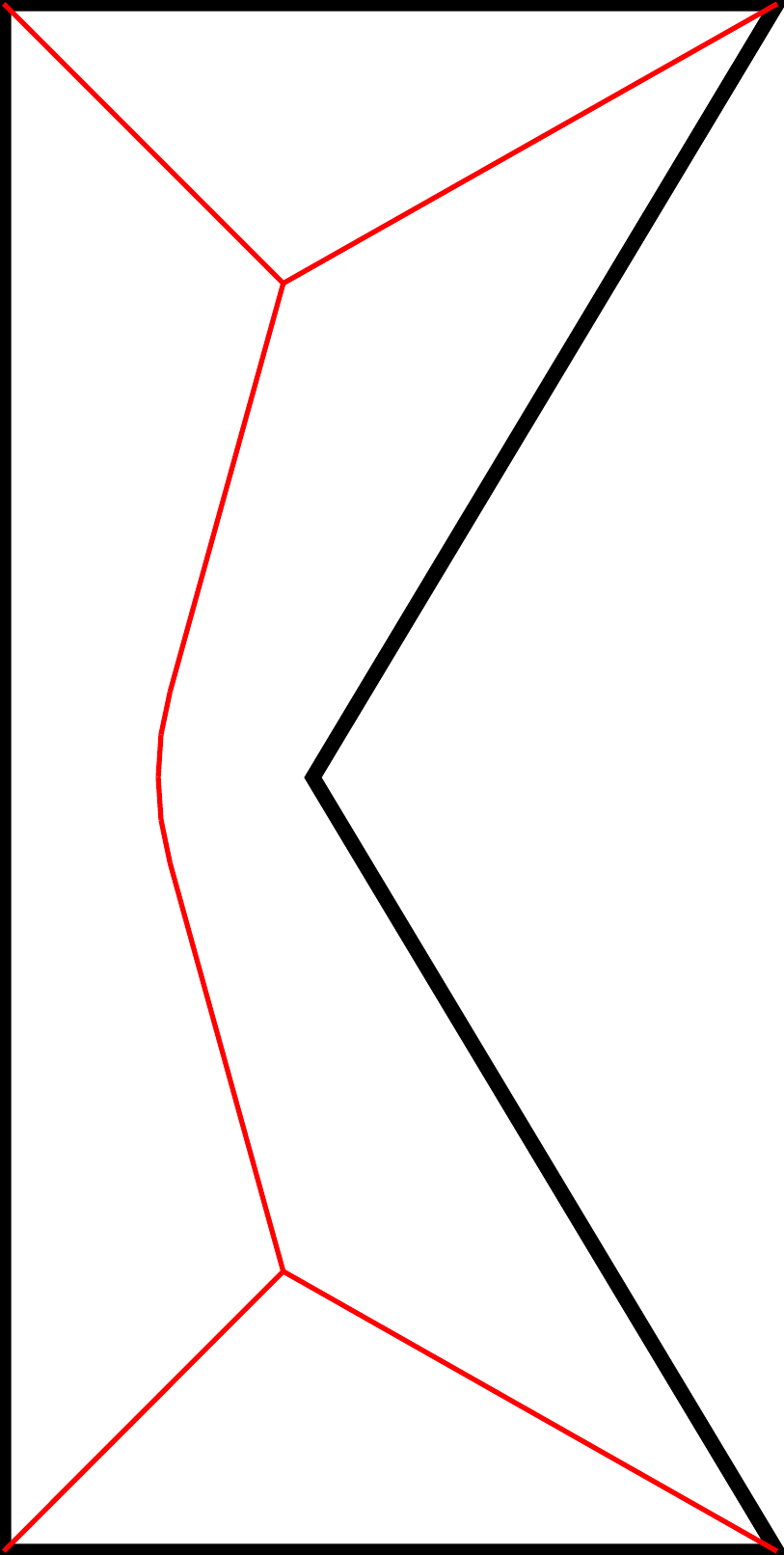}
\caption{Medial Axis}\label{shape_decomposition_mat}
\end{subfigure}
\begin{subfigure}[t]{\figwidth}\centering
\includegraphics[height=\figwidthTwo]{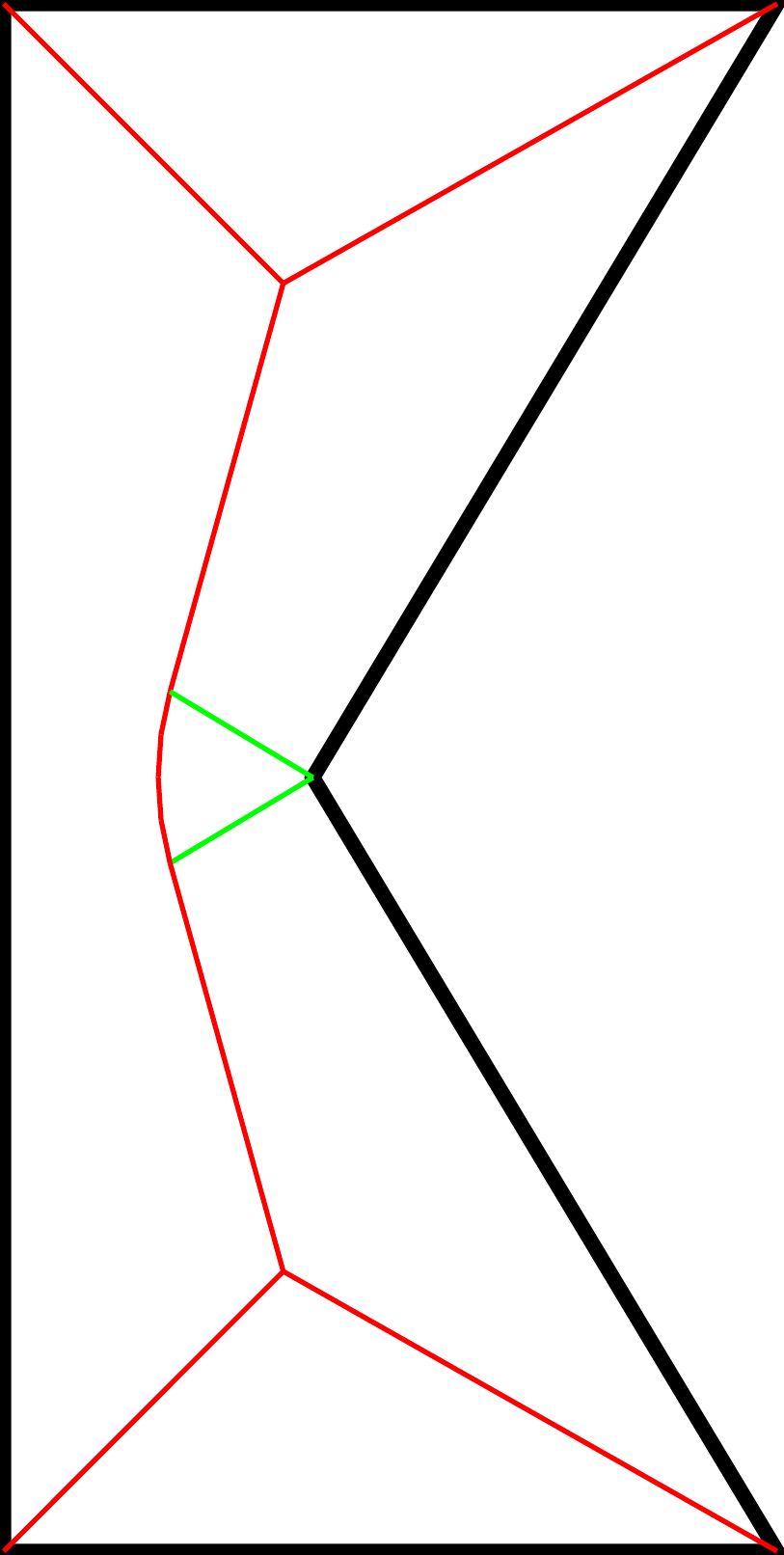}
\caption{Voronoi Diagram}\label{shape_decomposition_vd}
\end{subfigure}
\begin{subfigure}[t]{\figwidth}\centering
\includegraphics[height=\figwidthTwo]{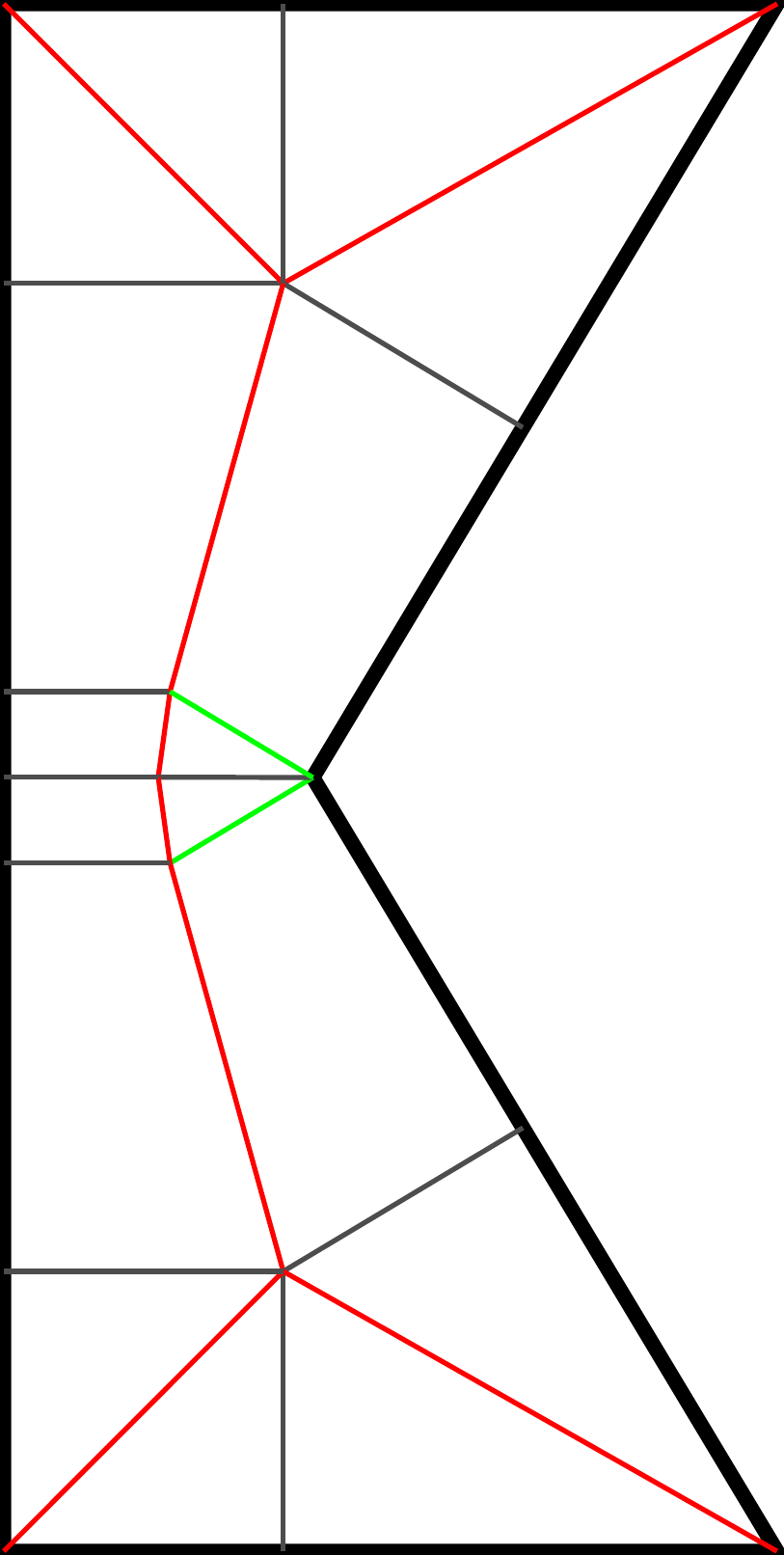}
\caption{Skeletal Trapezoidation}\label{shape_decomposition_st}
\end{subfigure}
\begin{subfigure}[t]{\figwidth}\centering
\includegraphics[height=\figwidthTwo]{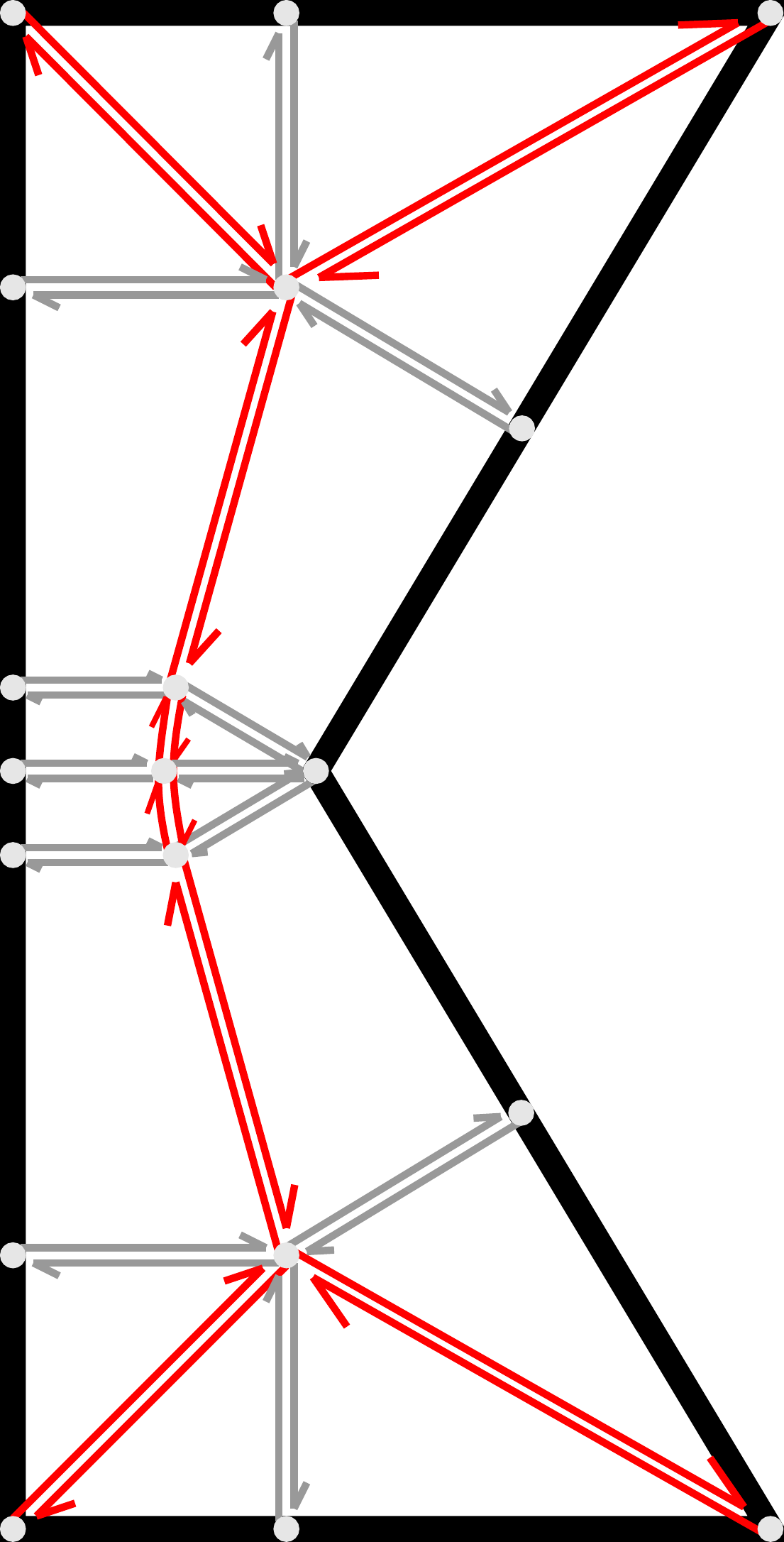}
\caption{Data structure}\label{shape_decomposition_datastructure}
\end{subfigure}
\caption{
Skeletonization of an outline shape (black).
Relation between the medial axis (red), the limited Voronoi Diagram (red and green) and the Skeletal Trapezoidation (red, green and gray): MAT $\subset$ Limited VD $\subset$ ST.
\subref{shape_decomposition_datastructure} The skeleton is represented using a half-edge data-structure.
}
\label{skeletonization_comparison}
\end{figure}

%\subsubsection{Medial axis transform}
\paragraph{Medial axis transform}
The medial axis is a \revise{compact and complete representation of}{representation commonly used to analyze} a shape.
It is defined as the set of positions where the inscribed circle meets the boundary in at least two locations~\cite{blum1967transformation,lee1982medial}.
The resulting skeleton consists of straight edges and parabolic edges.
An example is illustrated in \cref{shape_decomposition_mat}.
We call the set of points on the outline polygon $P$ closest to a skeletal point $v$ its \emph{support}:
\begin{equation}
    \text{sup}(v) = \argmin_{x\in P} |x - v|.
\end{equation}
The shortest distance for a point on the skeleton is called its feature radius, $R(v)$.
%See \cref{MAT_explanation_circles}.
%We therefore represent the skeleton in a standard half-edge data-structure.
%See \cref{shape_decomposition_datastructure}.
%Alternatively the medial axis can be defined as the locations where the UoC surface is discontinuous~\cite{blum1967transformation}
%The medial axis can therefore be seen as the space within which contour following toolpaths are generated.
%See \cref{MAT_explanation}.
The medial axis along with the feature radius values along the skeleton form a complete shape descriptor, known as the medial axis transform (MAT).
%Feature radius and node locations will be used below to analyse the shape locally.

By vertically raising the center of an inscribed circle to a height that equals the center's feature radius, a cone is formed. 
The union of all such cones forms a 3D solid volume. 
The medial axis can thus also be interpreted as ribs of the surface of the union of cones~\cite{blum1967transformation}.

%\subsubsection{Skeletal trapezoidation}
\paragraph{Skeletal trapezoidation}
Starting from the medial axis we decompose the input polygon into a set of quads and triangles, so that we can perform the slicing stage on simple shapes.
We employ a shape decomposition similar to the one proposed by \citeauthor{Ding2016a}~\cite{Ding2016a}. 
The basic idea is to add edges connecting each node $v$ on the medial axis to each of its support points $\text{sup}(v)$. 
The resulting skeleton decomposes the outline shape into trapezoids and triangles.
Considering the fact that the concept of trapezoidation conventionally allows for the degenerate case where a trapezoid resolves into a triangle~\cite{chazelle1984,fournier1984}, we call this shape decomposition the \emph{Skeletal Trapezoidation} (ST).

\begin{figure}\centering
\includegraphics[width=.75\columnwidth]{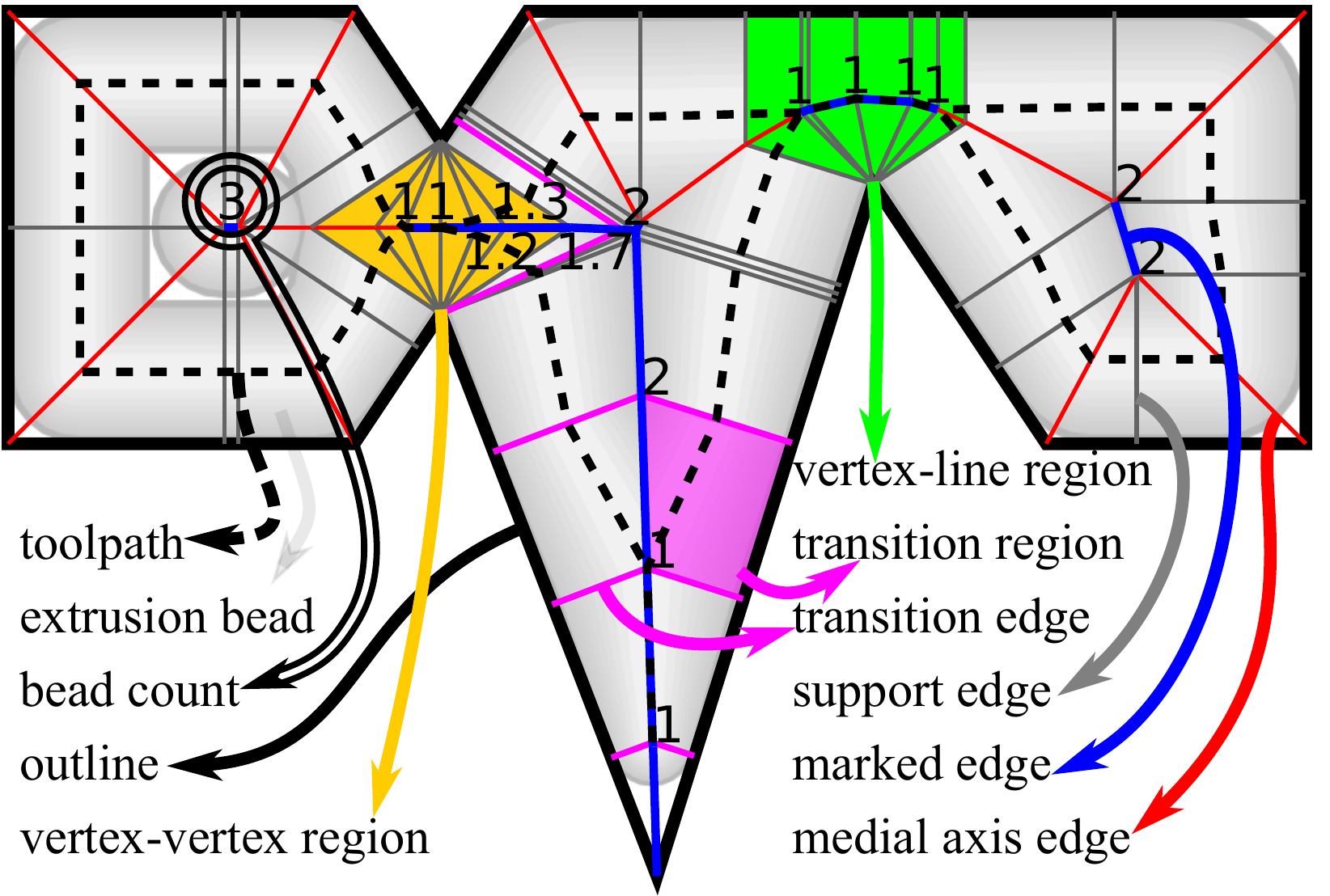}
\caption{Illustrative explanation of terms and color coding that are consistently used in this paper.}
\label{legend}
\end{figure}

The edges generated by the MAT are classified into three types:
\begin{enumerate*}
\item line-line edge -- straight edge generated from two line segments in the outline polygons,
\item vertex-line edge -- parabolic edge resulting from an outline vertex and a line segment in the outline, and 
\item vertex-vertex edge -- straight edge resulting from two outline vertices.
\end{enumerate*}
The vertex-line and vertex-vertex edges are discretized into pieces with a length up to \revise{$d^\text{discretization}$}{\SI{0.2}{\milli\meter}, which gives an approximation error of only $\pm$\SI{0.01}{\milli\meter}}.
This allows to approximate the feature radius between two discretized nodes $v_0$ and $v_1$ by linear interpolation. 
Again we connect the newly inserted nodes to their support, which results in vertex-line regions and vertex-vertex regions such as depicted in \cref{legend}.

% The ST contains all edges from the medial axis which are generated from two line segments in the outline polygons.
% The interaction between an outline vertex and a line segment results in a parabolic edge, which is discretized into pieces with a length up to $d^\text{discretization}$ (see the middle of \cref{shape_decomposition_st}).
% Additionally medial axis edges which are generated from two outline vertices are also discretized, so that the feature radius $R$ in between two discretized nodes $v_0$ and $v_1$ can be approximated by linear interpolation between $R(v_0)$ and $R(v_1)$.
% Again we connect the nodes which are introduced by the discretization to their support, which results in vertex-line regions and vertex-vertex regions such as depicted in \cref{legend}.

%\subsubsection{Union of cones}
\paragraph{Approximation of union of cones}
The skeletal trapezoidation (ST) provides a means to visualize the union of cones (UoC) approximated by a 3D surface mesh composed of quadrilateral and triangular patches.
%In order to visualize the ST as a 3D surface, we add a third dimension.
We assign each node in ST a (real number) height value measured in terms of beads, referred to as the \emph{bead count} $b$.
We define the bead count as the number of beads to fit along the \emph{diameter} of the inscribed circle centered at node $v$, i.e. $2R(v)$, by
\begin{equation}
    \tilde{b}_v = 2 R(v) / w^*
\label{eq:initialBeadCount}
\end{equation}
where $w^*$ is the nozzle size. 
%While the feature radii are measured along the support edges, the diameter is not measurable as such because the support edges are generally not collinear;
We divide the diameter rather than the radius as this allows to deal with an odd number of beads while using integer logic.
%The result is a mesh representing the union of cones (UoC) consisting entirely of quads and triangles.
Note that although the overview of the method was described geometrically in terms of the UoC, the actual toolpath generation relies on the two-dimensional ST;
the use of the bead count as a height value is only a visualization aid.

\paragraph{Implementation}
The medial axis of a polygonal shape is a subset of the Voronoi Diagram generated from the line segments and vertices of the shape~\cite{lee1982medial}. 
The edges of the Voronoi diagram that fall outside of the outline shape are irrelevant for our purpose and \revise{}{are }thus discarded.
Note that besides the full medial axis, the Voronoi diagram also contains edges connecting to concave vertices in the outline shape (see \cref{shape_decomposition_vd}). 
These extra edges are a subset of the edges connecting a node to its support, so we keep them in.
From the Voronoi diagram we add nodes to discretize parabolic edges and edges formed by two concave outline vertices, and then connect all nodes to their supports, forming a skeletal trapezoidation. 
We then assign each node the bead count values using \cref{eq:initialBeadCount}.
We compute the Voronoi diagram using the Boost \verb!C++! libraries~\cite{schaling2011boost}, which implements the algorithm proposed by \citeauthor{fortune1986sascg}~\cite{fortune1986sascg}.
A half-edge data-structure is used to represent the Voronoi diagram (\cref{shape_decomposition_datastructure}).

\subsection{Center classification}\label{sec_center_classification}
%The simple technique of performing uniform width offset produces large overfill and underfill areas on central regions of the ST.
%These central regions present themselves as upper ridges in the mountains of the UoC surface.
%In order to prevent the over- and underfill we will change the bead counts in the central regions.
%This sections covers what parts of the UoC will be marked as being central.
In order to prevent over- and underfill \revise{to occur}{from occurring} in central\revise{}{ regions}, \revise{we mark }{}parts of the ST \revise{}{are marked }as being central.
\revise{The remainder is called peripheral.}{}
Our framework will decide on a beading at all the marked nodes in `the center' and apply the beading outward to the unmarked nodes (\cref{sec_peripheral_height_adjustment}).

\revise{We mark a}{A} node in the ST \revise{}{is marked as }as central if its feature radius is larger than that of all its neighboring nodes, i.e. a local maxima.
%That is: we mark the local maxima, i.e. the mountain tops of the UoC as being central.
\revise{We also mark an}{An} edge \revise{}{and its two nodes are also marked }as being central if it is significant according to a significance measure.

\begin{figure}
\centering
\setlength{\figheight}{.3\columnwidth}
\begin{subfigure}{0.45\columnwidth} \centering
\includegraphics[height=\figheight,frame]{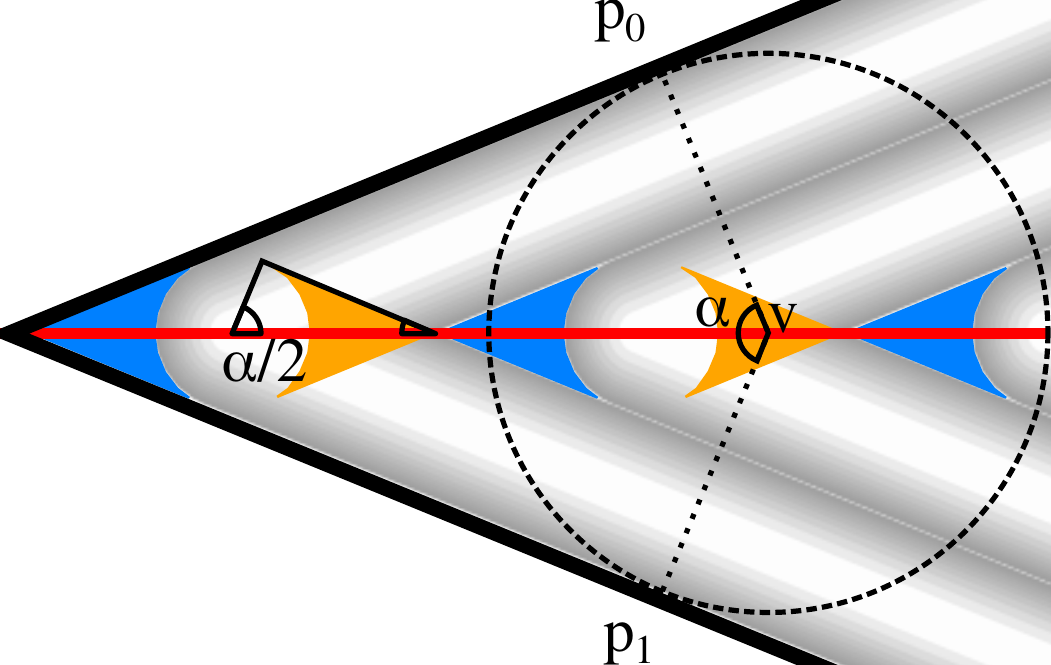}
\caption{Over- and underfill}\label{naive_overfill_underfill}
\end{subfigure}
\begin{subfigure}{0.45\columnwidth} \centering
\includegraphics[height=.8\figheight]{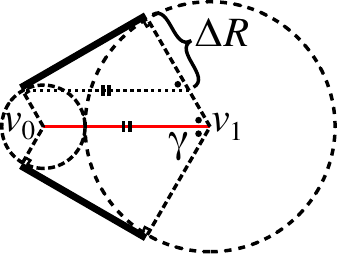}
\caption{Significance measure}\label{distance_based_angles}
\end{subfigure}
\caption{
Properties of the significance measure along a skeletal edge (red) generated from two polygon lines (black) using the properties of inscribed circles (gray) and their radii (dashed).
\subref{naive_overfill_underfill}
The size of overfill (orange) and underfill areas (azure) for the uniform toolpathing technique can be calculated from the bisector angle.
% $a = 180\degree - \beta$ is the angle between a location $v$ and its support $p_0$ and $p_1$.
\subref{distance_based_angles}
The significance measure can be simplified using $\alpha = 2 \gamma = 2 \cos^{-1} \Delta R / |v_1 - v_0|$.
}
\end{figure}

%\subsubsection{Significance measure}\label{sec:significance_measure}
\paragraph{Significance measure}
We make use of the \emph{bisector angle} as an indicator of significance which is commonly used in shape analysis.
%Centrality can be formalized by looking at a commonly used significance measure knows as the \emph{bisector angle}.
The bisector angle $\alpha$ is the interior angle $\angle{p_0lp_1} \leq \SI{180}{\degree}$, between any location $l$ on an edge of the ST and its two supporting points $ \{ p_0, p_1 \} = \text{sup}(l)$~\cite{attali1996modeling}. 
An edge is significant if the bisector angle on any location on the edge exceeds a prescribed $\alpha_\text{max}$. 
As illustrated in \revise{See }{}\cref{naive_overfill_underfill}, for a polygon with a pointy wedge area of an angle $\beta$, we have $\alpha = 180\degree - \beta$.
This corresponds to overfill areas and underfill areas the size of \revise{$\nicefrac12 (w^*)^2 \left( \nicefrac14 \tan ( \alpha / 2) - \alpha / 2 \right)$}{$\nicefrac14 (w^*)^2 \left( \tan ( \alpha / 2) - \alpha / 2 \right)$} when filled using the simple technique of uniform bead width $w^*$.
\revise{}{A too large $\alpha_\text{max}$ may leave a lot of under-/overfill, while a too small value may introduce toolpaths to fill in negligibly small underfills.
We therefore set $\alpha_\text{max} = \SI{135}{\degree}$.}
Although significance measures are commonly used as a heuristic for finding the parts of a skeleton which are in some sense relevant~\cite{attali1996modeling,Sud2007},
we use the bisector angle as an \emph{exact} indicator of the amount of overfill and underfill in the uniform toolpaths of constant width.
%Contrary to related literature we will \emph{keep} the non-significant regions of the skeleton.
%We mark all significant edges as such and contrary to related literature we keep the unmarked edges of the ST.

% The significance evaluation is performed efficiently by checking the ratio between feature radius $R$ and the Euclidean distance:
% if $ | R(v_1) - R(v_0) | / |v_1 - v_0| >  \cos(\alpha_\text{max} / 2)$ then $\alpha > \alpha_\text{max}$. See \cref{distance_based_angles}. This ratio has a clear geometrical interpretation as the slope of the ridge in the UoC surface.

To avoid evaluating the bisector angle at any location on all edges, we devise an efficient measure which operates only on the two nodes of an edge.
Because all locations along a line-line edge have the same bisector angle we can evaluate whether the edge is significant by checking whether 
\begin{align}\label{simple_significance_measure}
| R(v_1) - R(v_0) | / |v_1 - v_0| >  \cos(\alpha_\text{max} / 2)
\end{align}
(see \cref{distance_based_angles}).
This ratio has a clear geometrical interpretation as the slope of the ridge in the UoC surface.
For vertex-line edges and vertex-vertex edges only a portion of the edge is significant.
We therefore introduce nodes at the boundaries of the significant portion during the discretization of such edges (see Appendix~\ref{edge_discretization}).
The significance of all edges can then accurately be evaluated using \cref{simple_significance_measure}.

%\subsubsection{Marking filtering}
\paragraph{Marking filtering}
After initializing the marking at all edges and nodes, we filter out high frequency changes in the marking in order to ensure that the generated toolpath is smooth. 
The filtering is performed by additionally marking some unmarked elements, rather than the opposite since unmarking central regions reintroduce\revise{}{s} large over- and underfill areas.
From each marked node $v_0$ with an upward unmarked edge attached we walk along the upward edges;
if the total length traversed until we reach another marked node $v_1$ is shorter than some filter distance $d_\text{max}^\text{unmarked}$, we mark all edges encountered as being central.
\revise{}{We use $d_\text{max}^\text{unmarked} = w^*$ in order to filter out high frequency oscillations in the order of magnitude of the nozzle size, while keeping close to the significance measure.}

\subsection{Central height adjustment}\label{sec_central_height_adjustment}
\revise{Now that we have identified and marked}{After }the central regions\revise{}{ have been identified,} \revise{we quantize }{}their heights\revise{}{ are quantized}.
\revise{We first quantize}{First,} the initial bead count $\tilde{b}$ \revise{}{is quantized }into an integer bead count $\bar{b}$ at the marked nodes using a quantization operator $q$,
then \revise{find }{}the locations along the edges where $q$ makes a jump from one bead count $n$ to another $n+1$\revise{}{ are identified}
and then \revise{introduce }{}ramps \revise{}{are introduced }to smoothly transition from $n$ to $n+1$ using fractional bead counts $\hat{b}$ along the smooth transition.
%\subsubsection{Initial bead count}\label{sec_initial_bead_count}

\begin{figure*}
\centering
\setlength{\figwidth}{0.13\textwidth}
\begin{subfigure}{\figwidth}
\includegraphics[width=\columnwidth]{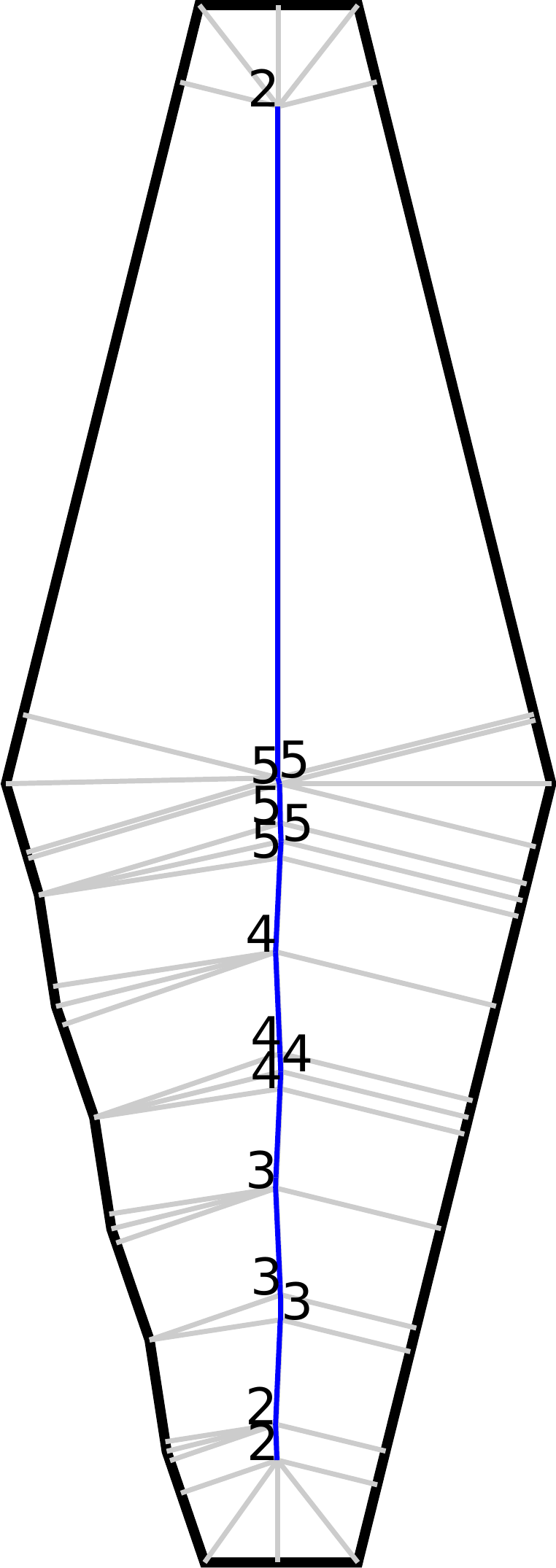}
\caption{Quantized bead counts}\label{beading_transitioning_filtering__bead_count}
\end{subfigure}
\begin{subfigure}{\figwidth}
\includegraphics[width=\columnwidth]{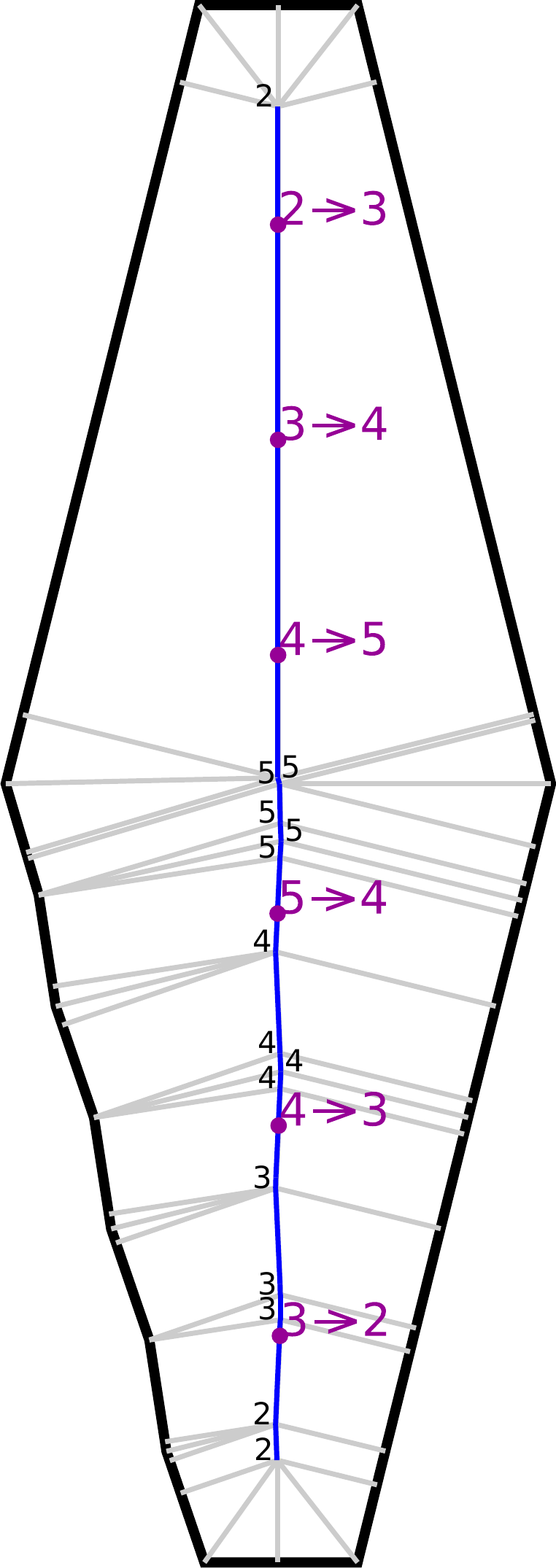}
\caption{Transition anchors}\label{beading_transitioning_filtering__transition_mids}
\end{subfigure}
\begin{subfigure}{\figwidth}
\includegraphics[width=\columnwidth]{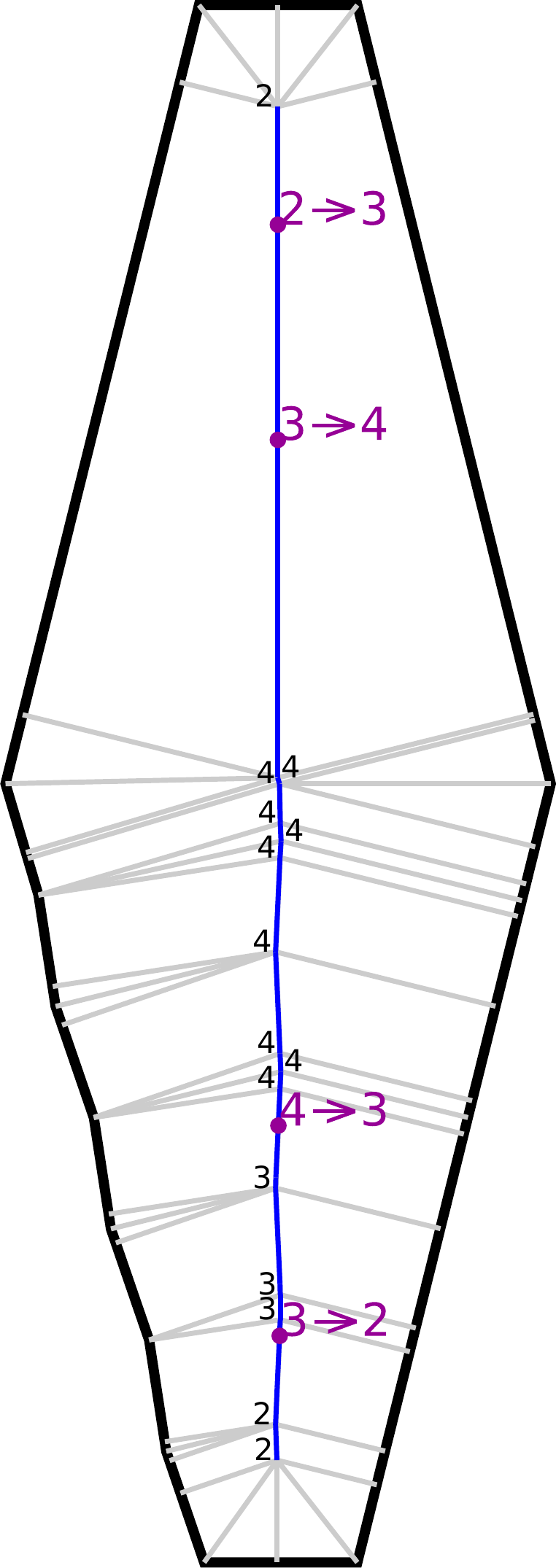}
\caption{Filtered anchors}\label{beading_transitioning_filtering__filtered}
\end{subfigure}
\begin{subfigure}{\figwidth}
\includegraphics[width=\columnwidth]{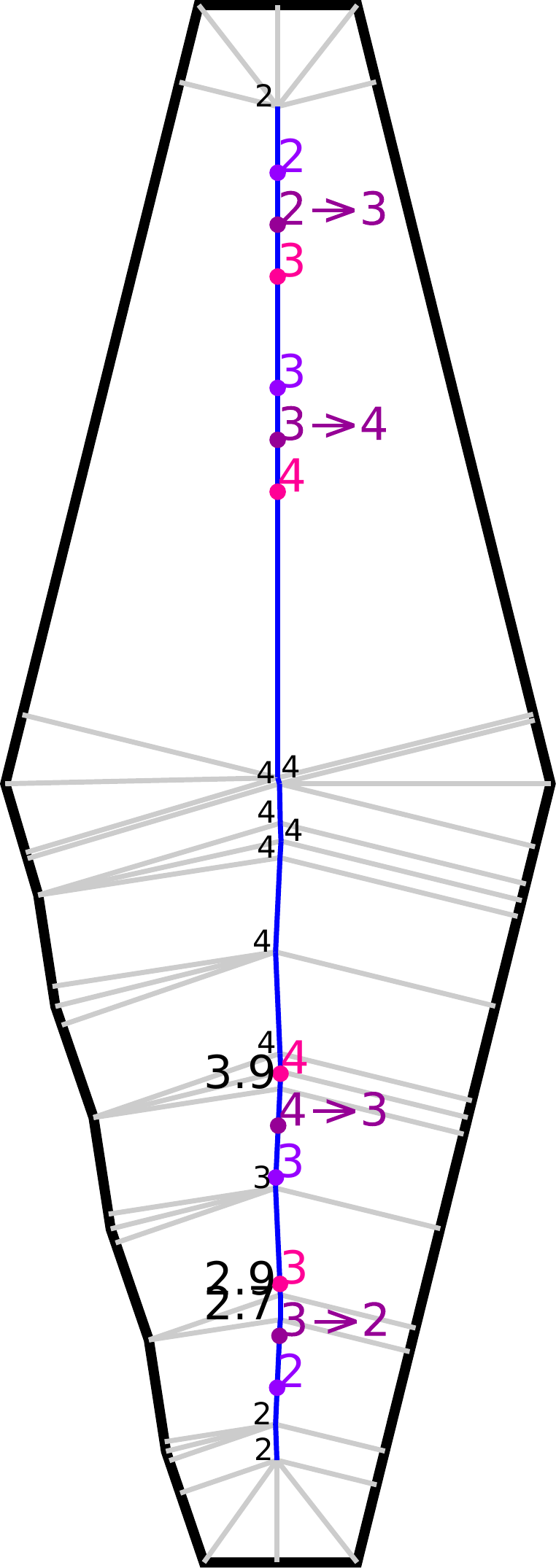}
\caption{Transition ramp ends}\label{beading_transitioning_filtering__transition_ends}
\end{subfigure}
\begin{subfigure}{\figwidth}
\includegraphics[width=\columnwidth]{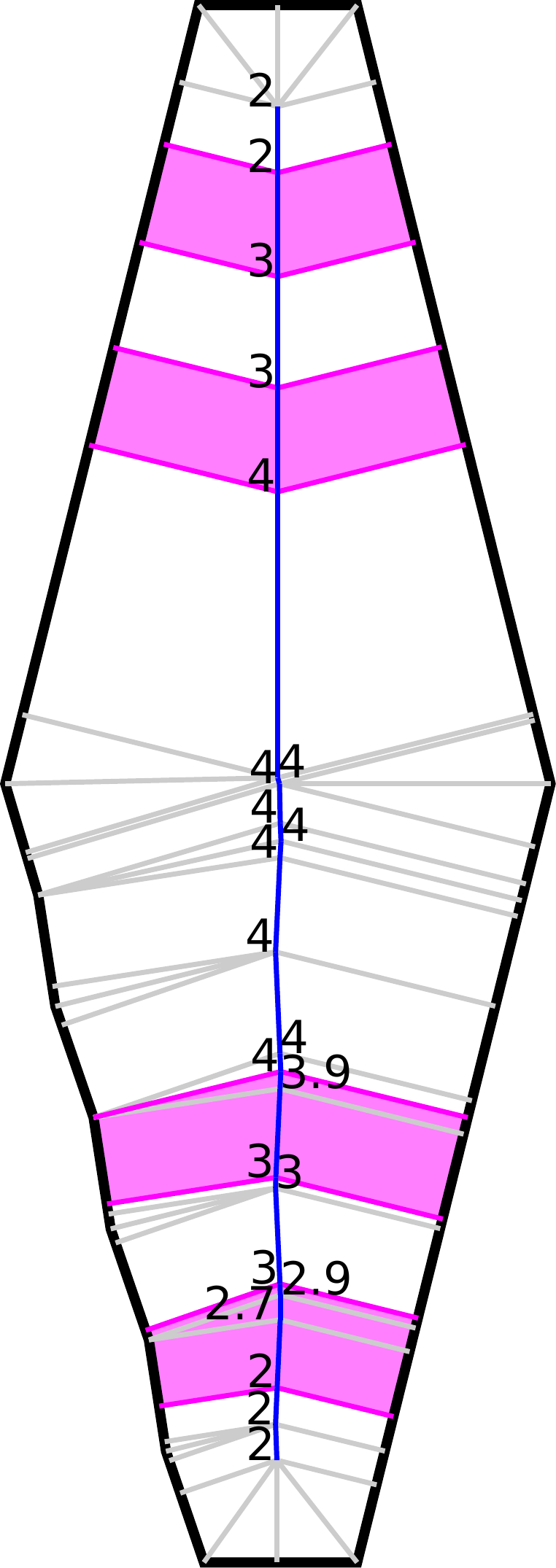}
\caption{Radial support edges}\label{beading_transitioning_filtering__transitions_applied}
\end{subfigure}
\begin{subfigure}{.27\textwidth}
\hspace*{-.5cm}\includegraphics[width=1.2\columnwidth]{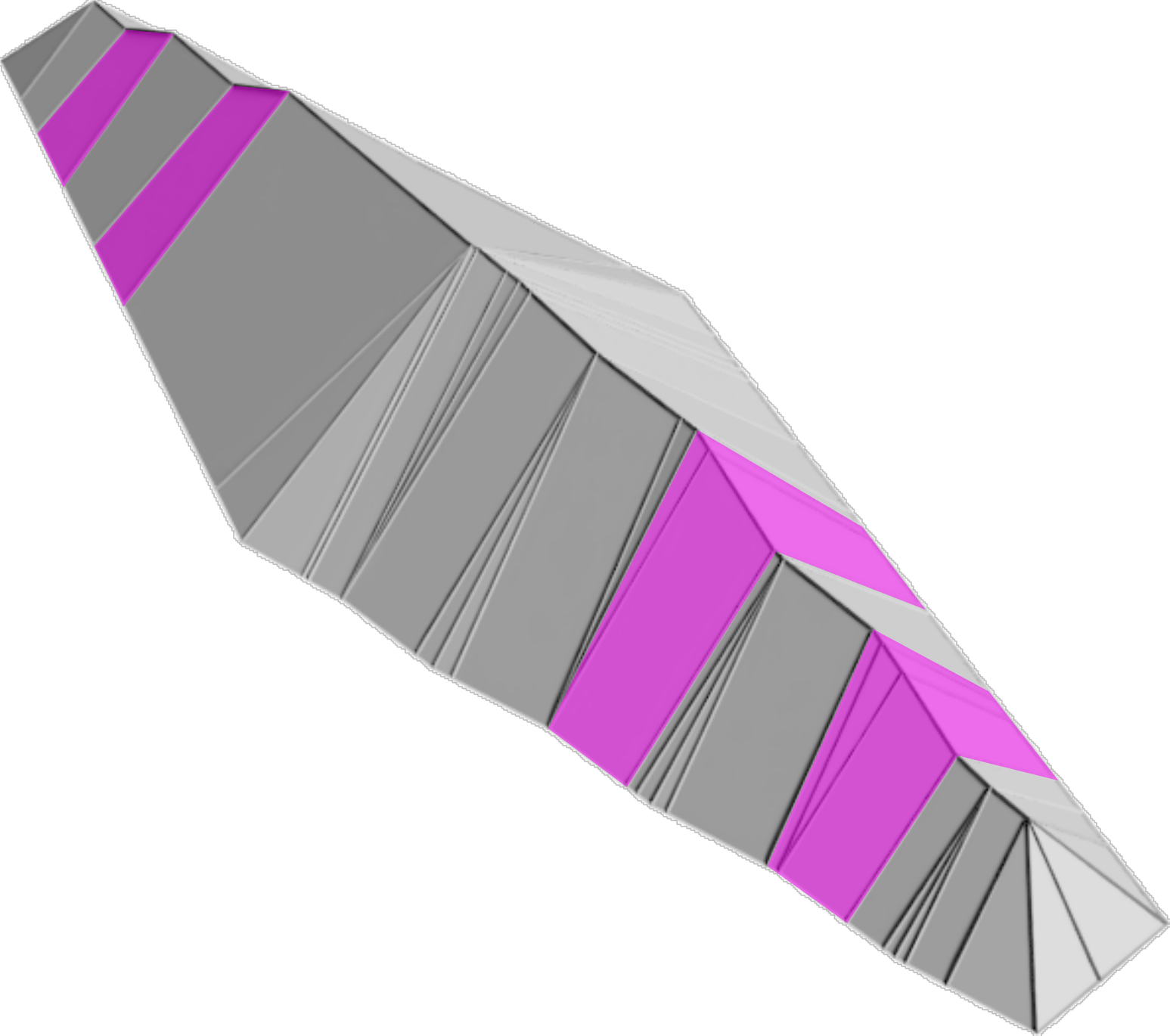}
\caption{Result UoC}\label{beading_transitioning_filtering__result_uoc}
\end{subfigure}
\caption{
Applying bead counts and transitioning on a shape showing the difference between a simple ST (top) and a mirrored version with small perturbations in the outline (bottom).
Outline in black, central edges marked in blue, radial edges in grey.
\subref{beading_transitioning_filtering__bead_count} First we initialize the bead counts (black) in the marked edges (blue).
\subref{beading_transitioning_filtering__transition_mids} We then extract the anchor locations (purple) where the bead count transitions.
\subref{beading_transitioning_filtering__filtered} We then filter out regions which exhibit frequent transition.
\subref{beading_transitioning_filtering__transition_ends} We then calculate the end locations (magenta, pink) of the transitions and modify the bead count at nodes in between to fractional values.
\subref{beading_transitioning_filtering__transitions_applied} Finally we introduce nodes at the ends and introduce radial edges (purple) as per the trapezoidation constraint.
The symmetry in the result shows that transitioning is robust against small perturbations in the outline shape.
}
\label{beading_transitioning_filtering}
\end{figure*}

%\subsubsection{Quantization}

\paragraph{Quantization}
We define a quantization operator $q$ to map a feature diameter ($d=2R(v)$) to a bead count: $q: \mathbb{R} \to \mathbb{N}$.
Because our quantization scheme should round to the nearest integer multiple of the nozzle size, we have
$q(d) = \left\lfloor d / w^* + \nicefrac12 \right\rfloor$.
Alternative quantization schemes are discussed in \cref{sec_generalization}.
By applying $q$ to the heights of central nodes we quantize the bead count:
\begin{equation}\label{eq_quantized_bead_count}
\bar{b}_v = q(2R(v)) = \left\lfloor \tilde{b}_v + \nicefrac12 \right\rfloor
\end{equation}

\iffalse
Along with this quantization, an amendment to the marking filtering is performed, in order to alleviate a problem that will be handled in \cref{section_beading_conflicts}.
Specifically, unmarked regions are marked as central if they lie between two nodes with the same integer bead count.
\fi
%We then filter out short unmarked regions where the bead count remains constant in order to limit the extent of a problem handled in \cref{section_beading_conflicts}.
% From each marked node $v_0$ with an upward unmarked edge attached we walk along the upward edges until we hit another marked node $v_1$.
% If the upper node has the same bead count $b^*_{v_1} = b^*_{v_0}$ we mark all edges and nodes $v$ in between, and set the bead count $b^*_v \leftarrow b^*_{v_0}$.

\paragraph{Transition anchors}
For a marked edge which connects nodes $v_0$ and $v_1$ with $\bar{b}_{v_0} \le n < \bar{b}_{v_1}$, we determine the \emph{transition anchor locations} at which the bead count transitions from $n$ to $n+1$.
To this end, we introduce the function 
\begin{equation}
    q^{-1}(n) := \argmax_d q(d) = n,
\end{equation}
which gives the feature diameter $d$ at which the bead count $q$ transitions from $n$ to $n+1$.
The location of the anchor $v_x$ is then computed by inversely interpolating $R(v_x) = q^{-1}(n)$, i.e. 
\begin{equation}
    v_x = v_0 + (v_1 - v_0) \frac{ q^{-1}(n) - R(v_0) }{ R(v_1) - R(v_0) }.
\end{equation}
%for each $n$ such that $b^*_{v_0}\le n<b^*_{v_1}$.
An illustration of the anchors is shown in \cref{beading_transitioning_filtering__transition_mids}.

We perform a filtering step to prevent frequently changing the bead count back and forth within a short distance.
For two consecutive anchors which transition to opposite directions, if the distance between them is smaller than some limit $d_\text{max}^\text{transition}$, the bead counts at all nodes in-between are set to the surrounding bead counts, and consequently these anchors are removed (See \cref{beading_transitioning_filtering__filtered}).
\revise{}{A value of $d_\text{max}^\text{transition} = \SI{1}{\milli\meter}$ seems to produce satisfactory results.}
\revise{}{This means that for some small regions we generate toolpaths with bead widths outside the typical range.}

% For each edge which contains a transition anchor we walk along the marked edges until we encounter another anchor.
% If the other anchor is a transition in the opposite direction and the traversed distance is within some limit $d_\text{max}^\text{transition}$ we remove the transitions and set the bead counts at all nodes in the filtered region to the surrounding bead counts.
% See \cref{beading_transitioning_filtering__filtered}.

% \subsubsection{Smooth transition}
\paragraph{Smooth transitions}

\begin{figure}
\centering
\setlength{\figwidth}{\columnwidth}
\begin{subfigure}{0.9\figwidth}\centering
\includegraphics[width=\columnwidth]{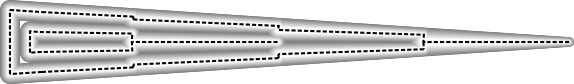}
\caption{Without transitioning}
\end{subfigure}
\begin{subfigure}{0.9\figwidth}\centering
\includegraphics[width=\columnwidth]{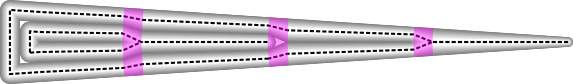}
\caption{Transitioning}
\end{subfigure}
\caption{
Sharp turns around regions where the bead count changes are prevented by transition regions (highlighted in cyan).
}
\label{transitions}
\end{figure}

A sharp transition from $n$ to $n+1$ beads at an anchor location creates sharp turns in the toolpath (see \cref{transitions} top).
We introduce a transition length $t(n)$ to ensure a smooth transition (see \cref{transitions}). 
The length of the transition is set to $t(n) = w^*$ and it is centered at the anchor, i.e. the distance from the lower end $v_0$ to the anchor position $v_x$ is set to
\revise{$t_0(n) = \Delta(v_0v_x) = \nicefrac12 w^*$,}
{$t_0(n) \equiv \Delta(v_0v_x) =  t(n) \left( q^{-1}(n) / w^*  - n \right)$,}
where $\Delta$ is the total distance along the edges between two nodes.
\revise{}{The transition length $t(n)$ ensures that the center beads don't overlap with the innermost transitioning beads, while keeping the amount of underfill low and the toolpath smooth.
The transition anchor position $t_0(n)$ ensures that the transitions never overlap with each other or with locations where all beads have the preferred width $w^*$.}

We discard any transition anchor which is too close to the end of a chain of marked edges for the smoothed transition to fully fit within the marked region.
In order to make the transition ramps robust against small perturbations in the outline shape which cause extra (support) edges in the skeleton,
we modify the nodes $v_x$ which are between the two ends $v_0$ and $v_1$ of the transition by (re-)assigning them a fractional bead count $\hat{b}$ which is linearly interpolated between the two ends of the transition (see \cref{beading_transitioning_filtering__transitions_applied}):
\begin{equation}
\hat{b}_{v_x} = n + {\Delta(v_0v_x)} / {\Delta(v_0v_1)}
\end{equation}
Note that although the ST is not stable against noise in the boundary shape, the distance field itself is, so by designing our algorithms such that they are stable against changes in the topology of the skeleton our method is stable against small perturbations in the outline.
%This modification makes the transition ramps robust against small perturbations in the outline shape (to be explained in \cref{section_beading_interpolation});
%compare the top and bottom of \cref{beading_transitioning_filtering}. 
Finally we update the ST by adding support edges at the transition ends. 
As shown in \cref{beading_transitioning_filtering__result_uoc}, the marked regions in the UoC mesh have become horizontal at integer multiples of $\nicefrac12 w^*$ for long stretches with ramps in between.

\subsection{Beading}\label{sec_peripheral_height_adjustment}
\begin{figure}
\centering
\setlength{\figheight}{.29\columnwidth}
\begin{subfigure}{0.4\columnwidth}\centering
\includegraphics[height=\figheight]{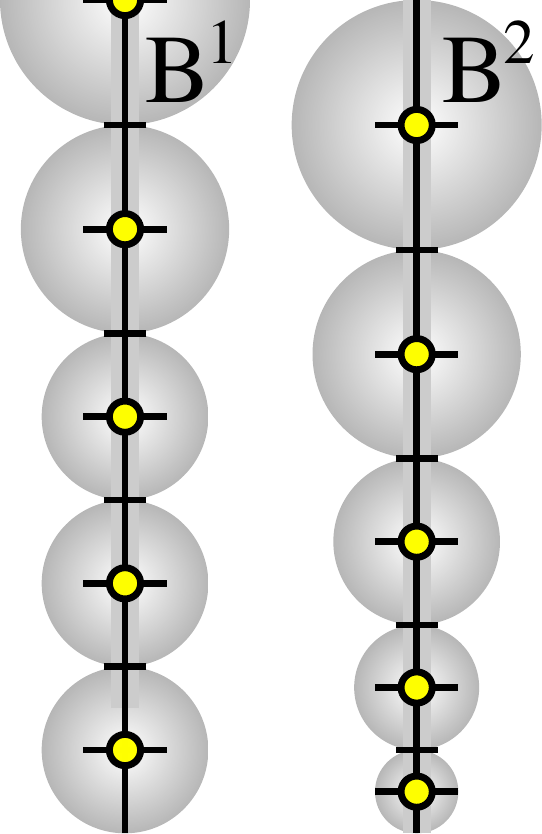}
\caption{Example beadings}\label{example_beading}
\end{subfigure}
\begin{subfigure}{0.4\columnwidth}\centering
\includegraphics[height=\figheight]{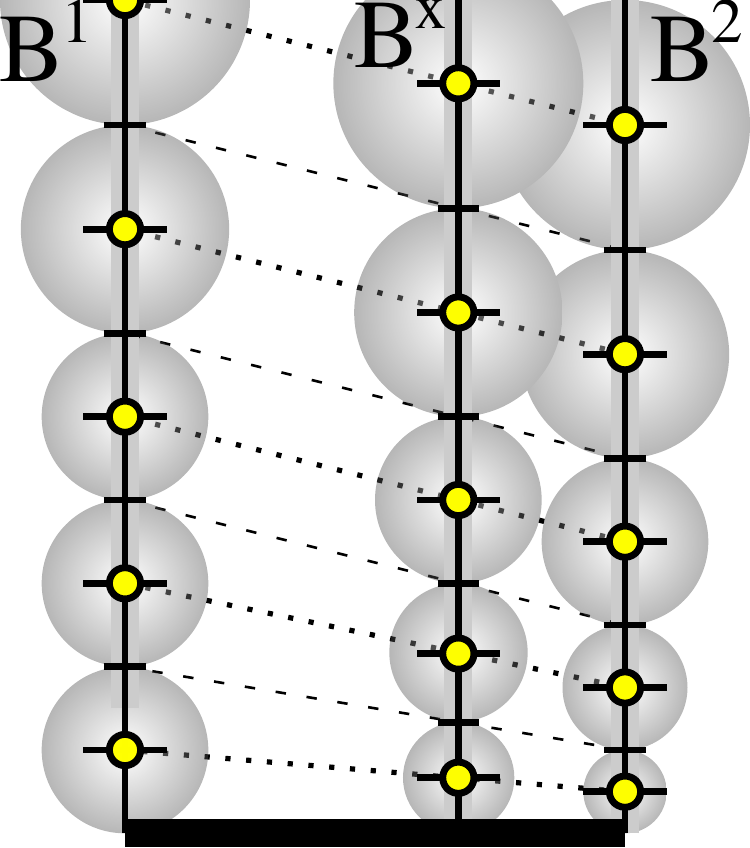}
\caption{Interpolation}
\end{subfigure}
\caption{
Interpolation between two beadings $B^1$ and $B^2$ with odd and even bead count resulting in a beading $B^x$ at $n+\nicefrac23$.
Bead indices are counted inward from the outline (thick black).
Interpolation of locations in dots, interpolation of widths in dashes.
}
\label{beading_interpolation}
\end{figure}

Now that we \revise{have determined}{know how to determine} the bead counts in the \revise{}{marked }central regions \revise{we will describe}{the question is} how the \revise{peripheral}{unmarked} regions are handled.
Determining bead count values for the unmarked nodes and interpolating linearly along the unmarked edges would mean that toolpath sites would be distributed evenly along each unmarked bone;
while that would suffice for the evenly distributed beading scheme, it wouldn't allow for more sophisticated, non-linear schemes.
Instead we determine the radial distance to the boundary at which each bead should occur from the boundary to the center.
Each central node is associated with a sequence of radial distances $L$ which control the locations of the beads, starting from the outer bead and ending in the center.
Together with a sequence of bead widths $W$\revise{}{,} these form what we call a \emph{beading} $B$.
For our distributed beading scheme we compute the beading for a central node $v$ with $n = \lfloor \hat{b}_v \rfloor$ beads and a diameter $r = R(v)$  as:
\begin{align*}
    B(n,r) &= \left( W(n,r), L(n,r) \right)   &=   \left( \left\{  w_0  \dots w_{\lceil n/2 \rceil-1} \right\}, \left\{ l_0 \dots l_{\lceil n/2 \rceil-1} \right\} \right) \\
    w_i &= r / n  & \text{for all $i \in \mathbb{N} : i < n / 2 $}\\
    l_i &= r / n (i + \nicefrac12) & \text{for all $i \in \mathbb{N} : i < n / 2 $}
\end{align*}
where
$w_i$ and $l_i$ are the width and location of the $i$th bead,
respectively, counting from the outline inward.
Example beadings for an odd and even bead count with arbitrary widths are visualized in \cref{example_beading}.

\begin{figure}\centering
\setlength{\figheight}{.2\columnwidth}
\begin{subfigure}{.45\columnwidth}\centering
\includegraphics[width=.95\columnwidth,frame]{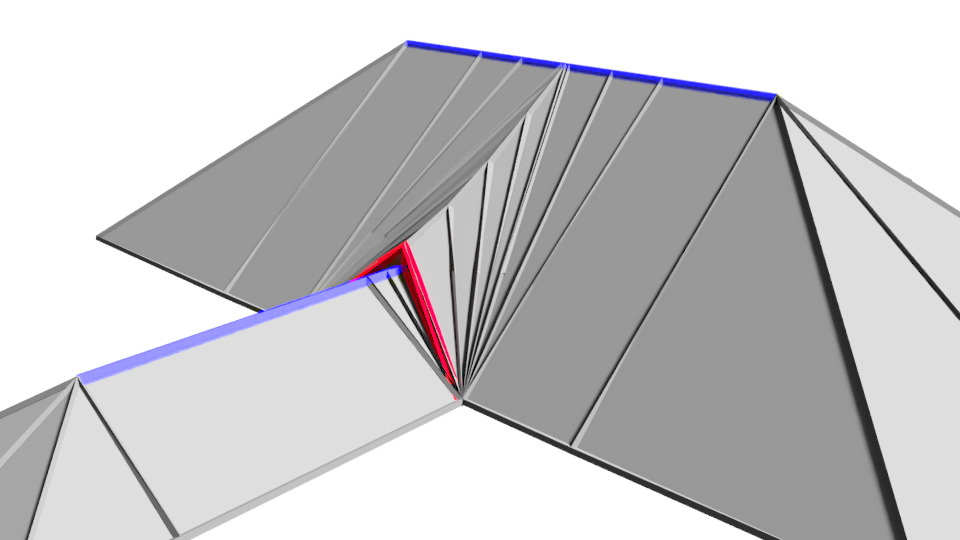}
\includegraphics[height=\figheight]{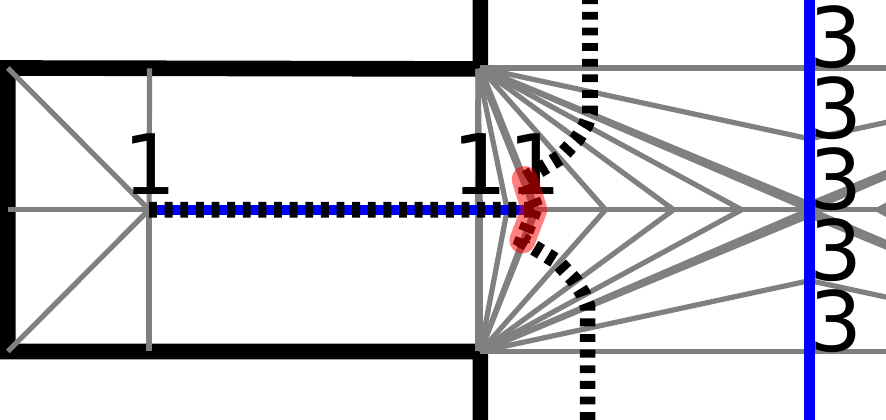}
\caption{Beading conflict}\label{beading_conflict}
\end{subfigure}
\begin{subfigure}{.45\columnwidth}\centering
\includegraphics[width=.95\columnwidth,frame]{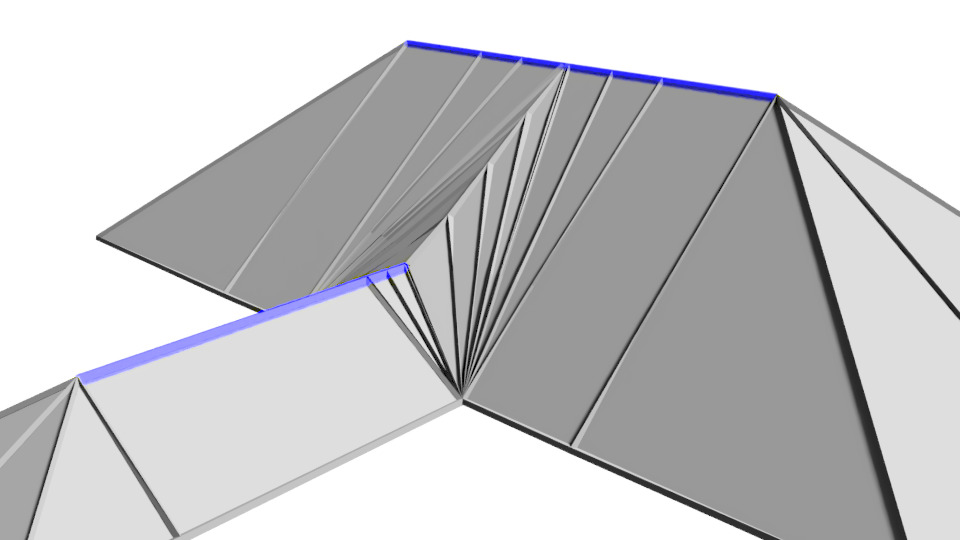}
\includegraphics[height=\figheight]{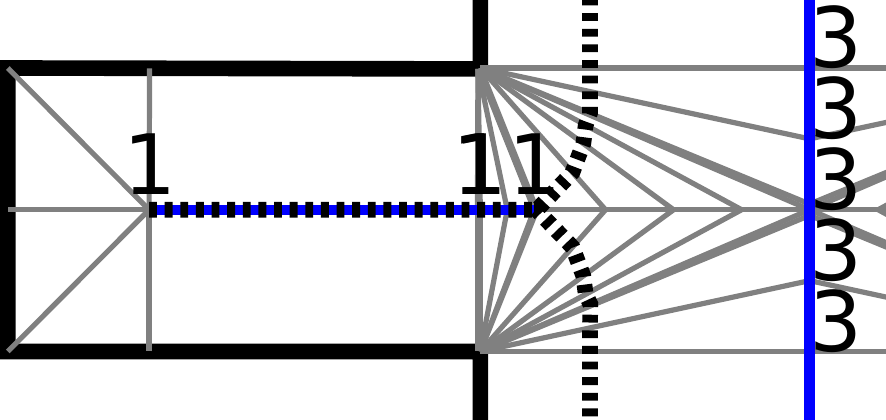}
\caption{Conflict resolution}\label{beading_conflict_solved}
\end{subfigure}
\caption{
\subref{beading_conflict} The beading propagated from above conflicts with the beading below.
\subref{beading_conflict_solved} The beading conflict is resolved by gradually interpolating between the two beadings.
The ramp to the upper ridge doesn't line up with the lower ridge, which means that
the toolpaths (dashed) resulting from the beading propagated from above doesn't align with the beading from the thin outline feature (highlighted in red).
}
\label{beading_conflict_problem}
\end{figure}

\paragraph{Beading interpolation}
The beading is defined in terms of an integer number of beads, while we have assigned a fractional bead count to nodes within a transition region.
In order to generate a beading for a node $v$ with $n < b^*_v < n+1 $ we linearly interpolate the bead widths and locations between a beading $B^1$ based on $n$ and a beading $B^2$ based on $n+1$ (see \cref{beading_interpolation}).
Such interpolation is also used to deal with beading conflicts (see \cref{beading_conflict_problem}).
There we also apply beading interpolation from a marked node $v_m$ upward along unmarked bones,
and interpolate between $v_m$ and the beading at the top of the slope over some distance $t_\text{beading}$ from the lower marked node\revise{.}{, which we set to $t_\text{beading} = w^* $, so that the transition is not too swift.}

\paragraph{Beading propagation}
The beading information is then broadcast throughout the ST from central regions outward,
so that each unmarked node $v$ knows the beading of the marked node on top of the ramp on which $v$ is placed.
We first broadcast the beading information upward from all marked nodes,
so that we can then deal with beading conflicts in a downward phase.
\revise{This the}{The} downward phase makes sure that all nodes have a beading associated with it, so that the slicing algorithm can efficiently slice the edges leading up to a marked or unmarked node.

\subsection{Toolpath extraction}\label{sec_toolpath_extraction}
\begin{figure}
\centering
\setlength{\figheight}{.29\columnwidth}
\begin{subfigure}{0.14\columnwidth}\centering
\includegraphics[height=\figheight]{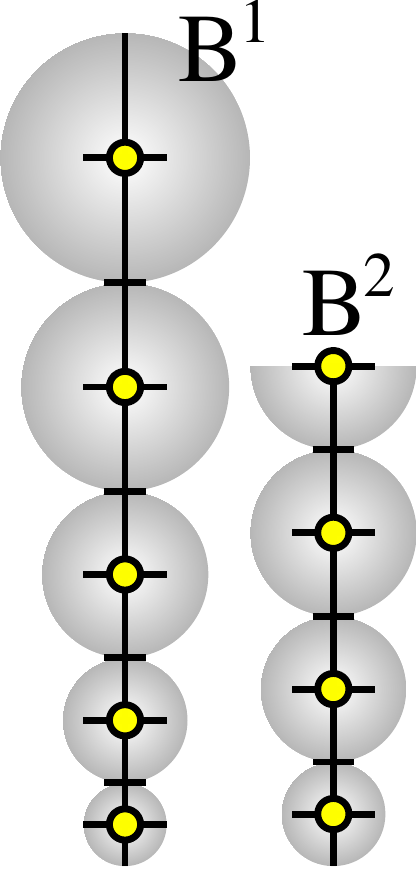}
\caption{Beadings}\label{trapezoid_beading_beading}
\end{subfigure}
\begin{subfigure}{0.5\columnwidth}\centering
\includegraphics[height=\figheight]{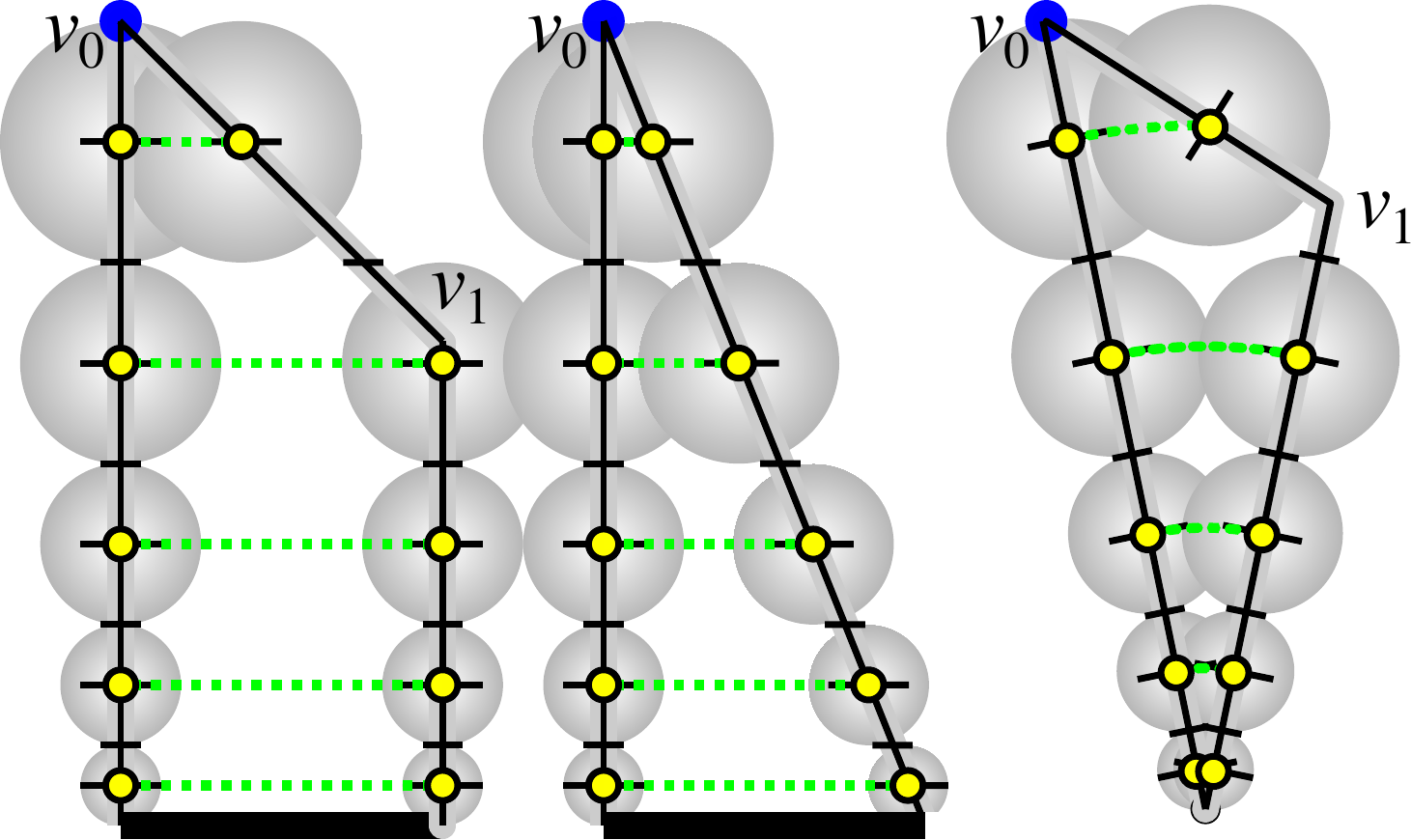}
\caption{Single beading propagated to all nodes}\label{trapezoid_beading_propagated}
\end{subfigure}
\begin{subfigure}{0.34\columnwidth}\centering
\includegraphics[height=\figheight]{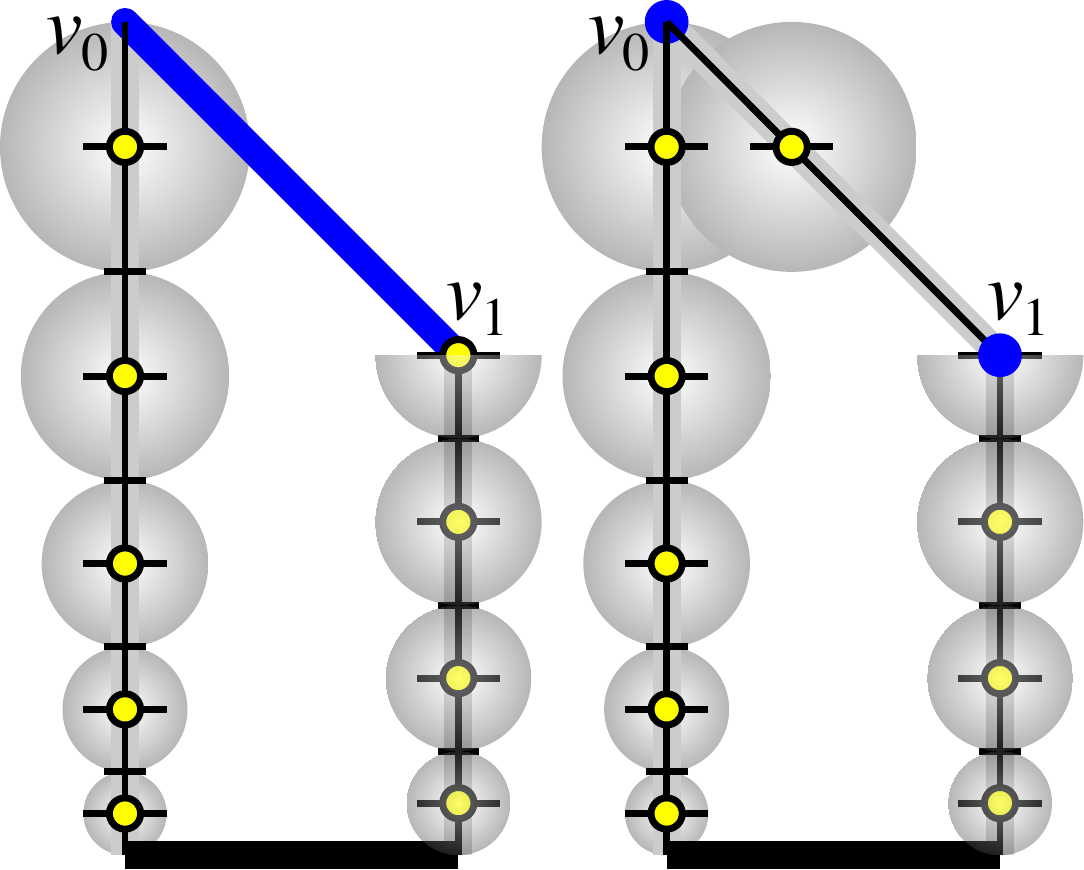}
\caption{Separate beadings at either top node}\label{trapezoid_beading_separate}
\end{subfigure}
\caption{
Applying beadings to generate sites along trapezoids.
\subref{trapezoid_beading_beading} shows the locations $l_i$ and widths $w_i$ of two arbitrary different beadings.
\subref{trapezoid_beading_propagated} shows the application of $B^1$ to the various types of trapezoid.
\subref{trapezoid_beading_separate} shows how a trapezoid with a marked edge will have two different beadings assigned, which will generate their respective sites along the support edges.
No sites will be generated along marked edges.
Wide black lines are outline segments, marked nodes and edges in blue, the sites in yellow and green wavefronts of equidistant radial distance at $R = l_i$.
}
\label{site_placement}
\end{figure}

\revise{Now that}{Once} each node has been assigned a beading\revise{}{,} we proceed to generate the toolpath sites along the edges of the ST.
A site $S$ consists of a location $v$ a width $w$ and an index $i$, which are computed for an edge $v_0v_1$ from the beading $B$ of the upper node $v_1$:
\begin{align*}
S &= \{ v, w, i \} \\ 
v &= v_1 + (v_0 - v_1) \frac{R(v_1) - l_i^B}{R(v_1) - R(v_0)} \\ 
w &= w_i^B
% \\ i^S &= i
\end{align*}
for any $i$ for which $R(v_0) < l_i^B \leq R(v_1)$.
See \cref{site_placement}.
We store all sites of an edge in a mapping from edge to a list of sites.

\begin{figure}
\centering
\setlength{\figheight}{.3\columnwidth}
\begin{subfigure}{.5\columnwidth}\centering
\includegraphics[width=\figheight,rotate=90]{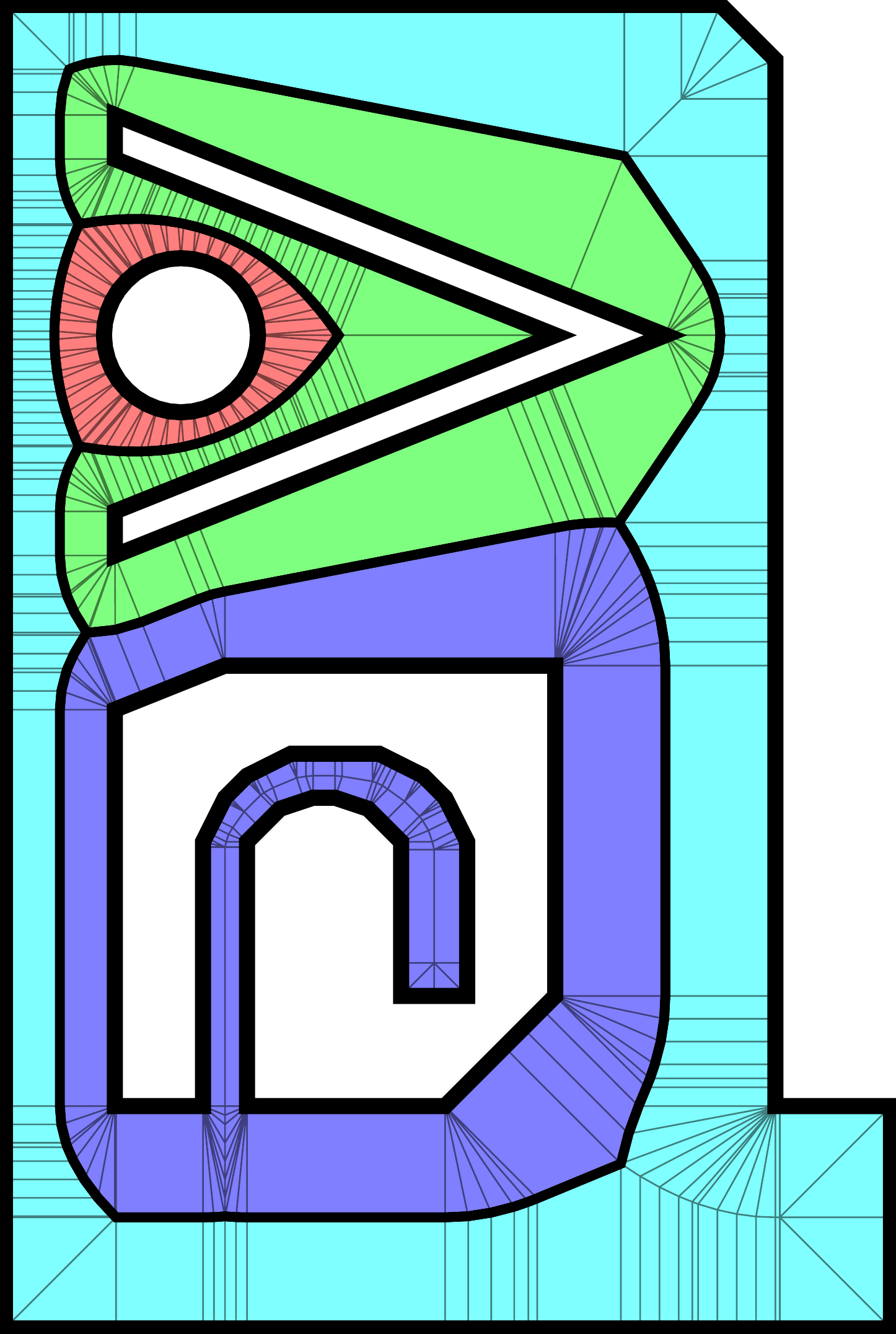}
\caption{Polygon domains}\label{shape_decomposition_domains}
\end{subfigure}
\begin{subfigure}{.45\columnwidth}\centering
\includegraphics[height=\figheight]{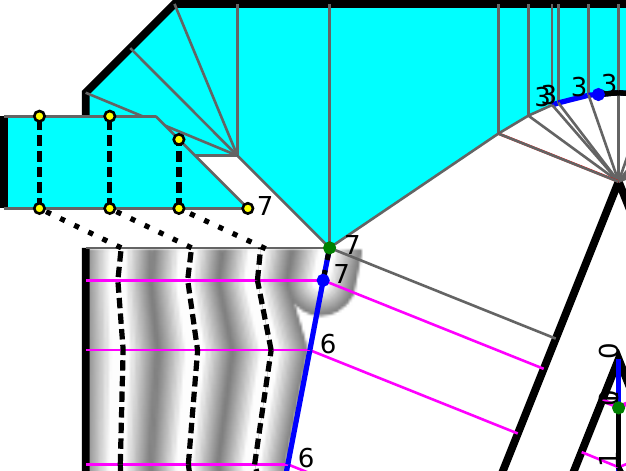}
\caption{Extrusion segment chaining}\label{segment_generation_chaining}
\end{subfigure}
\caption{
Generating toolpaths on a part of the test outline shape by chaining together extrusion segments along each polygon domain.
Each edge is assigned toolpath sites (yellow) which are connected together as shown in the singled out trapezoid.
By following the trapezoids along the domain (cyan) of a single outline polygon,
the extrusion segments can efficiently be connected into existing polyline toolpaths (light and dark gray).
}
\label{segment_generation}
\end{figure}

We then generate extrusion segments for each trapezoid by connecting together the sites of the same index.
See \cref{segment_generation}.
If the amount of sites on both sides of the trapezoid is not the same then this trapezoid is in a transition and we leave one inner site unconnected.

Because the bead count is defined in terms of the feature diameter rather than the radius, only some of the bead count values $\hat{b}$ in a central region coincide with a slicing height.
When the bead count $\hat{b}$ is even, the ridge is sliced as normal;
the intersection between a slicing plane and the mesh surface results in a polyline on both sides of the ridge, which are connected together into a polygonal toolpath.
When the bead count $\hat{b}$ is odd, the ridge will coincide exactly with a slicing height, which results in a single polyline toolpath being generated along the middle of the feature.
In that case we should prevent the algorithm from generating the center extrusion segment twice from the trapezoids on either side of that segment.
We therefore use some arbitrary condition to decide which one of the two to include based on the ordering of the coordinates of $v_0$ and $v_1$: $x_0 < x_1 \lor (x_0 = x_1 \land y_0 < y_1)$.

All trapezoids in the ST are assigned to separate domains, corresponding to which boundary polygon they are connected to (see \cref{shape_decomposition_domains})~\cite{Ding2016a}.
By traversing the trapezoids \revise{in }{}per domain in order we can efficiently connect all segments into polylines.
See \cref{segment_generation}.
In a final step we connect the ends of polylines together, so that the final toolpaths contain both polygons and polylines.

\begin{figure}
\centering
\setlength{\figwidth}{.35\columnwidth}
\begin{subfigure}{0.45\columnwidth}\centering
\includegraphics[width=\figwidth,frame]{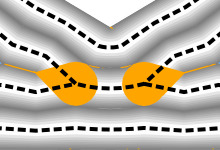}
\caption{No reduction}
\end{subfigure}
\begin{subfigure}{0.45\columnwidth}\centering
\includegraphics[width=\figwidth,frame]{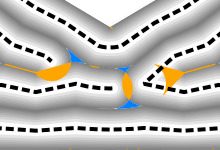}
\caption{Reduction}
\end{subfigure}
\caption{
Reducing polyline toolpaths away from intersections in order to prevent overfill.
Toolpath locations in black, underfill in azure and overfill in orange.
}
\label{polyline_reduction}
\end{figure}

Around the transition locations and around nodes with odd bead count and more than two marked edges attached there will be intersections in the toolpaths.
Such intersections cause overfill because the nozzle passes the location multiple times.
We deal with this special case by forcing a new polyline when traversing the trapezoids, and in the final polyline connection step we greedily connect the first two polylines ending in the same location and retreat all other polylines ending in that same location in order to prevent the overfill.
In order to retreat a polyline which ends in a site $S$\revise{}{,} we remove part of the polyline paths up to the intersection by a distance of $w^S d_\text{max}^\text{intersection}$\revise{, where $d_\text{max}^\text{intersection}$ is some ration between $0$ and $1$.}{.
We set $d_\text{max}^\text{intersection} = \SI{75}{\percent}$ in order to slightly favor overfilling over underfilling.}
This ratio effectively deals with the balance between overfill and underfill generated at that location after the retreat has been applied.
See \cref{polyline_reduction}.

%% file: 5_generalization.tex
\section{Beading schemes}\label{sec_generalization}
A critical component in toolpath generation is how to distribute the beads over the feature radius.
While the framework presented in the previous section takes evenly distributed beads as an example, it allows to apply different beading schemes to configure the bead distribution to cater for specific requirements from the application, 3D printer or material.

\revise{
\begin{definition}\label{beading_scheme_definition}
The beading scheme is configured by the following set of functions and constants:
\begin{align*}
\{
&q(d),
B(n, r),
t(n),
t_0(n),
\\
&\alpha_\text{max}, 
t_\text{beading}, 
d_\text{max}^\text{transition}, 
d_\text{max}^\text{unmarked}, 
d_\text{max}^\text{intersection},
d^\text{discretization}
 \}
\end{align*}
where, as introduced in previous section,
$q(d)$ the quantization operator,
$B(n,r)$ is the beading function consisting of $\left( W(n,r), L(n,r) \right)$,
$t(n)$ the transition length,
$t_0(n)$ the transition anchor position,
$\alpha_{\text{max}}$ is the limit bisector angle,
$t_\text{beading}$ is the beading conflict transition length,
$d_\text{max}^\text{unmarked}$ is the filter distance for unmarked regions,
$d_\text{max}^\text{transition}$ is the filter distance for transition anchors,
$d_\text{max}^\text{intersection}$ is the reduction ratio at 3-way intersections
and
$d^\text{discretization}$ is the discretization length in vertex-vertex VD edges and vertex-line VD edges.
%$W(n, r)$ and $L(n, r)$ give the sequence of bead widths $\left\{ w_i \right\}$ and radial locations $\left\{ l_i \right\}$, respectively.
\end{definition}
}{}

\revise{
The following restrictions hold:
\begin{enumerate}
	\begin{minipage}{\columnwidth}
	\setlength\intextsep{0pt}
	\begin{wrapfigure}[4]{o}{.4\columnwidth} %\centering
	\includegraphics[width=.3\columnwidth,frame]{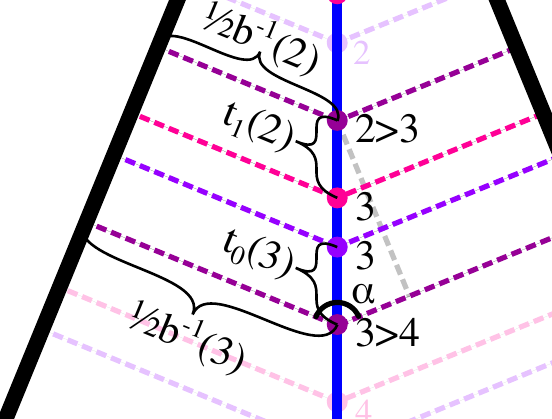}
	%\caption{
	%%Transition ends (magenta and pink) shouldn't cross each other.
	%Placement of transition ends (magenta and pink) with respect to the anchor positions of the transitions (purple) on a ST edge (blue) with bisector angle $\alpha \approx \SI{135}{\degree}$.
	%The distance between the anchor position and the upper end ($t_1$) and the distance between the anchor position and the lower end of the transition ($t_0$) should add up to less than the total distance between the anchor positions, which is limited by $\alpha_\text{max}$.
	%}
	\label{transition_placement}
	\end{wrapfigure}
	\item Transitions shouldn't overlap each other: \\ $t_1(n) + t_0(n+1) < \frac{ q^{-1}(n + 1) - q^{-1}(n) }{ \cos \nicefrac12 \alpha_\text{max}}$ \\ for each $n \in \mathbb{N}$ \\ (see \cref{transition_placement})
\end{minipage}
\item An odd beading should produce an odd single toolpath exactly in the center: $B(n, r)_{\lfloor n/2 \rfloor} = r$ in case $n$ is odd
\item For the smoothness and continuity of toolpaths we require that $W_n$ is monotonic and continuous at each bead index $n$ for constant bead count $c$: $0 \leq \frac{\partial W(c, r)_n}{\partial r} \leq 1$
\end{enumerate}
}{}

\revise{}{
\begin{definition}\label{beading_scheme_definition}
A beading scheme is defined by the quantization operator $q$ and the beading operator $B$: $\{q(d), B(n ,r)\}$.
The beading function $B(n,r)$ consists of $\left( W(n,r), L(n,r) \right)$,
which provides sequences of $n$ bead widths and of $n$ distances from the outline to fill up a radial distance $r$.
\end{definition}
}

% the condition that transitions don't overlap is already caught by the way in which we defined $t_0$ in section 3.4 Smooth transitions

\revise{}{
For the smoothness and continuity of toolpaths we require that $W_n$ is monotonic and continuous at each bead index $n$ for constant bead count $c$: $0 \leq \frac{\partial W(c, r)_n}{\partial r} \leq 1$.
We further ensure
that beads don't overlap,
that beads are extruded from the center of where they end up
and that odd bead counts produce a single polyline toolpath exactly in the center
by determining the bead locations from the widths:
}
\begin{align*}
L(n,r)_i = 
\begin{cases}
-\frac12 W(n,r)_i + \sum_{j=0}^i W(n,r)_j & \text{ if } i < \frac12 (n -1) \\
r & \text{ if } i =  \frac12 (n -1) \\
%d - W(n,r)_{n-1-i} & \text{ otherwise }\\
\end{cases}
\end{align*}

% \subsection{Beading schemes}
We introduce several beading schemes which determine the bead count and their widths in various ways.
We can emulate a variety of toolpath generation methods from related literature by defining new beading schemes.
We also introduce new beading schemes which produce toolpaths with less extreme widths compared to techniques from existing literature.

\revise{
A beading scheme is defined as a set of some particular variables and functions (\cref{beading_scheme_definition}).
Our beading schemes are based on a preferred width $w^* = \SI{0.4}{\milli\meter}$, which is equal to the diameter of the printing nozzle.
Most of the beading schemes we introduce share a common ground:
\begin{align*}
L(n,r)_i = 
\begin{cases}
-\frac12 W(n,r)_i + \sum_{j=0}^i W(n,r)_j & \text{ if } i < \frac12 (n -1) \\
r & \text{ if } i =  \frac12 (n -1) \\
%d - W(n,r)_{n-1-i} & \text{ otherwise }\\
\end{cases}
\\
\begin{array}{rlrl}
t_\text{beading} &= w^* 
&
d_\text{max}^\text{transition} &= \SI{1}{\milli\meter}
\\
t(n) &= w^*
&
d_\text{max}^\text{unmarked} &= w^*
\\
t_0(n) &=  t(n) \left( q^{-1}(n) / w^*  - n \right)
&
d^\text{discretization} &= \SI{0.2}{\milli\meter}
\\
\alpha_\text{max} &= \SI{135}{\degree} 
&
d_\text{max}^\text{intersection} &= \SI{75}{\percent}
\end{array}
%\end{align*}
%\begin{align*}
\end{align*}
}{}

\revise{
The toolpath locations $L$ ensure 
that the beads don't overlap,
that beads are extruded from the center of where they end up
and that the symmetry restrictions are met.
The transition anchor position $t_0$ ensures that the transitions never overlap with the locations $v$ where $R(v) = \nicefrac12 n w^*$ for $n \in \mathbb{N}$.
The transition length $t$ ensures that the center beads don't overlap with the innermost transitioning beads, while keeping the amount of underfill low and keeping the toolpath smooth.
The limit bisector angle $\alpha_\text{max}$ ensures that we don't employ transitioning in shallow wedge regions, which would result in a lot of short odd single bead polylines, which would break up the semi-continuous nature of polygonal extrusion paths.
}{}

\begin{table}%\centering
\caption{
\revise{Made figure into table}{
Beading schemes.
}
} % revise
\label{beading_schemes}
\setlength{\figwidth}{.45\linewidth}
\captionsetup[subtable]{justification=justified,singlelinecheck=false}
\begin{tabular}{|l|l|}
\hline
\begin{subtable}[t]{\figwidth}
\smallskip
\caption{Uniform scheme}\label{formula_uniform}
$\begin{aligned}
q^-(d) &= 2 \left\lfloor \frac{d}{ 2w^*} + \frac12 \right\rfloor \\
W(n,r)_i &= w^* \text{ for all } i 
%\\
%L(n,r)_i &= w^* \left(i + \frac12 \right) \text{ for all } i < \frac12 n
\end{aligned}$
\end{subtable}
&
\begin{subtable}[t]{\figwidth}
\smallskip
\caption{Outer bead}\label{formula_outer_bead}
$\begin{aligned}
q(d) &=
\begin{cases}
1 & \text{ if } d < w^* \\
2 & \text{ otherwise } \\
\end{cases}
 \\
W(n,r)_i &= 
\begin{cases}
2r & \text{ if } n = 1 \\
w^* & \text{ otherwise } \\
\end{cases}
%\\
%L(n,r)_i &= 
%\begin{cases}
%2r / 2 & \text{ if } n = 1 \\
%w^* / 2 & \text{ otherwise } \\
%\end{cases}
\end{aligned}$
\end{subtable}
 \\ \hline
\begin{subtable}[t]{\figwidth}
\smallskip
\caption{Constant bead count}\label{formula_constant_bead_count}
$\begin{aligned}
q(d) &= C \\
W(n,r)_i &= 2 r / n \text{ for all } i 
%\\
%L(n,r)_i &= 2r / n \left(i + \frac12 \right) \text{ for all } i < \frac12 n
\end{aligned}$
\end{subtable}
&
\begin{subtable}[t]{\figwidth}
\smallskip
\caption{Evenly distributed}\label{formula_evenly_distributed}
$\begin{aligned}
q(d) &= \left\lfloor \frac{d}{ w^*} + \frac12 \right\rfloor \\
W(n,r)_i &= 2 r / n \text{ for all } i 
%\\
%L(n,r)_i &= 2r / n (i + \frac12) \text{ for all } i < \frac12 n
\end{aligned}$
\end{subtable}
 \\ \hline
\multicolumn{2}{|l|}{
\begin{subtable}[t]{\linewidth}
\smallskip
\caption{Centered}\label{formula_centered}
$\begin{aligned}
%q^-(d) &= 2 \left\lfloor \frac{d}{ 2w^*} + \frac12 \right\rfloor \\
q(d) &= q^-(d) +
\begin{cases}
-1 & \text{ if } q^-(d) w^* - d > w^* - \revise{r}{d}_\text{max} \\
1  & \text{ if }  q^-(d) w^* - d < w^* - \revise{r}{d}_\text{min} \\
0 & \text{ otherwise}
\end{cases}
\\
W(n,r)_i &= 
\begin{cases}
2 r - (n-1) w^* &\text{ if } i = \frac12 (n-1) \\
w^* &\text{ otherwise }
\end{cases}
%\\
%L(n,r)_i &= 
%\begin{cases}
%2r / 2 & \text{ if } i = \frac12 (n-1) \\
%w^* \left(i + \frac12 \right) & \text{ otherwise }
%\end{cases}
\end{aligned}$
\end{subtable}
}
 \\ \hline
\multicolumn{2}{|l|}{
\begin{subtable}[t]{\linewidth}
\smallskip
\caption{Inward distributed}\label{formula_inward_distributed}
$\begin{aligned}
q(d) &= \left\lfloor \frac{d}{ w^*} + \frac12 \right\rfloor \\
W(n,r)_i &= w^* + E(n,r) \frac{\omega(n,r)_i}{\sum_{j=0}^{n-1} \omega(n,r)_j} \text{ for all } i \\
E(n,r) &= 2r - n w^* \\
%L(n,r)_i &= 2r / n (i + \frac12) \text{ for all } i < \frac12 n
\omega(n,r)_i &= \max(0, 1 - N^{-2} (i - (n-1)/2)^2 )
\end{aligned}$
\end{subtable}
}
 \\ \hline
\end{tabular}
\end{table}

\paragraph{Uniform beading scheme}
We can define a beading scheme which emulates the uniform width offsetting technique by disabling the marking of edges, so that we never employ transitioning.
\revise{See}{We can simply set $\alpha_\text{max} = \SI{180}{\degree}$ and supply a simple beading scheme given by} \cref{formula_uniform}.

\paragraph{Outer bead}
We can emulate the method from \citeauthor{Moesen2011} by carefully choosing how the beading scheme functions deal with the outermost bead.
Also we turn off the reduction of toolpaths near 3-way intersections $d_\text{max}^\text{intersection} = \SI{0}{\percent}$, so that the polygonal toolpaths emulate the remaining area to be filled by another path planning technique similar to their technique.
We don't need transitioning, so we also set $t(n) = 0 $.
See \cref{formula_outer_bead}.

\paragraph{Constant bead count}
We can emulate the method from \citeauthor{Ding2016a} by dividing the feature radius over the widths of a constant number of beads.
Additionally in order to emulate their definition of ``branches'' we mark all ST edges \revise{}{($\alpha_\text{max} = \SI{0}{\degree}$) and }we unmark the outer edges connected to the outline shape in a separate algorithm.
Note that this deviation from the proposed framework violates the robustness against small perturbations in the outline polygon, since this marking depends on the topology of the graph of the ST.
See \cref{formula_constant_bead_count}.

\paragraph{Centered}
We can emulate the method from \citeauthor{Jin2017JMS} by transcribing how they deviate from the uniform width toolpaths.
We therefore base the beading scheme on the bead count $q^-(d)$ defined by the uniform beading scheme.
\citeauthor{Jin2017JMS} replace two beads from the uniform toolpaths by a single one when the \revise{radius}{distance} between the center \revise{and either }{}of those beads falls short of \revise{$r_\text{min} = 0.8 w^*$}{$d_\text{min} = 0.8 w^*$}\revise{}{, which gives us $w_\text{max}=d_\text{min} + 2 \cdot \frac12 w^* = 1.8 w^*$}.
Conversely, they place an extra bead when the \revise{radius between the center and either of the innermost beads}{distance} exceeds \revise{$r_\text{max} = 1.25 w^*$}{$d_\text{max} = 1.25 w^*$}~\cite{Jin2017JMS}\revise{}{, which gives us $w_\text{min}= d_\text{max} - 2 \cdot \frac12 w^* = 0.25 w^*$}~\cite[p.~72]{Jin2017JMS}.
We emulate the rounded polygonal path rerouting they define by supplying a transition length \revise{}{$t(n) = \frac12 w^*$} which results in a discretized version of their rounded polygon segment.
See \cref{formula_centered}.

\paragraph{Evenly distributed}
By taking the advantages of the above two schemes we can define a beading scheme which constitutes a novel toolpathing technique.
We can evenly divide the local feature diameter over the widths of all beads, but choose a local bead count better matching the local feature size.
We determine the local bead count by dividing the diameter by the preferred bead width and rounding to the nearest integer.
This reduces the demands on the system and deviation from mechanical properties caused by beads with extreme deviations from the preferred width.
See \cref{formula_evenly_distributed}.

\begin{figure}
\centering
\setlength{\figwidth}{.8\columnwidth}
\setlength{\figheight}{.3\columnwidth}
\begin{minipage}[b]{0.8\linewidth}
% -p test_geometry/wedge.svg -o wedge -s nrdi -n 1.5 -a --scale 1.5
\begin{subfigure}[t]{\figwidth}\centering
\includegraphics[width=\figwidth]{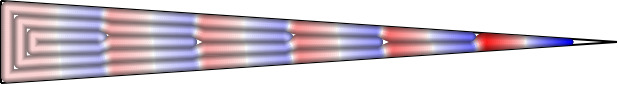}
\caption{Evenly distributed}
\end{subfigure}
\begin{subfigure}[t]{\figwidth}\centering
\includegraphics[width=\figwidth]{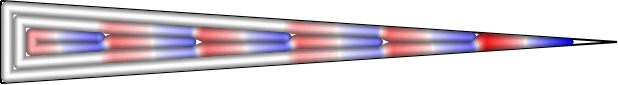}
\caption{Inward distributed}
\end{subfigure}
\end{minipage}
\begin{subfigure}[t]{.1\columnwidth}\centering
\includegraphics[height=\figheight]{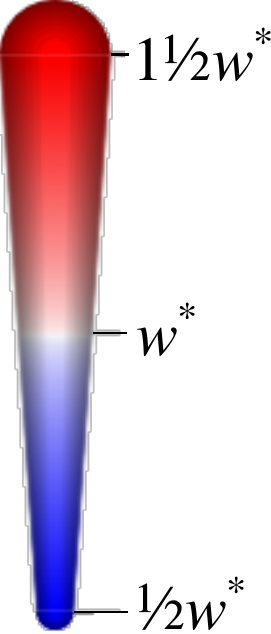}
\end{subfigure}
\caption{
Closeup of toolpaths generated with the distributed \revise{}{and inward ($N=1.5$)} beading schemes for a large wedge shape.
Colors represent bead widths.
}
\label{distributed_comparison}
\end{figure}

\paragraph{Inward distributed scheme}
The evenly distributed scheme can be conceptualized as calculating the total discrepancy $E$ between the actual feature diameter $d$ and the total preferred width $n w^*$, dividing the total discrepancy by the number of beads and setting the width of each bead to 
$w^* + E / n$.
However, depending on the application we might want a different distribution of widths.
We therefore supply a beading scheme which supports an arbitrary distribution of the discrepancy.
The distribution is determined by some weighing function $\omega(n,r)$, which defines the portion of the discrepancy to distribute to each bead.
%\paragraph{Inward distributed}
See \cref{formula_inward_distributed}.
For example, we can choose an $\omega$ which distributes the discrepancy over the innermost $N$ beads, and distribute most of it to the inner beads.
See \cref{distributed_comparison}.
That way we limit the region of impact of the distributed scheme to a central region and have the preferred bead width $w^*$ in regions farther away.
This limits the impact of transitioning regions so that transitions keep the toolpaths smooth farther away from the central regions. % and it forces most of the beads to have exactly the preferred width.
%Moreover, the bead widths equal the preferred bead width for large regions meaning that mechanical properties derived for prints using the uniform scheme will still hold approximately.

\paragraph{Widening \revise{}{meta-scheme}}
Complementary to any of these schemes we can enforce a minimum feature size \revise{at no extra cost}{and minimum bead width} in our framework.
Regions where the model is narrower than \revise{the nozzle size}{some $r_\text{min}$} can be printed with a bead width \revise{}{$w_\text{min}$} larger than the model thickness.
See \cref{min_feature_size,widening}.
We can simply override
\begin{align*}
q'(d) &= 
\begin{cases}
0 & \text{ if } 0 \leq d < 2 r_\text{min} \\
1 & \text{ if }  2 r_\text{min} \leq d < w^*  \\
q(d) & \text{ otherwise}
\end{cases}
\\
W'(n,r)_0 &=
\begin{cases}
\max \left( w_\text{min}  ,  2 r \right) & \text{ if } 2 r < w^* \\
W(n,r)_0 & \text{ otherwise}
\end{cases}
\end{align*}

\revise{}{
\paragraph{Shell meta-scheme}
The industry standard of FDM is to generate only a limited contour-parallel perimeters and to fill the remainder using a direction-parallel strategy.
We therefore provide a meta-scheme to generate adaptive bead width toolpaths only in narrow regions and generate the limited number of perimeters $M$ using the preferred width in regions which are wide enough.
We also take care not to leave gaps which are too small to be filled using the direction-parallel strategy:

\begin{align*}
q'(d) &= \min(M, q(d))
\\
W'(n,r)_i &= 
\begin{cases}
W(n, M w^*)_i & \text{ if } 2 r > q^{-1}(M) \\
W(n,r)_i & \text{ otherwise}
\end{cases}
\end{align*}

These meta-schemes introduce non-linearities in the quantization function.
Because the beading is only evaluated at nodes in the skeleton, we need to make sure that there are nodes at the locations along the skeleton where the non-linearities happen.
We therefore insert extra nodes along with their ribs at locations $v$ with a radial distance $R(v) = r_\text{min}$ for widening and at $R(v) \in \left\{ M w^*, q^{-1}(M), q^{-1}(M) + \nicefrac12 w^* \right\} $ for the transition from narrow shell to unconstrained shell.
Combining all meta-schemes functionality we can generate results such as depicted in \cref{shell}.

\begin{figure}
\centering
\setlength{\figwidth}{.35\columnwidth}
\setlength{\figwidthTwo}{.6\columnwidth}
\begin{minipage}[b]{\figwidthTwo} \centering
\begin{subfigure}[b]{\figwidthTwo}\centering
\includegraphics[width=\columnwidth]{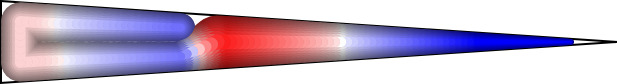}
\caption{Widening meta-scheme using $w_\text{min} = 2 r_\text{min} = 0.1$
\\ on top of the Distributed scheme}
\label{min_feature_size}
\end{subfigure}
\begin{subfigure}[b]{\figwidthTwo}\centering
\includegraphics[width=\columnwidth]{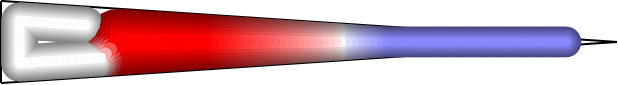}
\caption{Widening meta-scheme using $w_\text{min} = 0.5$, $2 r_\text{min} = 0.2$ 
\\ on top of the Centered scheme}
\label{widening}
\end{subfigure}
\end{minipage}
\begin{subfigure}[b]{\figwidth}\centering
\includegraphics[width=\figwidth]{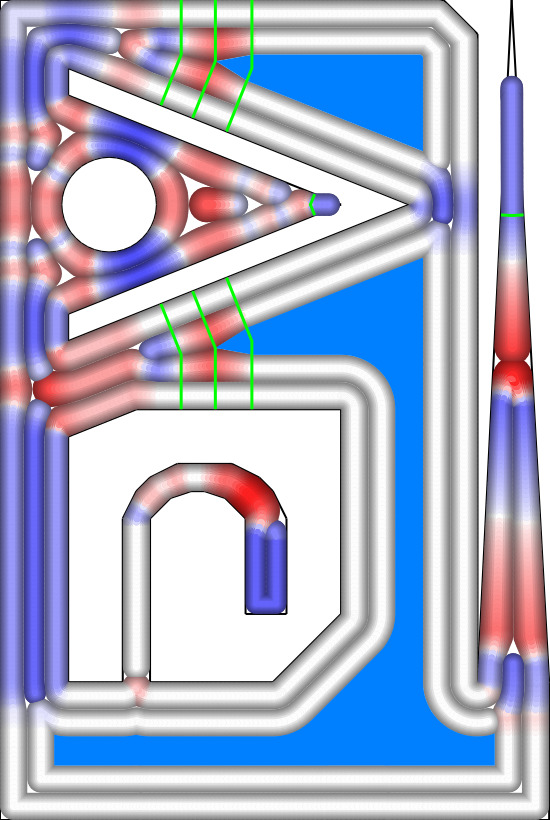}
\caption{Shell and widening}
\label{shell}
\end{subfigure}
\caption{
Toolpaths using the widening and shell meta-schemes.
\subref{min_feature_size} and~\subref{widening} show widening.
\subref{shell}
show toolpaths generated with the inward distributed strategy ($N=1.5$) in conjunction with the shell meta-scheme ($M=4$) and the same widening as in~\subref{widening}.
% -p test_geometry/gMAT_example.svg -b 4 -s i -n 1.5 -a --mins 0.2 --minw .5
Widening and shell require extra edges (green) at key locations in the skeleton.
The azure area is to be filled using some direction-parallel toolpaths.
}
\label{widening_shell}
\end{figure}

}

%% file: 6_2_printing_results.tex
\revise{}{
\section{Fabrication}
In order to accurately manufacture adaptive width toolpaths using an off-the-shelf 3D printing system,
we need a model which relates the required width to process parameters such as movement speed and filament extrusion speed.
A different approach might be appropriate depending on whether the filament feeder is mounted directly on the print head (a.k.a. \emph{direct drive}) or the filament fed from the back of the printer to the print head via a \emph{Bowden tube}.
Because Bowden style 3D printing systems have the filament feeder relatively far away from the nozzle, changing the internal pressure in the system requires a large amount of filament movement, which requires a prohibitive amount of time.

\subsection{Back pressure compensation}
Because changing the internal pressure is difficult in our setup,
we keep the internal pressure constant, and vary the movement speed instead.
%In order to accurately realize a varying bead width we vary the movement speed, while keeping the internal pressure in the system constant.
One approach would be\revise{}{ to} keep the filament inflow $f$ (in \si{\milli\meter\cubed\per\second}) constant by varying movement speed accordingly \cite{Kuipers2018}.
However, that doesn't result in the intended filament outflow variation - see \cref{zero_back_pressure}.
We conjecture that the filament outflow is related to the total pressure in the system,
which depends not only on the amount of filament in between the feeder wheel and the nozzle (which we keep constant), 
but also depends on the back pressure that the previous layer exerts on the filament protruding from the nozzle.
The amount of back pressure is most likely monotonically related to the requested line width.
We compensate for the back pressure using a simple linear model:

\begin{align}
 v(w) &= \frac{f(w)}{h w} \\ 
% f &\sim p \\
% p &= p_\text{in} + p_\text{ext} \\
% p_\text{in} &= C \\
% p_\text{ext} &\sim w \\
% p_\text{ext} &= w / w^* - 1 \\
% f &= f^* - k p_\text{ext} \\
 f(w) &= f_0 - k \left( w / w_0 - 1 \right)
 % f_0 &= v_0 w_0 h 
% v &= \frac{f^* - k p_\text{ext}}{h w} \\ 
% v &= \frac{v^* w^* h - k (w / w^* - 1)}{h w}
\end{align}
where
$v(w)$ is the movement speed as a function of requested bead width $w$,
$f(w)$ is the filament outflow,
$f_0$ is a constant reference flow,
$w_0$ is a constant reference bead width
and
$k$ is the amount of back pressure compensation.

% adapted from 5.5 Discussion on implications
% limitations of back pressure compensation
Our back pressure compensation method effectively changes the speed to realize adaptive width,
but this approach is limited, since the movement speed is constrained by acceleration considerations near bends in the toolpath~\cite{Ertay2018}.
Moreover, as the layer height is decreased the back pressure becomes larger compared to the internal pressure, which might cause the back pressure compensation method to demand prohibitively slow movement speeds.
Furthermore, the shape and filling of the previous layer might influence the amount of back pressure.
%
% direct drive & pressure advance
Accurate flow control can be further enhanced by using a direct drive hardware system and by employing \emph{pressure advance algorithms} which dynamically change the internal pressure \cite{tronvoll2019investigating}.
Conversely such a setup might benefit from some form of back pressure compensation as well.

\subsection{Print results}\label{print_results_section}
Using increments of $0.1$ we established that using a factor of $k=1.1$ yields satisfactory bead width variation for our setup where we use
$f_0 = v_0 w_0 h $
with
$v_0=\SI{30}{\milli\meter\per\second}$, 
$w_0=\SI{0.4}{\milli\meter}$
and
$h=\SI{0.1}{\milli\meter}$.
See \cref{back_pressure}.
The fact that the printed lines are wider than intended is compensated for using a flow reduction to \SI{90}{\percent}.
Test prints were performed on an unmodified Ultimaker S5 system,
with a standard  \SI{0.4}{\milli\meter} nozzle
and PLA filament.
The printing order is determined greedily by choosing the closest point of a polygonal extrusion path, or the closest of either end point in case of an open polyline extrusion path.
Because the machine instructions file format \emph{G-code} doesn't natively support adaptive width beads,
we discretize adaptive width extrusions into \SI{0.2}{\milli\meter} long segments of the average width.
The print results can be viewed in \cref{prints}.

\begin{figure}
\centering
\setlength{\figwidth}{0.32\columnwidth}
\setlength{\figheight}{0.5\columnwidth}
\begin{subfigure}[t]{\figwidth}\centering
\includegraphics[angle=90,height=\figheight]{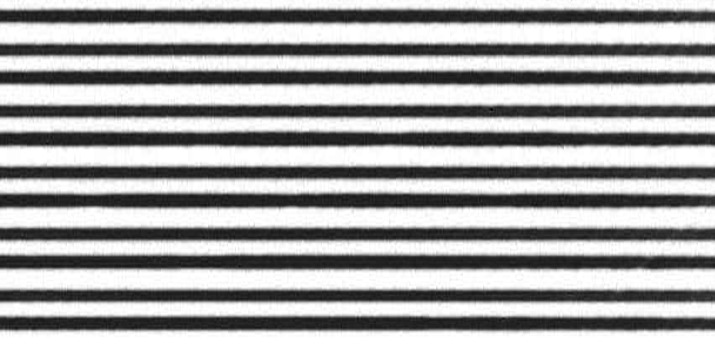}
\caption{$k=0$}\label{zero_back_pressure}
\end{subfigure}
\begin{subfigure}[t]{\figwidth}\centering
\includegraphics[angle=90,height=\figheight]{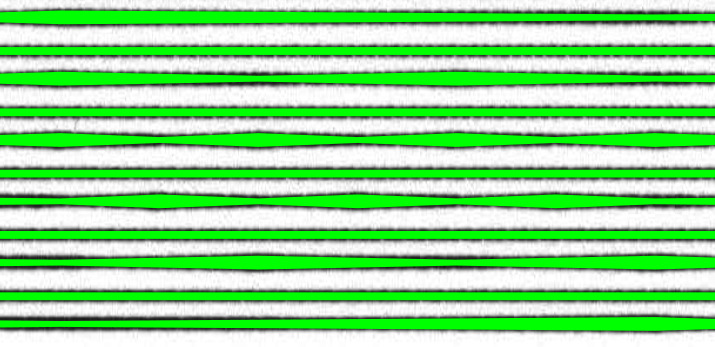}
\caption{$k=1.1$}\label{back_pressure}
\end{subfigure}
\begin{subfigure}[t]{\figwidth}\centering
\includegraphics[angle=90,height=\figheight]{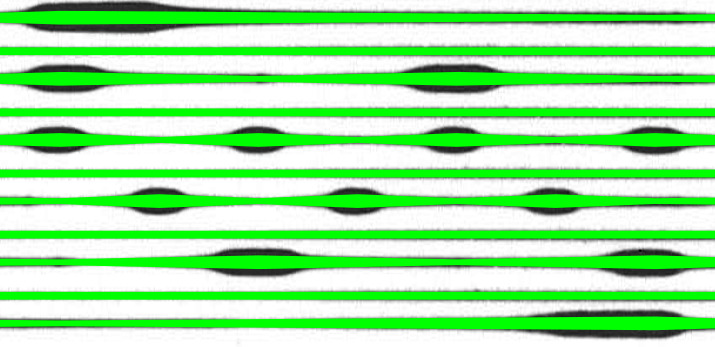}
\caption{$k=2.0$}\label{too_much_back_pressure}
\end{subfigure}
\caption{
Print results (black) of the varying width test on top of a dense white raft.
Target widths in green.
\subref{zero_back_pressure} Simple flow equalization without back pressure compensation results in nearly constant bead widths.
\subref{back_pressure} A value of $k=1.1$ seems to produce good results.
}
\label{back_pressure_compensation}
\end{figure}

% discussion of print results
In \cref{print_naive} the underfill problem of the naive uniform offset approach is most prevalent for the Ultimaker word mark, which negatively impacts the visual quality and the stiffness of the part.
Moreover, in the case of the spatially graded honeycomb\revise{}{,} there are several fully disconnected hexagons, which means the object falls apart when picked up.
The honeycomb print is also missing all parts which are slightly more thin than the preferred bead width $w^*$.
\Cref{print_center} still shows some underfill, but considerably less than the uniform approach.
These prints also exhibit dark regions where the translucency of the layer is less because the bead is higher.
This can be explained by inaccuracies in the back pressure compensation method, which arise for bead widths which deviate from the preferred width by a large amount.
\Cref{print_inward} diminishes the underfill nearly completely and the visual quality of these prints is more homogenous than those of the other methods.
Moreover, the absence of dark regions signifies that our proposed method is more robust against inaccuracies in the deposition system.
However, both the centered and inward distributed approach introduce transitions to a different bead count in the word `Delft', which reduces the dimensional accuracy on the outline around those locations.

\begin{figure}
\centering
\setlength{\figwidth}{\columnwidth}
\begin{subfigure}{\figwidth}\centering
\includegraphics[width=\figwidth]{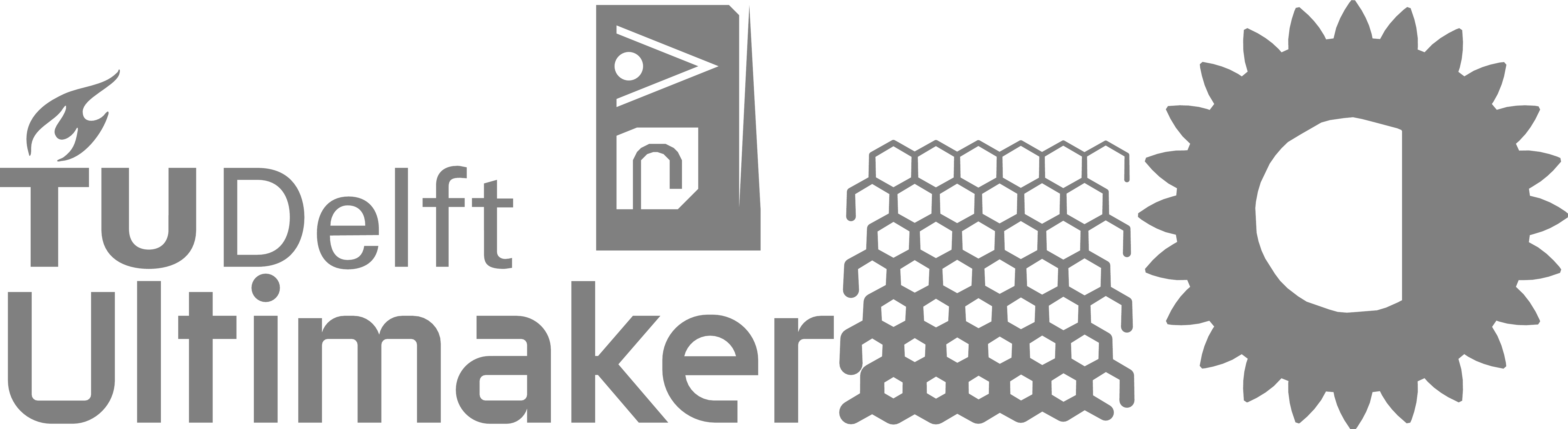}
\caption{Outlines}\label{print_outlines}
\end{subfigure}
\begin{subfigure}{\figwidth}\centering
\includegraphics[width=\figwidth]{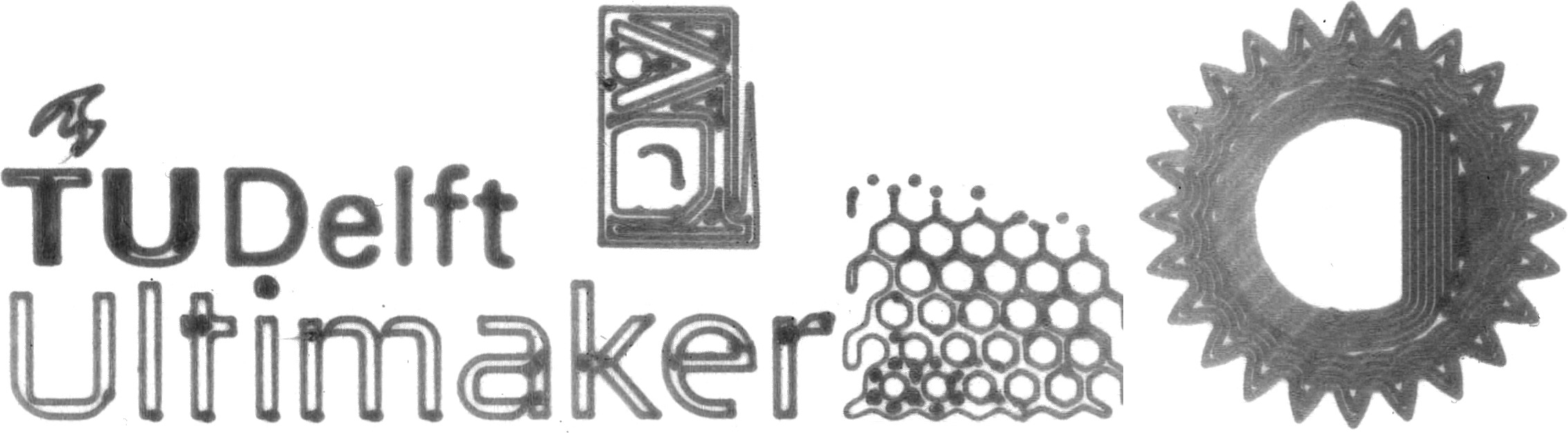}
\caption{Uniform}\label{print_naive}
\end{subfigure}
\begin{subfigure}{\figwidth}\centering
\includegraphics[width=\figwidth]{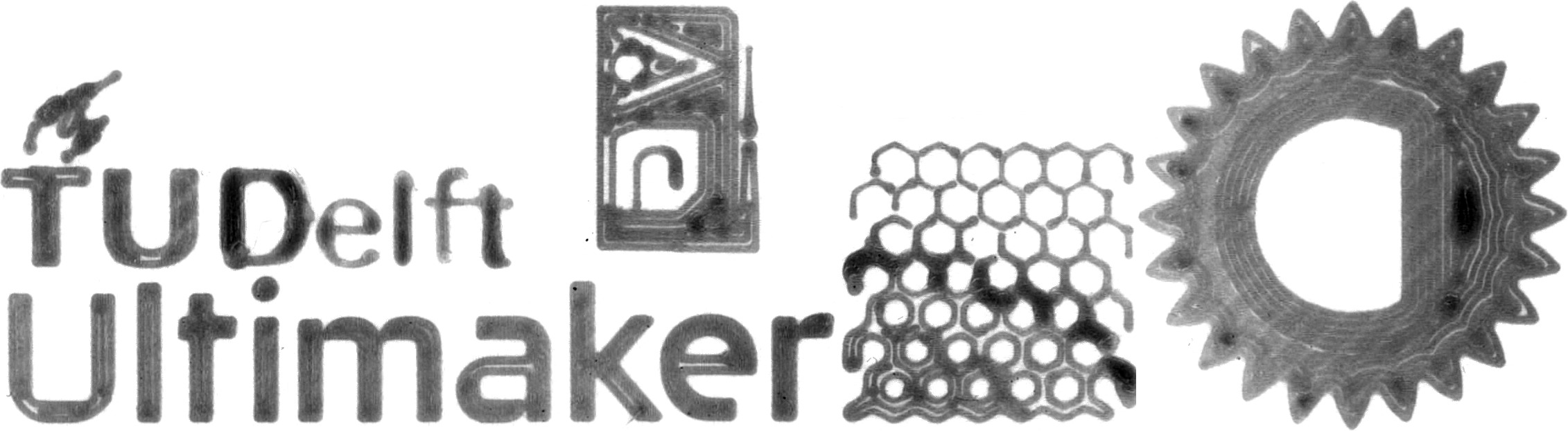}
\caption{Centered}\label{print_center}
\end{subfigure}
\begin{subfigure}{\figwidth}\centering
\includegraphics[width=\figwidth]{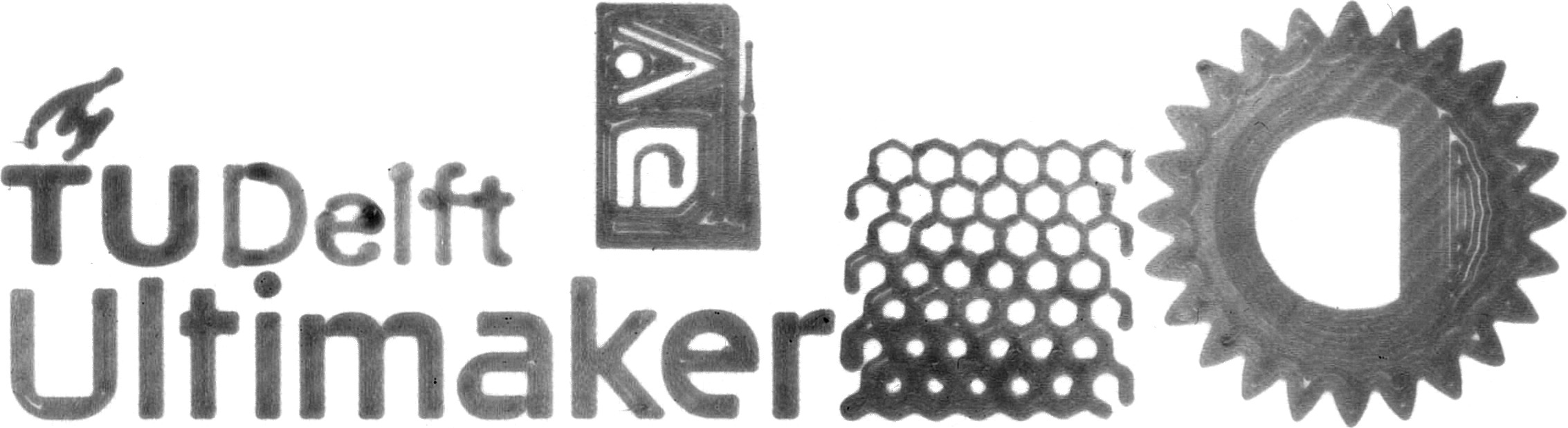}
\caption{Inward distributed}\label{print_inward}
\end{subfigure}
\caption{
Test shapes printed using the uniform scheme, centered scheme and the inward distributed scheme.
The uniform technique produces distinct underfill areas.
The centered scheme shows some defects due to inaccurate control of extreme deposition widths.
The inward distributed scheme produces the least defects.
}
\label{prints}
\end{figure}

}

%% file: 6_validation.tex
\section{Results and discussion}

We evaluate the proposed framework and the various beading schemes on a set of different types of 3D models, ranging over various applications and various types of geometry. 
The data set is described in Appendix~\ref{dataset}.
We sliced all models in the data set and selected 300 random slices for analysis.
Toolpaths of these 300 outline shapes are generated using the uniform technique as implemented by Clipper~\cite{johnson2014clipper} -- a state-of-the-art polygon offset library,
and by our framework using four beading schemes, i.e. the constant bead count scheme with a bead count of $C=4$, the centered, the evenly distributed, and the inward distributed beading scheme using \revise{$N=3$}{$N=2$}, all with a preferred bead width\revise{s}{} of \revise{$w^* = \SI{0.4}{\milli\meter}$}{$w^* = \SI{0.5}{\milli\meter}$ and using the widening meta-scheme to enforce a minimum printed feature size of $w_\text{min}=2r_\text{min}=\SI{0.3}{\milli\meter}$}.
The tests were performed on a desktop PC equipped with an Intel Core i7-7500U CPU @ \SI{2.70}{\giga\hertz} (a single core is used) and \SI{16.3}{\giga\byte} memory.
\revise{}{We report on the total statistics summed over the whole data set, because averaging would be biased.}

\subsection{Computational results}\label{sec_computational_results}

\subsubsection{Accuracy}
\begin{figure*}
\centering
\setlength{\figwidth}{0.19\textwidth}
\setlength{\figheight}{0.283\textwidth}
\begin{subfigure}{\figwidth}\centering
\includegraphics[height=\figheight]{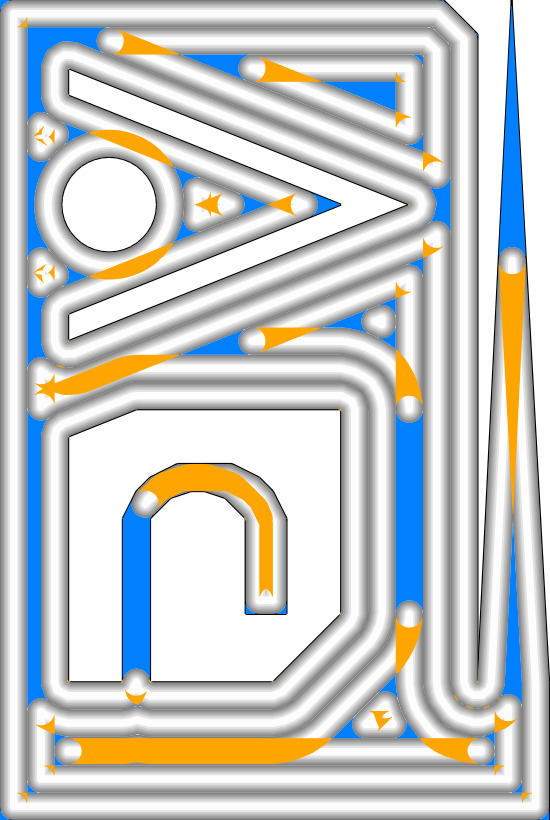}
\includegraphics[height=\figheight]{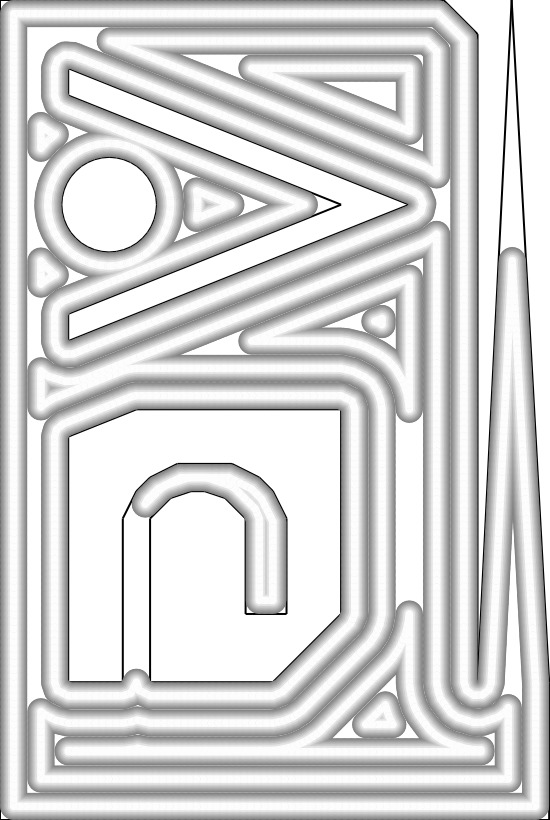}
\caption{Uniform}\label{TEST_naive_accuracy}
\end{subfigure}
%\begin{subfigure}{\figwidth}\centering
%\includegraphics[height=\figheight]{sources-validation-gMAT-example-TEST-SingleBead-accuracy}
%\includegraphics[height=\figheight]{sources-validation-gMAT-example-TEST-SingleBead-widths}
%\caption{Single}\label{TEST_SingleBead_accuracy}
%\end{subfigure}
\begin{subfigure}{\figwidth}\centering
\includegraphics[height=\figheight]{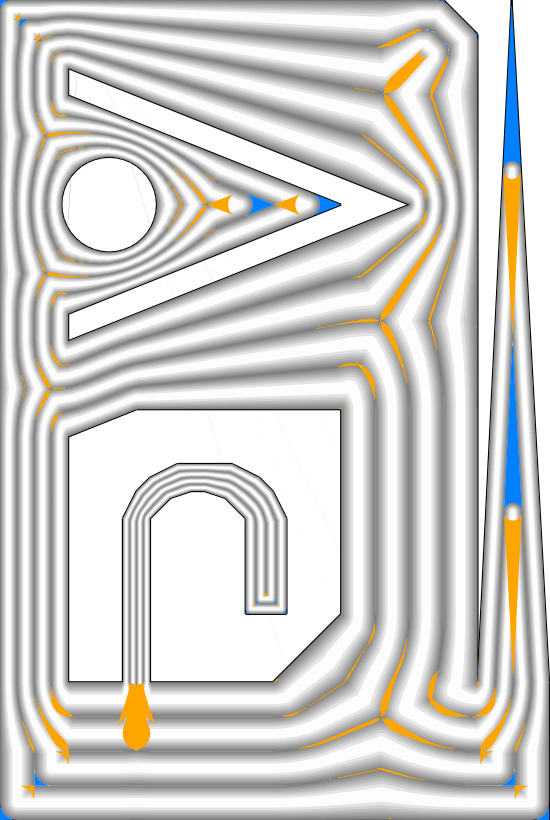}
\includegraphics[height=\figheight]{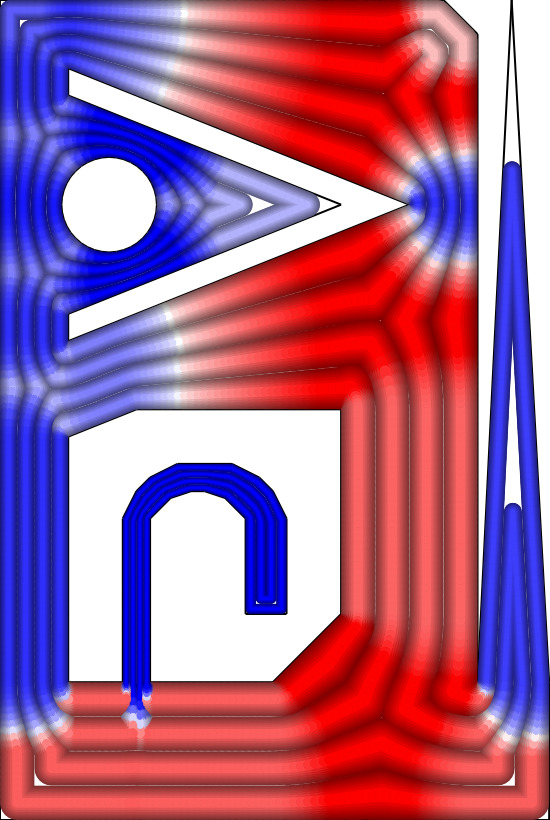}
\caption{Constant}\label{TEST_Constant_accuracy}
\end{subfigure}
\begin{subfigure}{\figwidth}\centering
\includegraphics[height=\figheight]{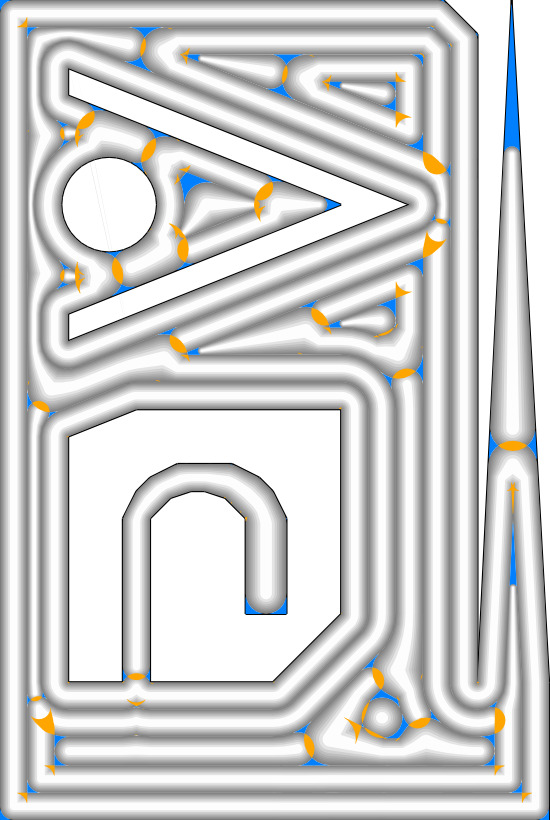}
\includegraphics[height=\figheight]{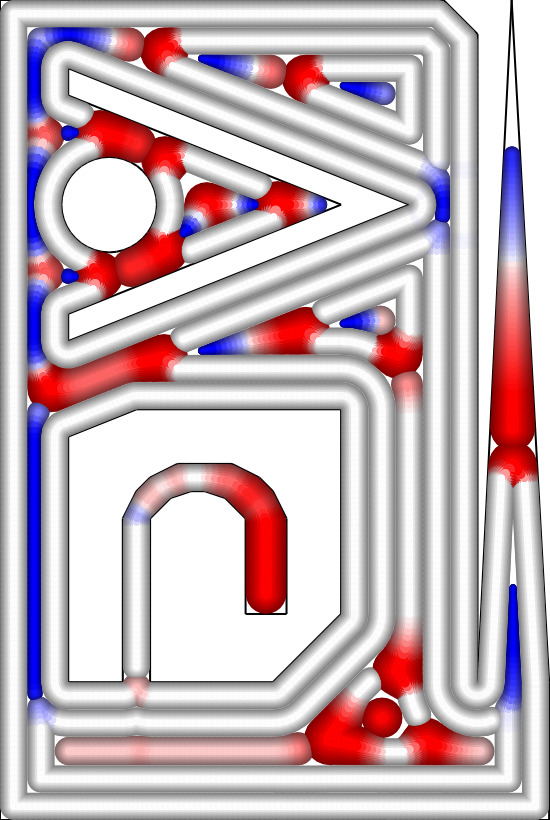}
\caption{Centered}\label{TEST_Center_accuracy}
\end{subfigure}
\begin{subfigure}{\figwidth}\centering
\includegraphics[height=\figheight]{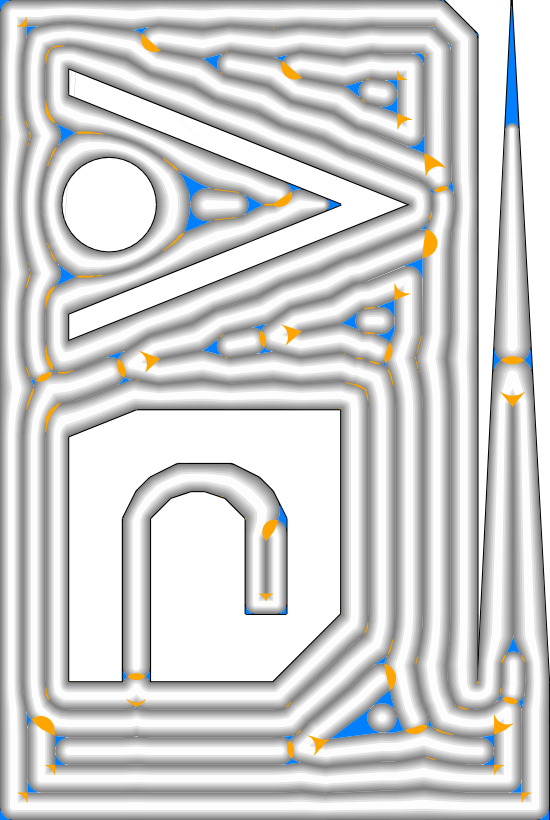}
\includegraphics[height=\figheight]{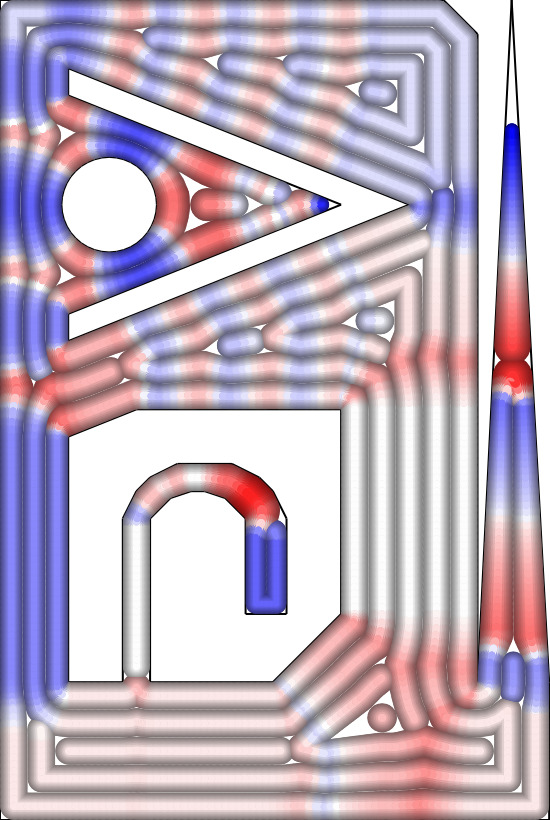}
\caption{Distributed}\label{TEST_Distributed_accuracy}
\end{subfigure}
\begin{subfigure}{\figwidth}\centering
\includegraphics[height=\figheight]{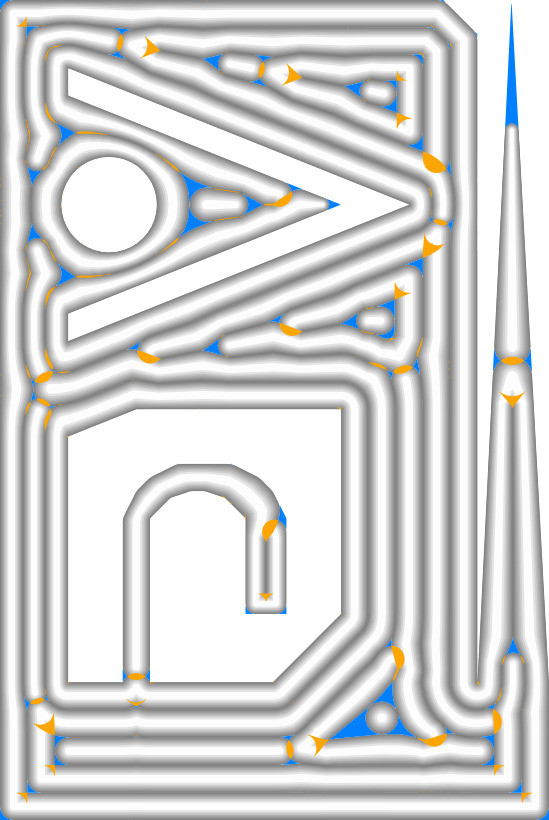}
\includegraphics[height=\figheight]{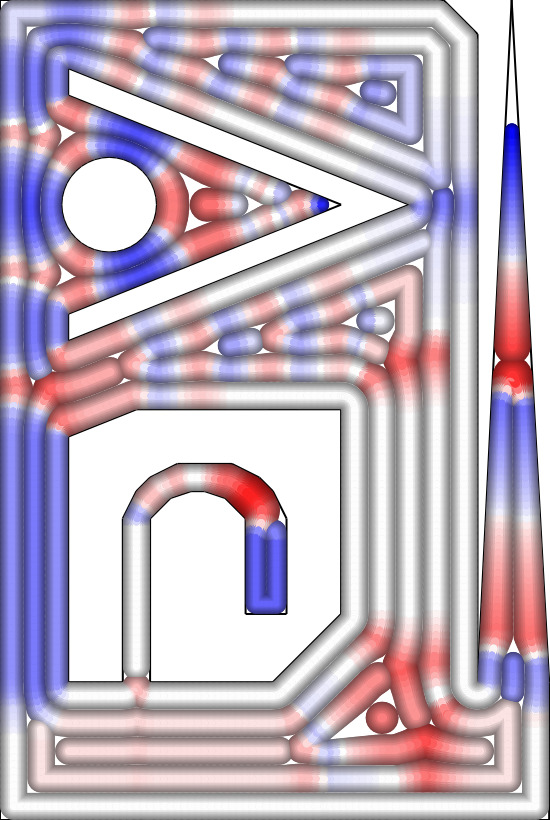}
\caption{Inward \revise{}{($N=2$)}}\label{TEST_InwardDistributed_accuracy}
\end{subfigure}
\begin{subfigure}{.04\columnwidth}\centering
\vspace{4.7cm}
\includegraphics[height=\figheight]{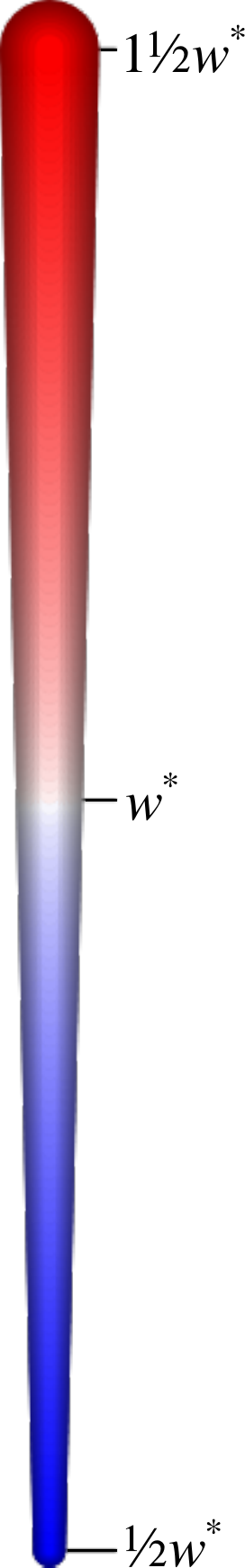}
\end{subfigure}
\caption{
Visualization of the overfills and underfills (top) and the widths (bottom) for various beading schemes.
Extrusion beads in gray tones,
overfill in orange,
underfill in azure,
narrow beads in blue
and wide beads in red.
\revise{New shape contains pointy wedge to see impact on minimal feature size}{In order to distinguish clearly from the Distributed scheme the Inward is limited to $N=2$.}
}
\label{visualized_accuracy}
\end{figure*}

\begin{figure*}
\centering
\setlength{\figheight}{0.25\textwidth}
\setlength{\figwidth}{0.32\textwidth}
\begin{subfigure}{\figwidth}\centering
\includegraphics[height=\figheight]{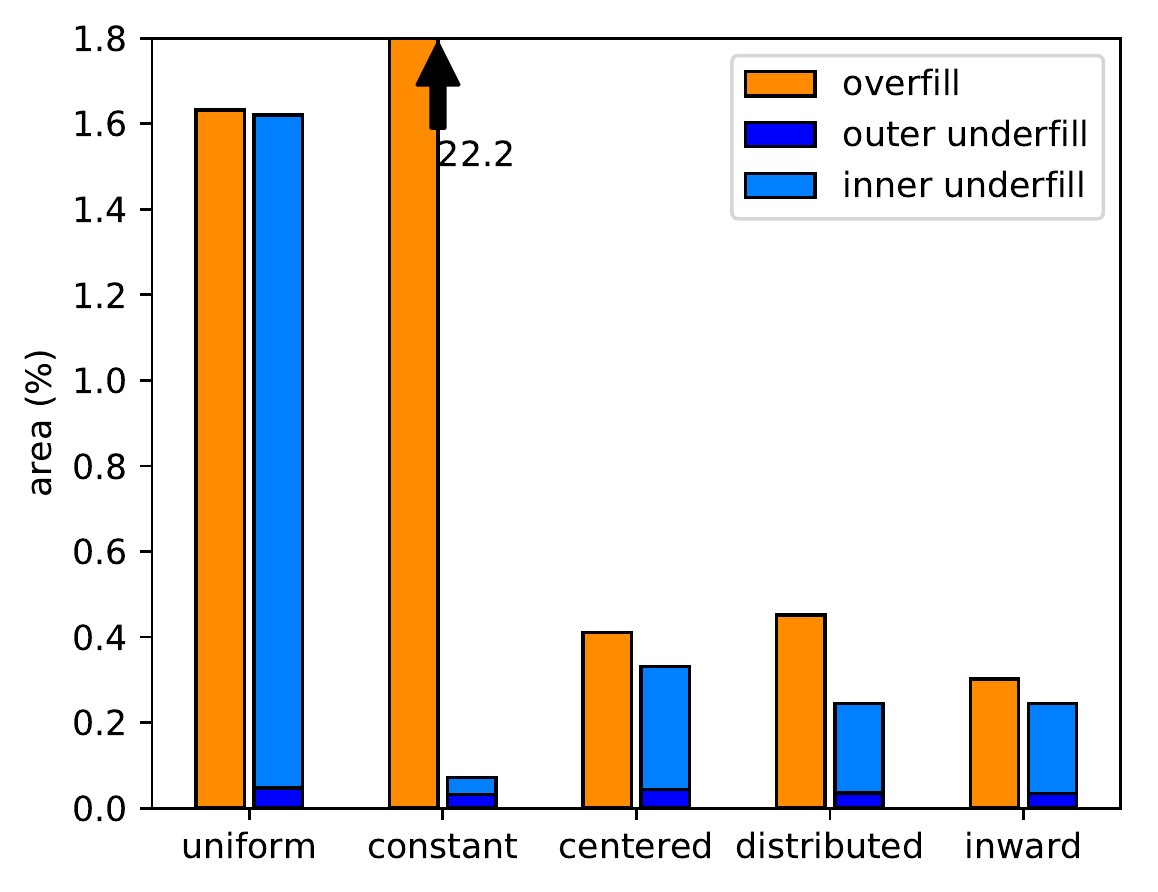}
\caption{\revise{}{Over- and underfill}}
\label{over_underfill}
\end{subfigure}
\begin{subfigure}{\figwidth}\centering
\includegraphics[height=\figheight]{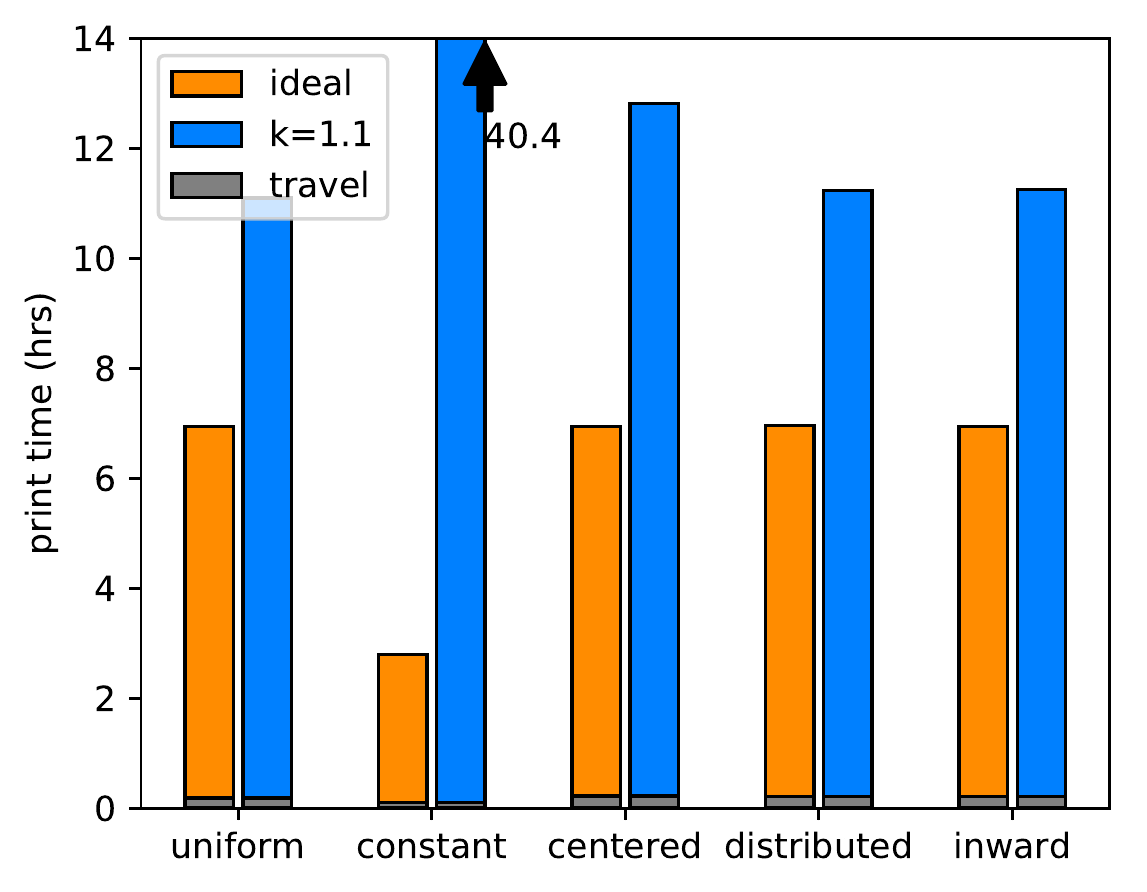}
\caption{\revise{}{Print time}}
\label{printtime}
\end{subfigure}
\begin{subfigure}{\figwidth}\centering
\includegraphics[height=\figheight]{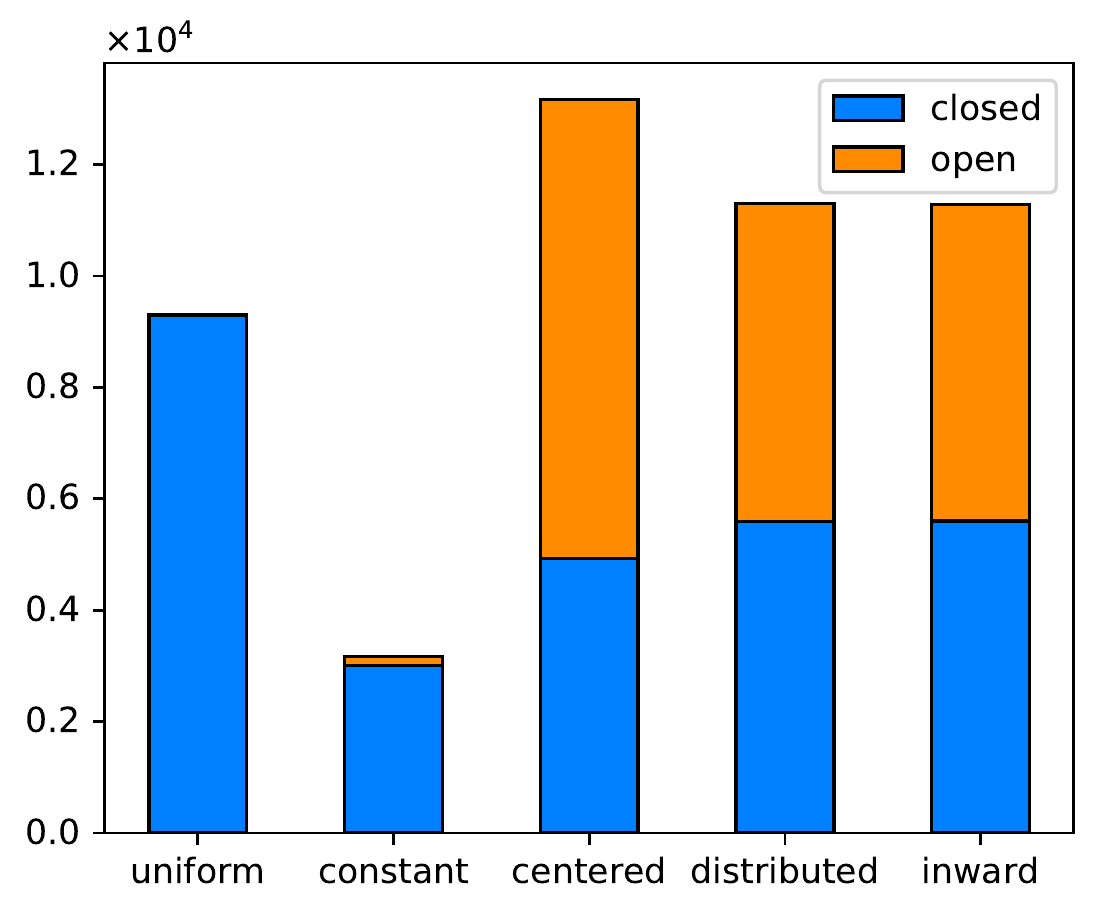}
\caption{\revise{}{Path counts}}
\label{pathcounts}
\end{subfigure}

\begin{subfigure}{\figwidth}\centering
\includegraphics[height=\figheight]{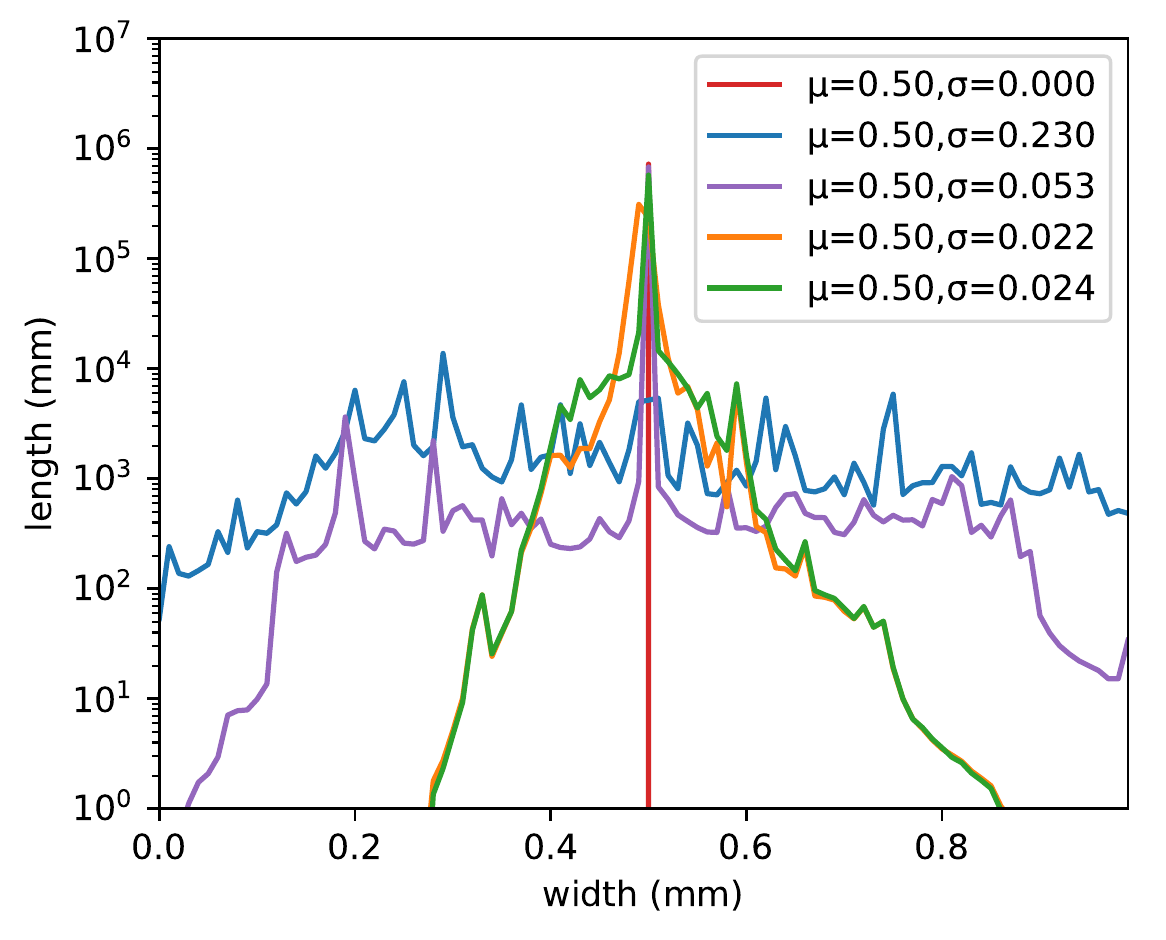}
\caption{\revise{}{Extrusion widths}}
\label{widthHistogram}
\end{subfigure}
\begin{subfigure}{\figwidth}\centering
\includegraphics[height=\figheight]{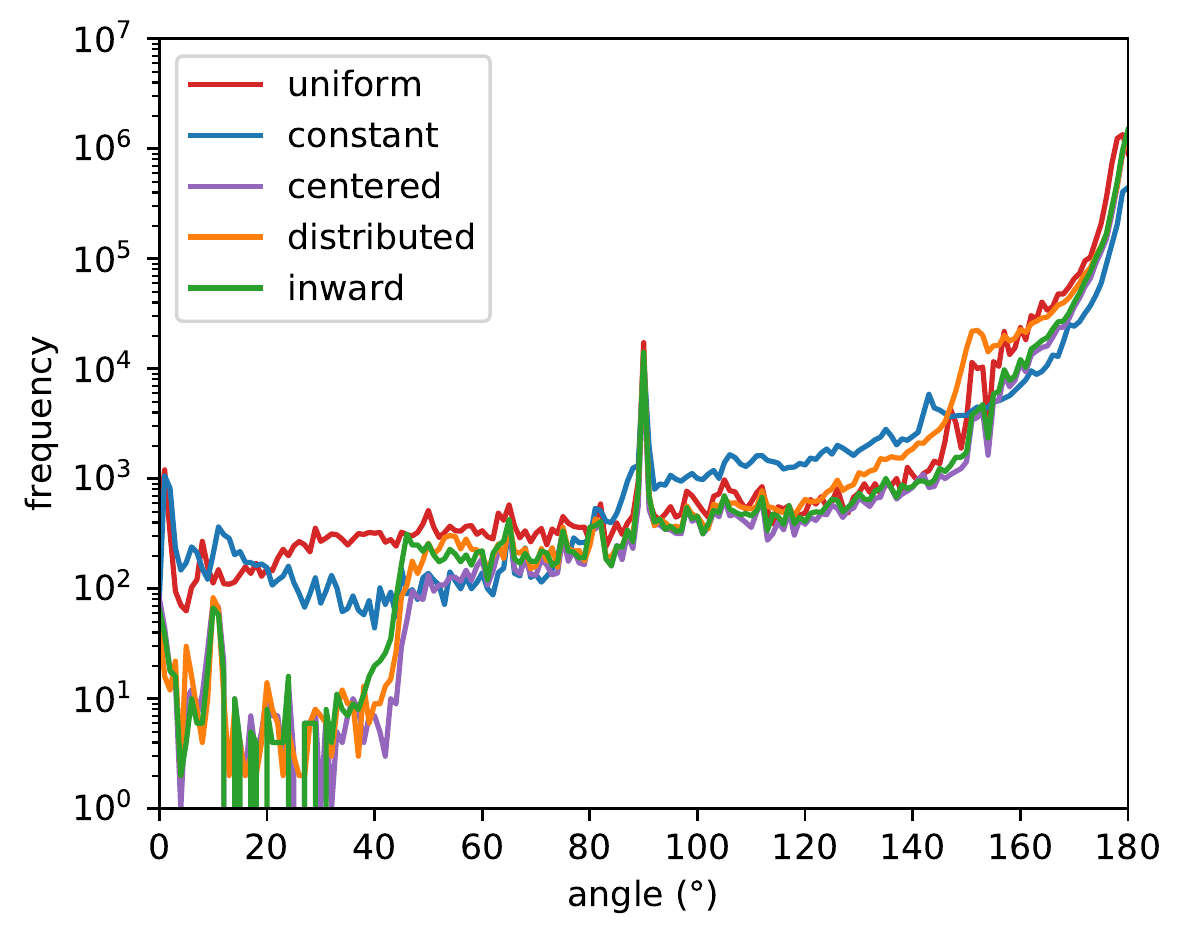}
\caption{\revise{}{Site angles}}
\label{smoothness}
\end{subfigure}
\begin{subfigure}{\figwidth}\centering
\includegraphics[height=\figheight]{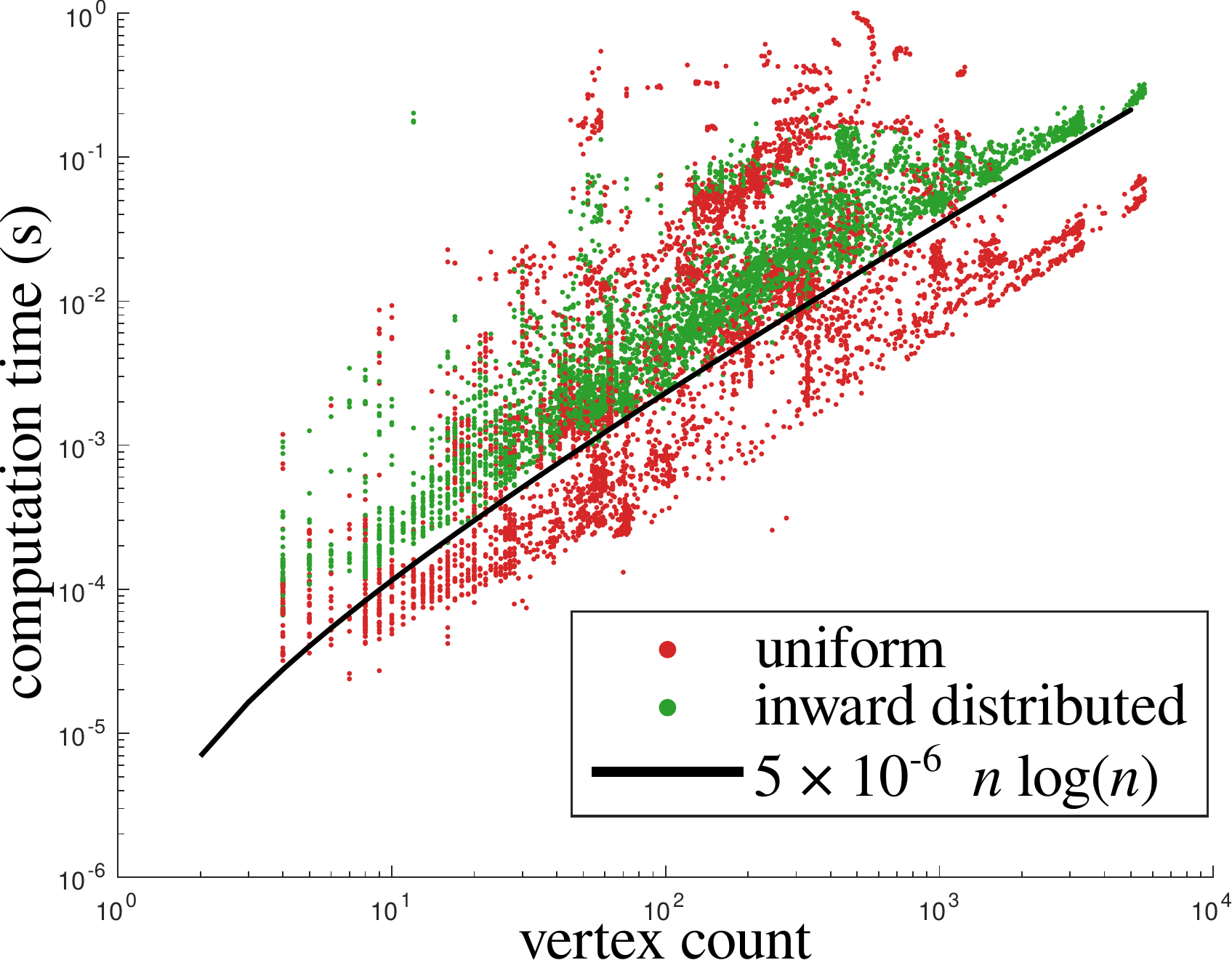}
\caption{Computation time}
\label{computime}
\end{subfigure}

\caption{
Statistical analysis of the toolpaths from applying the uniform width technique and various beading schemes using our framework to a data set of 300 slices.
Note the use of a logarithmic scale \revise{}{in the bottom graphs }on the Y-axes and \revise{on some of}{for \subref{computime} on} the X-axes as well.
}
\label{statisticsfig}
\end{figure*}

We first evaluate the accuracy of different beading schemes in terms of the relative amount of the overfill and underfill. 
We construct the over- and underfill area by comparing the shapes covered by each extrusion move \revise{(\cref{segment_visualization_simple})}{} with each other and with the total shape of the boundary polygons. (For implementation details see Appendix~\ref{accuracy_calculation}.)
This results in polygonal shapes such as visualized in the top half of \cref{visualized_accuracy}:
there are orange shapes where the beads overlap and azure shapes in the voids in between the beads.
We compare the total area in \si{\milli\meter\squared} of these overfill and underfill shapes to the total area of the boundary for each sample in the data set
and report the average percentages in \revise{see }{}\cref{over_underfill}.
The inward distributed scheme has a calculated overfill of \revise{\SI{0.24}{\percent}}{\SI{0.30}{\percent}} and an underfill of \revise{\SI{0.17}{\percent}}{\SI{0.24}{\percent}}.
This is lower compared to the uniform scheme, which results in \revise{\SI{1.2}{\percent}}{\SI{1.63}{\percent}} overfill and \revise{\SI{1.7}{\percent}}{\SI{1.62}{\percent}} underfill in the data set.
\iffalse
names ['naive', 'Constant', 'Center', 'Distributed', 'InwardDistributed']
overfill [  1.63223926  22.21073295   0.37059604   0.45119725   0.30226112]
outer_underfill [ 0.04664107  0.03232884  0.04357603  0.03494708  0.03475748]
underfill [ 1.62009421  0.0712721   0.76886316  0.2440984   0.24461937]
\fi

\subsubsection{Uniformity}
We visualize the bead widths resulting from the different schemes in the bottom of \cref{visualized_accuracy}.
\revise{
We calculated the mean and standard deviation of the bead width, sampled at \SI{1}{\micro\meter} along the toolpaths.
\cref{widthHistogram} shows the distribution of extrusion width for each scheme binned at intervals of \SI{10}{\micro\meter}.
}{
We binned the toolpaths into width bins at \SI{0.01}{\milli\meter} increments and determine the total toolpath length pertaining to each bin.
From these statistics we calculate the mean and standard deviation and report them in \cref{widthHistogram}.
}
We found that the mean width of the inward and evenly distributed schemes is close to \revise{}{the }preferred bead width of \revise{\SI{400}{\micro\meter}}{\SI{0.5}{\milli\meter}}, while their standard deviation is lower than for the centered and constant bead count scheme. 
These results show that, while causing less overfill and underfill, inwards distributed and evenly distributed schemes deviate less or less often from the preferred bead width compared to the other schemes.
\revise{
We compared the width uniformity of the 6 outer beads for the inward and evenly distributed schemes, the distribution of these extrusion widths is shown in \cref{widthIndexedHistogram}. 
The outer beads of the inward distributed scheme deviate less from the preferred width compared to the evenly distributed scheme.
}{}

\revise{
\subsubsection{Smoothness of toolpaths}
In order to maintain a high printing speed, it is desirable that toolpaths have fewer and less sharp corners. 
We therefore measured the angle between consecutive extrusion segments generated by each scheme
and report on the occurrence of each angle in \cref{smoothness}.
All schemes show a higher number of corners for smaller angles with a peak towards \SI{180}{\degree} (straight).
We observe that compared to the uniform method our framework produces less acute angles which, but more obtuse angles.
The inward distributed scheme produces an order of magnitude less corners around \SI{150}{\degree}, compared to evenly distributed. 
We also investigated the effect of the transition regions on the smoothness of the toolpaths. 
\cref{smoothnessNoTransition} shows that introducing the transitions greatly reduces the number of corners around \SI{90}{\degree}. 
}{
\subsubsection{Print time}
The total time it takes to print a part is influenced not only by our back pressure compensation scheme, but also by the geometry of the toolpaths.
In order to separate these effects we report on the total print time when using back pressure compensation and when using a constant (maximum) movement speed in \cref{printtime}.
We estimate print times using a simulation of the Marlin firmware using the default movement settings of the setup described in \cref{print_results_section}.
While the idealized print time is predominantly determined by the total toolpath length, the print time using back pressure compensation is predominantly determined by the occurrence of wide beads, because they have a reduced the flow in \si{\milli\meter\cubed\per\second}.
Because of acceleration constraints imposed by the hardware the maximum movement speed is not reached near sharp corners.
We therefore also report on the angles of the bends in the toolpaths in \cref{smoothness}.
Furthermore, the print time is negatively affected by discontinuities in the extrusion process.
Between extrusions the printer needs to stop extrusion, travel to the next extrusion path and restart the extrusion process, which may introduce defects and incurs extra print time.
For closed polygonal toolpaths we can start anywhere within the path, so we can optimize the starting location so as to minimize the travel time.
We therefore report both on the open and closed path count in \cref{pathcounts}.
}

\subsubsection{Computational performance}
\cref{computime} plots the computation time against the vertex count of the layer for the full data set, comparing the uniform technique implemented using Clipper~\cite{johnson2014clipper} to our framework with the inward distributed scheme.
For polygonal shapes with as many as $10^4$ vertices, the computation for both approaches is less than 1 second, with our method being approximately five times that of the uniform technique.
These results could be improved upon by utilizing the locality inherent in our algorithms for parallelization on the GPU.

The computational complexity is limited by the generation of the Voronoi Diagram, which is $O(n \log n)$, where $n$ is the number of vertices in the input shape.
The other steps in our framework have a complexity of $O(m)$, where $m$ is the number of elements in the ST.
Therefore, the total running time of our algorithm is $O(n \log n)$.
Results in \cref{computime} confirm that both our framework and the uniform technique have an expected running time of approximately $5 \times 10^{-6} n \log n$ seconds.

\revise{
Test prints were performed on a custom FDM hardware setup, with a standard \SI{0.4}{\milli\meter} nozzle and a filament extrusion drive directly mounted on the print head.
The firmware of the printer employs \emph{linear advance} for accurately realizing adaptive deposition width:
gaining the extra pressure required to change to a wider bead is realized by 'advancing' an extra amount of filament into the physical system~\cite{tronvoll2019investigating}.
We set the preferred width to $w^* = \SI{0.6}{\milli\meter}$, to avoid fluttered printing of lines narrower than the nozzle size.
We used a layer thickness of \SI{0.2}{\milli\meter} and a movement speed of \SI{10}{\milli\meter\per\second}.

The prints are shown in \cref{prints}.
Because of inaccuracies in the deposition control system some of the prints show defects.
Such defects are less prevalent for the inward distributed beading scheme than the other schemes.
The prints which employ the uniform offsets technique show a lot of underfill, which impacts the visual quality of the print.
Moreover, in the case of the regular honeycomb there are several fully disconnected hexagons, which means the object falls apart.
The difference between the centered and the inward distributed schemes is less pronounced.
We can still see some loosely connected extrusion segments in both, which is attributable to inaccuracies in the extrusion system.
However, the middle of \cref{wedge_print} exhibits more defects in the regions where the centered scheme produces extreme bead widths.
The bottom print shows that the inward distributed beading scheme produces smoother prints with less defects.

\begin{figure}
\centering
\begin{subfigure}{\columnwidth}\centering
\setlength{\figwidth}{\columnwidth}
\includegraphics[width=\figwidth]{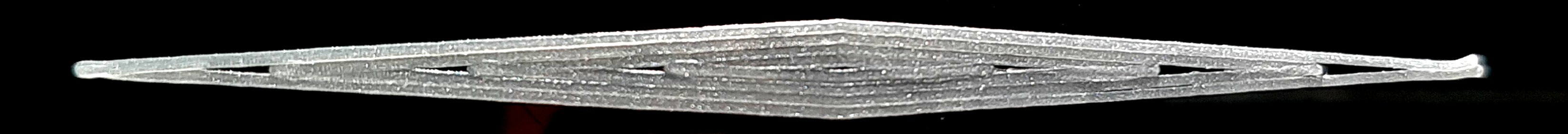}
\includegraphics[width=\figwidth]{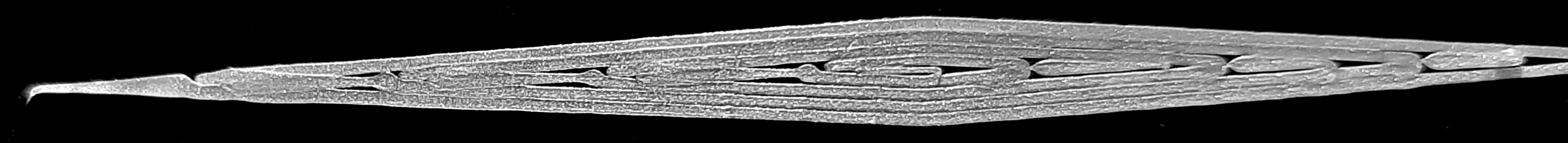}
\includegraphics[width=\figwidth]{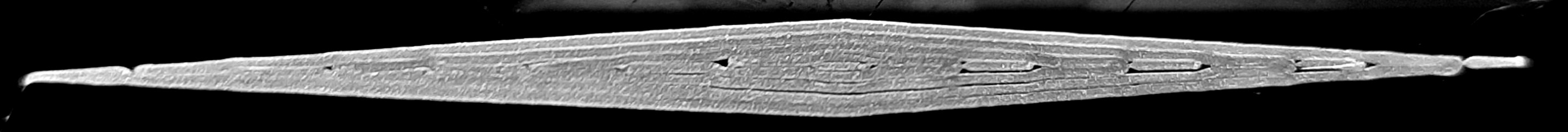}
\caption{Uniform (top), centered (mid) and inward distributed (bottom)}\label{wedge_print}
\end{subfigure}
\setlength{\figheight}{.38\columnwidth}
\setlength{\figheightTwo}{.47\columnwidth}
\setlength{\figwidth}{0.32\columnwidth}
\begin{subfigure}{\figwidth}\centering
\includegraphics[height=\figheight]{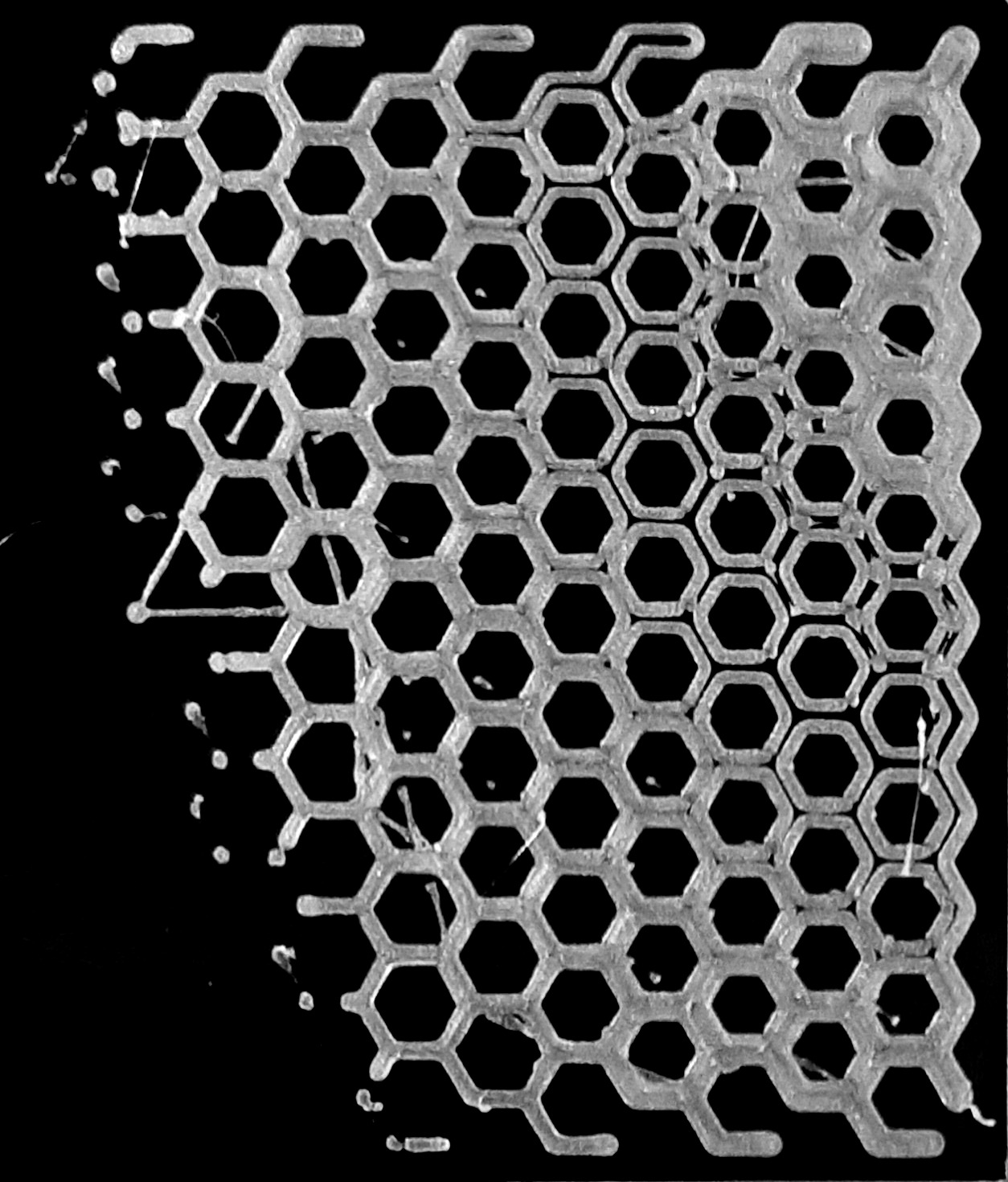}
\includegraphics[width=\figwidth]{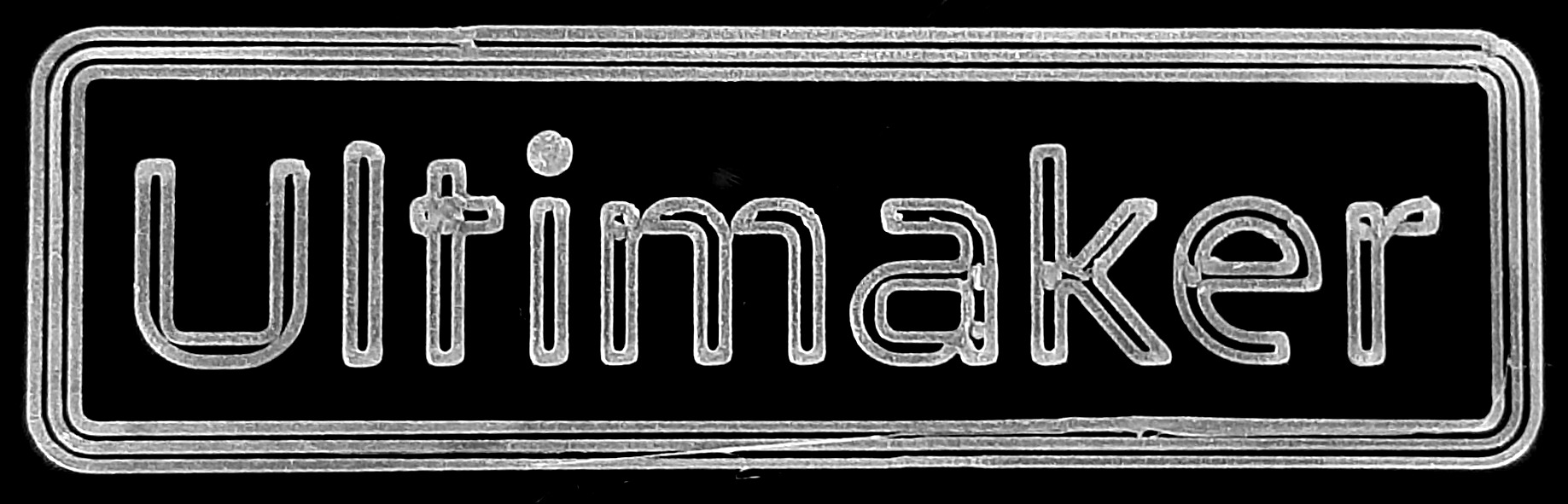}
\caption{Uniform}\label{print_naive}
\end{subfigure}
\begin{subfigure}{\figwidth}\centering
\includegraphics[height=\figheight]{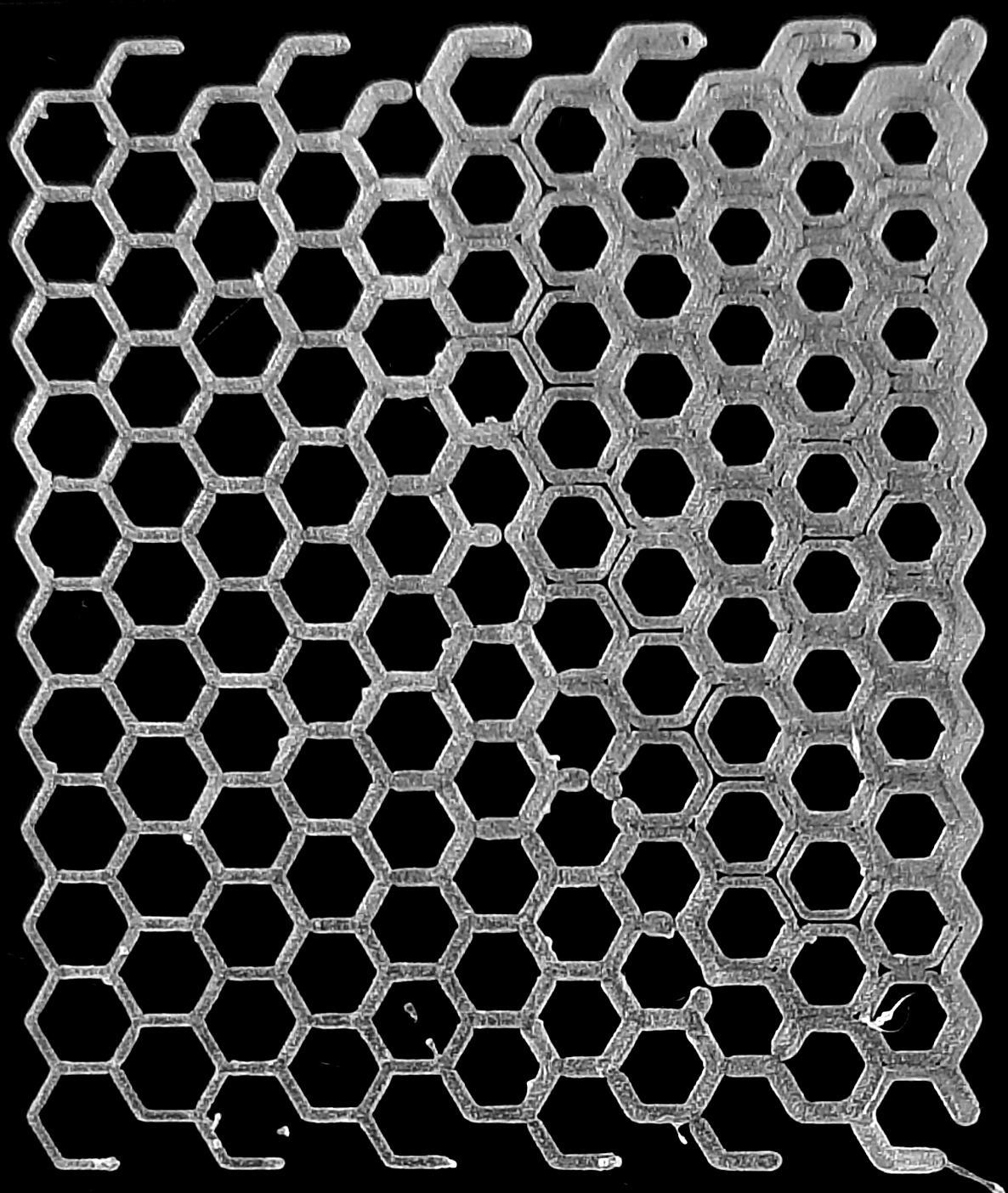}
\includegraphics[width=\figwidth]{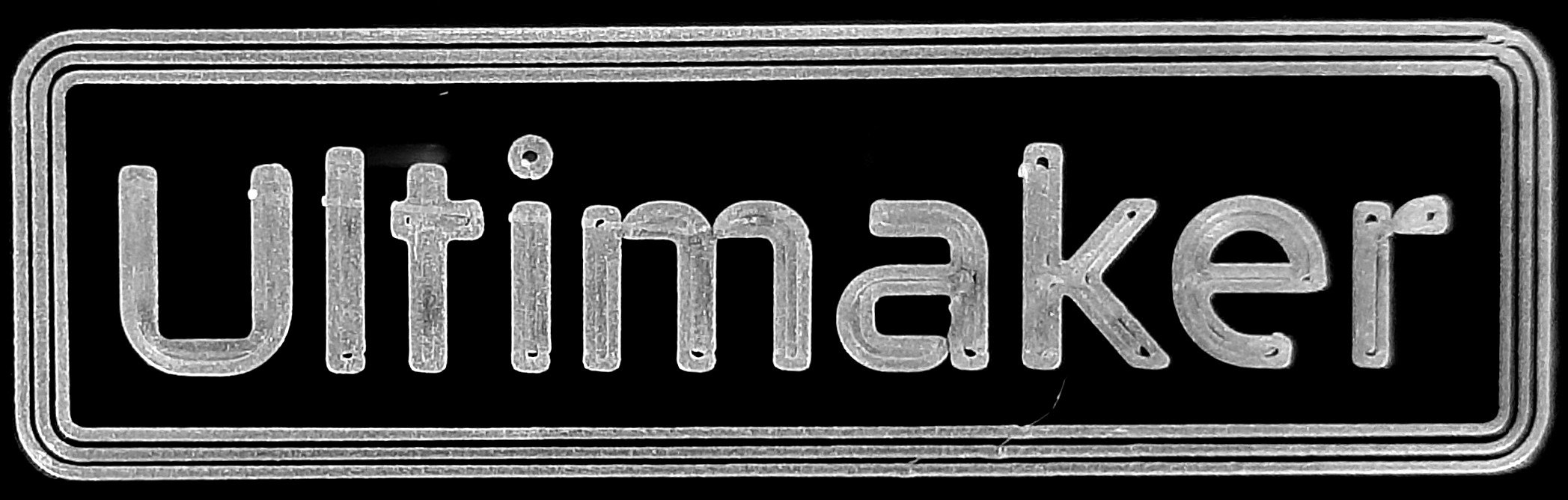}
\caption{Centered}\label{print_center}
\end{subfigure}
\begin{subfigure}{\figwidth}\centering
\includegraphics[height=\figheight]{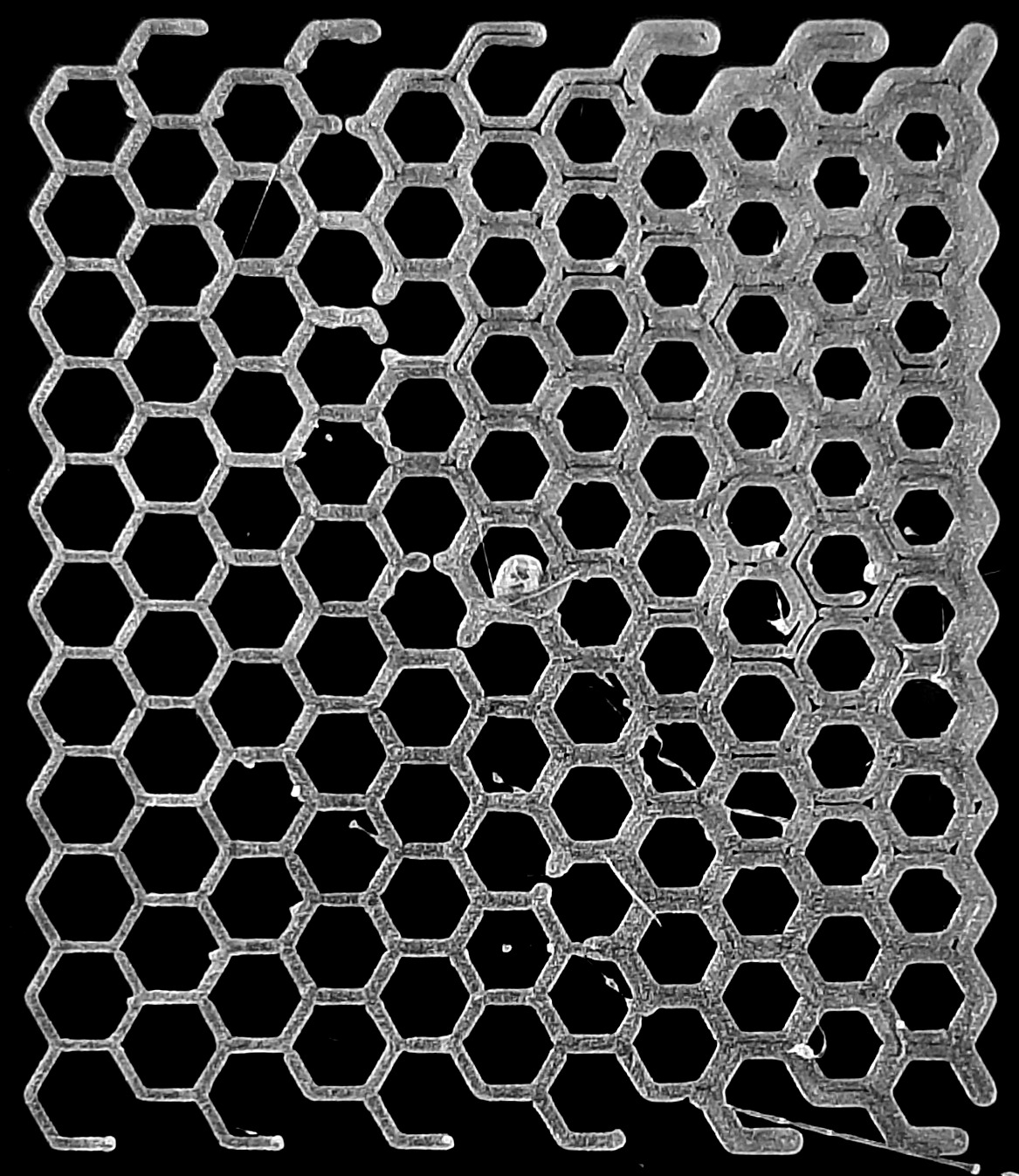}
\includegraphics[width=\figwidth]{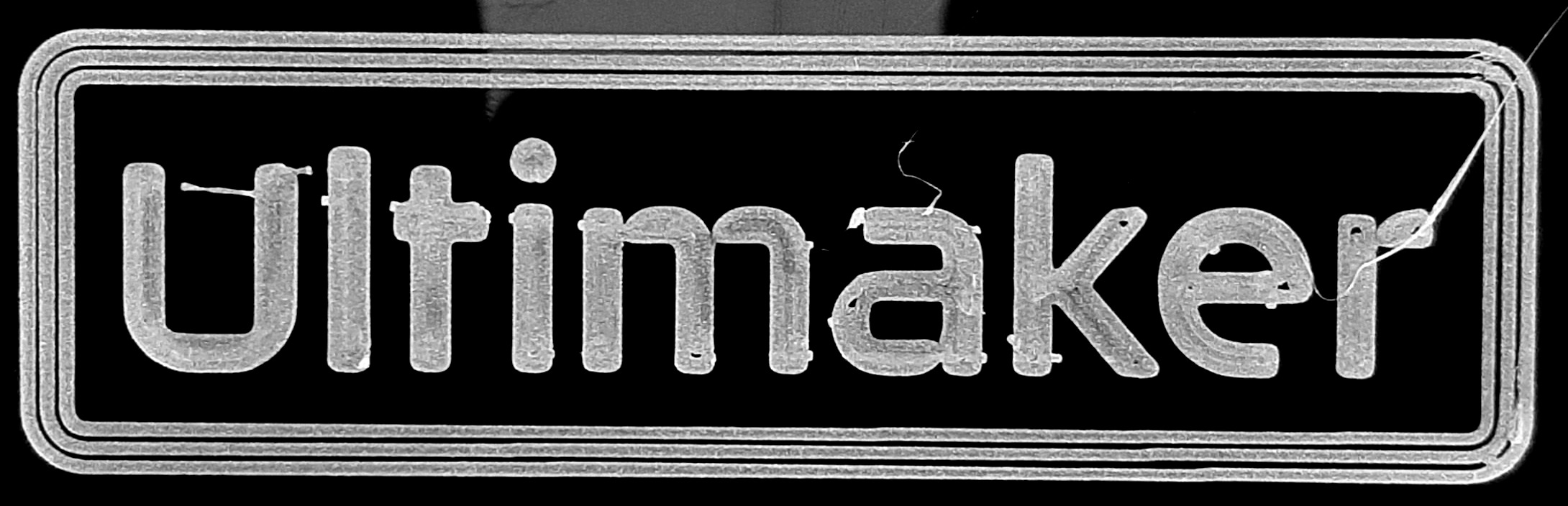}
\caption{Inward distributed}\label{print_inward}
\end{subfigure}
\caption{
Test shapes printed using the uniform scheme, centered scheme and the inward distributed scheme.
The uniform technique produces distinct underfill areas.
The centered scheme shows some defects due to inaccurate control of extreme deposition widths.
The inward distributed scheme produces the least defects.
}
\label{prints}
\end{figure}
}{}

%We therefore visualize include a semi-circle with a diameter equal to the starting width in the one end, and exclude it at the other end, because it will be included in the next extrusion segment.

%% file: 8_discussion.tex
%\section{Discussion}

\subsection{Comparison of beading schemes}
We can see from \cref{TEST_naive_accuracy}(top) and~\ref{over_underfill} that the uniform technique causes a lot of overfills and underfills: on average \revise{approximately}{} \revise{\SI{1}{\percent}}{\SI{1.6}{\percent}} of the total target area is covered by underfill and likewise for overfill.
To our knowledge, the uniform beading scheme, as well as the outer beading scheme, is of little use to FDM printers.

The constant bead count scheme effectively deals with underfills, but generates orders of magnitude more overfills compared to the other schemes. 
Also, the scheme comes at the cost of greatly varying bead widths and an average bead width that is not close to the preferred bead width.
Note that most overfill areas occur near regions of alternating bead width. 
\revise{}{While the scheme results in short toolpaths, as indicated by the idealized print time, it also results in a wide range of bead widths, which cause the back pressure compensation print time to be very large.
See \cref{statisticsfig}.}
For an input outline shape which contains both very small and very large features, the constant bead count scheme produces bead widths which can fall outside of the range of manufacturable bead widths.
Moreover the centrality marking is not robust against small perturbations in the outline; adding a small chamfer in a corner causes the unmarked ST to be very small at that location, which results in tiny bead widths.
\revise{}{See top right of \cref{TEST_Constant_accuracy}.}

In \cref{TEST_Center_accuracy} we can see that
the centered beading scheme effectively deals with \revise{both }{}overfill \revise{and underfill }{}and produces desired bead widths in all locations, except for the extrusion paths in the center, where the bead widths \revise{are within a factor 2 off from the desired bead width.}{range between $0.25 w^*$ and $1.8w^*$.}
\revise{}{However, it does produce some narrow underfill regions.}
\revise{%gs
According to \cref{over_underfill} the overfill and underfill for the centered, the evenly distributed and the inward distributed scheme are all approximately \SI{0.2}{\percent}, which is a considerable improvement over the uniform technique.
}{%sg
Compared to the uniform technique the centered technique increases the (open) path count, but considerably reduces over- and underfill and decimates the number of toolpath angles below \SI{45}{\degree}.
See \cref{statisticsfig}.%afs
}%zgds

However, according to \cref{widthHistogram} the centered scheme exhibits a wider range of bead widths than the distributed schemes:
the standard deviation of the bead widths in the centered scheme is approximately \revise{\SI{39}{\micro\meter}}{\SI{53}{\micro\meter}}, while that of the distributed schemes is approximately \revise{\SI{14}{\micro\meter}}{\SI{23}{\micro\meter}}.%_
\revise{}{\footnote{\revise{}{Although the standard deviation $\sigma$ of the inward distributed scheme is slightly higher than that of the evenly distributed scheme, the mean absolute deviation is lower (i.e. \SI{9}{\micro\meter} versus \SI{11}{\micro\meter}), because its distribution is more peaked.}}}
\revise{}{Moreover, because the quantization operator rounds to the nearest number of beads, in the worst case where we switch from a single to two beads the widths switch from $0.75w^*$ to $1.5w^*$, which is a considerably smaller range than in the centered scheme.}
We therefore conclude that the distributed schemes \revise{result in bead widths closer to the preferred widths}{exhibit a lower bead width variation and lower (open) path count} compared to the centered scheme.
%This is desirable for the manufacturability of the beads and can therefore have a positive effect on the mechanical properties and surface quality of the 3D prints. 

\revise{It is hard to visually identify the difference between the evenly and the inward distributed scheme in \cref{visualized_accuracy}, because that particular example shape does not have wide features.}
{\Cref{distributed_comparison,TEST_Distributed_accuracy,TEST_InwardDistributed_accuracy} show that in the inward distributed scheme the outer toolpaths have the preferred width more often than in the evenly distributed scheme, which means that the outline accuracy of the inward distributed beading is less affected by inaccuracies in the adaptive width control system.}
%{\Cref{TEST_InwardDistributed_accuracy} shows that in the inward distributed scheme the outer beads are more often equal to the preferred width compared to the evenly distributed scheme in \cref{TEST_Distributed_accuracy}, but that effect is more pronounced for wider geometry such as in \cref{distributed_comparison}.}
%While the difference between the evenly and inward distributed scheme in \cref{visualized_accuracy} results only in the outer bead being the preferred width in some locations, the effect of the inward distributed scheme is more pronounced in larger geometry, as can be seen in \cref{distributed_comparison}.}
\revise{However, \cref{distributed_comparison} and  \cref{widthIndexedHistogram} confirm that the outer toolpaths have the preferred bead width more often.}{}% _
Furthermore, \revise{from \cref{smoothness} }{}we find that compared to evenly distributed, the inward distributed scheme produces \revise{smoother toolpaths overall}{less corners with angles above \SI{130}{\degree} and less overfill, because the area of influence that bead count transitions have is limited in the inward distributed scheme}.
Thus the inward distributed scheme prevents over- and underfill, generates smooth toolpaths with more homogeneous width and affects smaller more centered parts of the print than the other schemes\revise{}{, while incurring little to no extra print time}.

\revise{}
{
\subsection{Limitations}
% design considerations
Because the performance of the various toolpathing techniques depends on the geometry of a model, they have ramifications for the practice of design for additive manufacturing.
Because the naive method produces under- or overfill for parts of an outline with a constant diameter $d \neq 2 i w^*$ it is best practice to design a model such that horizontal cross-sections have a feature diameter of an even integer multiple $i$ of the bead width.
For the center\revise{ed} scheme and for the current state of the art one should only avoid parts for which $(2i + 1.8) w^* < d < (2i + 0.25) w^*$ in order to avoid underfill.
For the distributed schemes however, there is no diameter at which the framework produces under- or overfill for a part with a constant diameter $d$.
The design consideration therefore reduces to limiting the diameter of your parts to be within the range $[w_\text{min}, \infty)$,
where $w_\text{min}$ is a configurable parameter when using the widening meta-scheme.

% single line segments & continuity
The default limit bisector angle $\alpha_\text{max} = \SI{135}{\degree}$ ensures that we don't employ transitioning in shallow wedge regions, which would result in a lot of short odd single bead polylines, which would break up the semi-continuous nature of polygonal extrusion paths;
$\alpha_\text{max} = \SI{135}{\degree}$ corresponds to $w^* / \cos \nicefrac12 \alpha_\text{max} \approx \SI{0.4}{\milli\meter}$ long segments
and under-/overfill areas of $\nicefrac14 (w^*)^2 \left( \tan ( \alpha / 2) - \alpha / 2 \right) \approx \SI{0.05}{\milli\meter\squared}$.
However, future work might be aimed at reducing under-/overfill in regions with a low bisector angle without the introduction of short single polyline extrusion segments.
If the over-/underfill problem is also solved for non-significant regions we might be able to increase $\alpha_\text{max}$ and reduce the discontinuity introduced by short extrusion segments.

% single beads
Another limitation of our method is that in a location $v$ with locally maximal $R(v) \approx (i + \nicefrac12) w^*$ the odd bead count will result in a single polyline extrusion segment consisting of only a single point.
This can be viewed in the bottom right of \cref{TEST_InwardDistributed_accuracy} for example.
In order to print such a dot, we make it into a \SI{10}{\micro\meter} long extrusion segment, with an altered width such that the total volume remains correct.
A more principled way of dealing with such a situation remains future work.

% emulation troubles
Finally it should be noted that although our framework can accurately emulate the constant bead count approach by \citeauthor{Ding2016a}, its emulation of the centered approach by \citeauthor{Jin2017JMS} is imperfect.
The transitions resulting from out framework introduce sharper corners and there is more width variation in those corners.
Whereas the width of the connecting segment in the approach by \citeauthor{Jin2017JMS} is the preferred width $w^*$, the bead widths closer to the center resulting from our framework will be twice the local radius, which is larger than $w^*$.
However, this inflated bead width variation is expected to have an insignificant impact on the measured bead width variation.
}

\input{10_applications}

\revise{
\subsection{Discussion on implications}
Note that the current industry standard in FDM printing employs little to no bead width variation.
Properly performing bead width variation calls for adaptations and developments in printers and firmware.
In the beading schemes we set a transition length of $t(n) = w^*$.
That will demand changes in cross-sectional area of the bead up to \SI{200}{\percent} over a small distance that is comparable to the nozzle size, which is challenging for some hardware systems.
Varying the movement speed can be utilized to change the cross-sectional area, but this approach is limited, since the movement speed is constrained by acceleration considerations near bends in the toolpath~\cite{Ertay2018,Kuipers2018}.
Our schemes require a more accurate control of the volumetric flow rate in \si{\milli\meter\cubed\per\second}.
Using a filament feeder directly mounted on the print head (a.k.a. direct drive) can control the flow more dynamicaly then FDM printers where the material is fed through a Bowden tube from a feeder mounted on the frame.
Still direct drive printers require some control system in order to accurately change the volumetric flow rate such as pressure advance algorithms~\cite{tronvoll2019investigating}.
Yet inaccuracies in direct drive systems employing advance algorithms might arise due to the changes in back-pressure required by changing bead size.
We expect that developments in printing hardware and firmware will address these challenges in the future.

Another limiting factor for adopting adaptive bead width is the format of G-code which stores machine instructions.
G-code does not support moves with varying cross-sectional area.
A typical extrusion move \lstinline!G1 X$x$ Y$y$ E$v$! only specifies the total amount of volume $v$ to be extruded in the move, not how that total amount should be distributed along the extrusion move.
A workaround is to approximate a variable width extrusion segment by smaller segments with constant width.
However, this introduces errors nevertheless.
Ideally the G-code language would be expanded in some way to allow for extrusion segments with varying cross-sectional area.
}{}

% Taking a broader perspective, we note that our proposed inward distributed scheme is a pragmatic solution.
% Rather than deriving some optimal beading scheme from a clear specification of the objective, we propose some arbitrary inward distributed beading scheme and show that it is better than the other beading schemes.
% An optimal beading scheme can be derived if the objective is formalized terms of a unambiguous fitness function, but that would depend on the specific hardware setup and application for which toolpaths are generated.
% This manuscript is therefore limited to showing the flexibility and versatility of the framework, rather than deriving an optimal beading scheme.

%% file: 10_applications.tex
\subsection{Applications}
Toolpath\revise{}{s} with varying width is particularly meaningful for narrow parts, since there the negative effect of under- and overfill is more pronounced than in wide parts.
In extreme cases, thin features will not be filled at all.
Therefore, our framework, while working for wide parts as well, shows most of its potential for objects which contain thin parts.

\begin{figure*}
\centering
\setlength{\figwidth}{0.099\textwidth}
\setlength{\figheight}{0.099\textwidth}
\begin{subfigure}{\textwidth}\centering
\includegraphics[width=\textwidth]{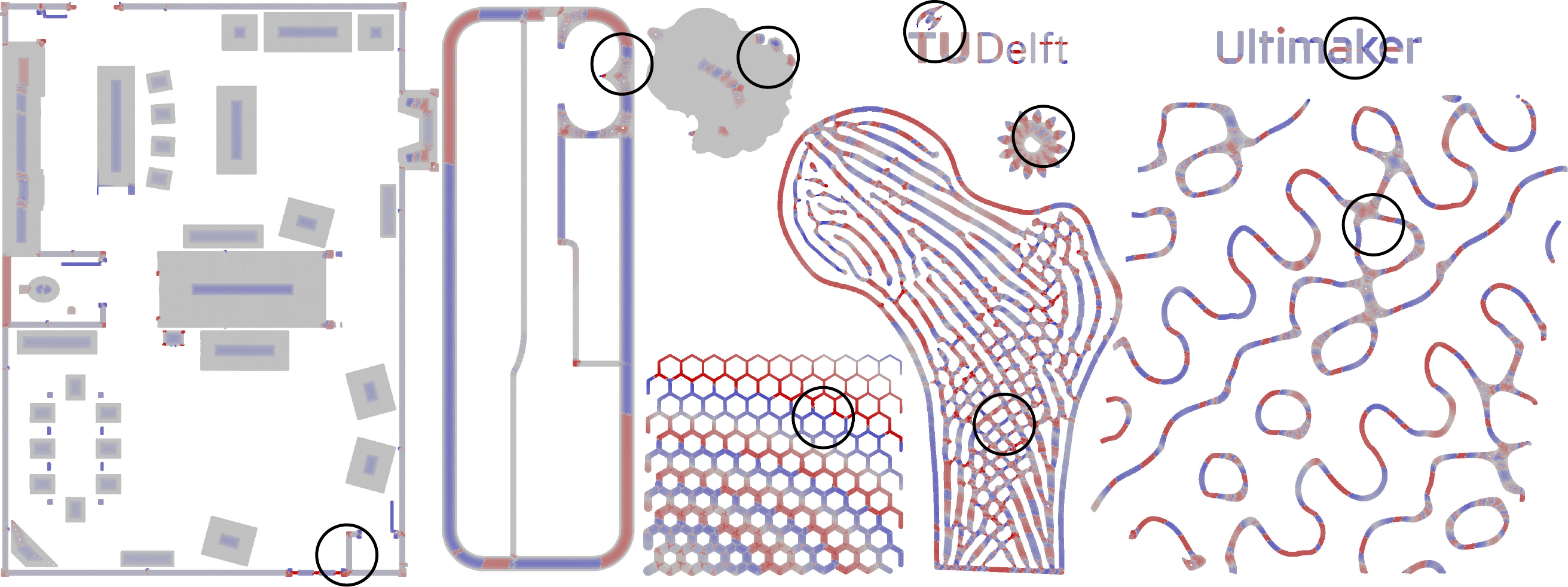}
%\caption{Overview}\label{applications_overview}
\end{subfigure}
\begin{subfigure}[t]{\figwidth}\centering
\includegraphics[height=\figheight]{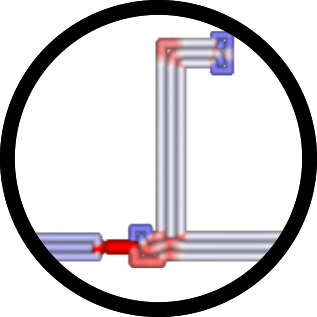}
\caption{House}\label{applications_house}
\end{subfigure}
\begin{subfigure}[t]{\figwidth}\centering
\includegraphics[height=\figheight]{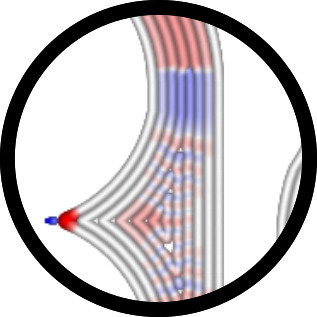}
\caption{Case}\label{applications_case}
\end{subfigure}
\begin{subfigure}[t]{\figwidth}\centering
\includegraphics[height=\figheight]{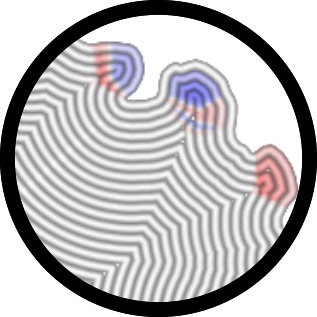}
\caption{Statue}\label{applications_statue}
\end{subfigure}
\begin{subfigure}[t]{\figwidth}\centering
\includegraphics[height=\figheight]{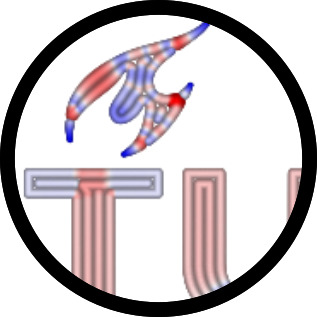}
\caption{TUD}\label{applications_tud}
\end{subfigure}
\begin{subfigure}[t]{\figwidth}\centering
\includegraphics[height=\figheight]{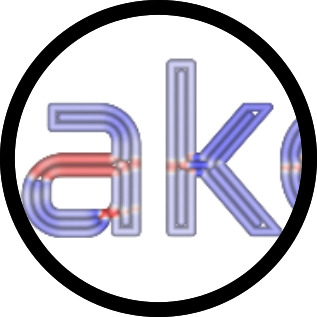}
\caption{UM}\label{applications_um}
\end{subfigure}
\begin{subfigure}[t]{\figwidth}\centering
\includegraphics[height=\figheight]{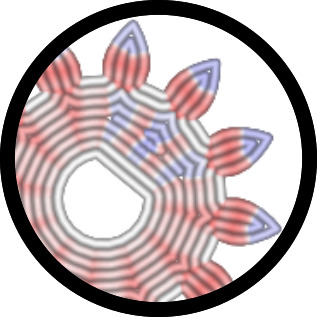}
\caption{Gear}\label{applications_gear}
\end{subfigure}
\begin{subfigure}[t]{\figwidth}\centering
\includegraphics[height=\figheight]{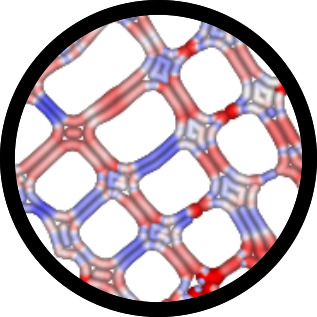}
\caption{Bone}\label{applications_bone}
\end{subfigure}
\begin{subfigure}[t]{\figwidth}\centering
\includegraphics[height=\figheight]{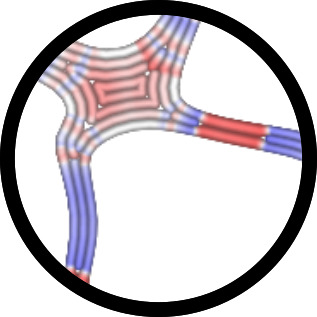}
\caption{Gyroid}\label{applications_gyroid}
\end{subfigure}
\begin{subfigure}[t]{\figwidth}\centering
\includegraphics[height=\figheight]{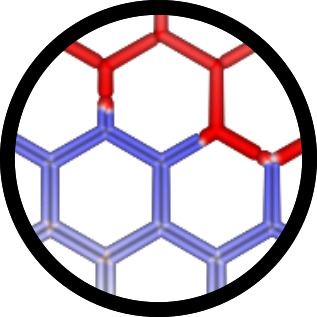}
\caption{Hex}\label{applications_hex}
\end{subfigure}
\begin{subfigure}[t]{.3\figwidth}
\includegraphics[height=\figheight]{sources-validation-widths-legend-small.pdf}
\end{subfigure}
\caption{
Visualization of the widths for the output toolpaths of the inward distributed beading scheme \revise{}{($N=3$) }applied to various example application objects.
From left to right and top to bottom: a house, a case for electronics, a statue, two common logos, a gear, a topologically optimized bone structure, a tilted homogeneous gyroid structure and a heterogeneous thickness hexagonal grid.
}
\label{applications_overview}
\end{figure*}

\Cref{applications_overview} collectively shows the application of the proposed inward distributed scheme for various types of 3D model, including both thin parts (architectural models, casings, embossed text, gears and microstructures) and wide parts (\cref{applications_case}) and organic shapes (\cref{applications_statue})).

For architectural models and casings, preventing over- and underfill is expected to make them stronger. 
For embossed text, preventing underfill reduces the various holes in the top surfaces, which is detrimental to the visual quality of those top surfaces.
For gears and similar mechanical parts that are designed with finite element analysis, the less variation in extrusion widths is closer to the assumptions under fast analysis (e.g. using homogenization~\cite{Liu2016CAD}).

Of particular interest are microstructures that could be uniquely fabricated by 3D printing.
For example, topology optimized bone-like structures~\cite{wu2017infill} contain filaments of varying thickness that follow a varying stress direction (\cref{applications_bone}).
An angled Gyroid structure with uniform thickness also results in outline shapes with varying width (\cref{applications_gyroid}). 
These structures are accurately densely filled using our framework.
Another class of microstructures consists of parameterized patterns with varying thickness to achieve functional gradation.
\Cref{applications_hex} shows the contour-parallel toolpath\revise{}{s} with varying width of a hexagonal grid neatly switches between different bead counts over the volume, preventing the jagged moves a direction-parallel toolpath\revise{}{s} would create for such a case~\cite{bates2018compressive}.

%% file: 12_conclusions_future_work.tex
\section{Conclusion}
In this paper we have introduced a framework for computing contour-parallel toolpaths employing adaptive bead width in order to minimize underfill and overfill areas.
\revise{}{%af
We introduced beading schemes which improve on the state of the art,
and we have introduced a back pressure compensation method for accurate fabrication of adaptive width.
}

Our framework is flexible, demonstrated by the several beading schemes which emulate existing techniques.
The computation times of our framework are on par with the state-of-the-art library for performing offsets of non-adaptive bead width.
Our framework is stable: small local changes in the outline shape cause only small changes in the toolpath.

\revise{We presented}{Compared to the state of the art,} the inward distributed beading scheme
\revise{It}{} reduces the amount of beads with a width deviating extremely from the preferred bead width \revise{}{by changing the width of several beads near the center instead of only the center-most bead}.
\revise{, and thus it is}{It is therefore} expected to limit the impact of varying the bead width in terms of production accuracy and homogeneity of material properties,
which in turn is helpful to efficiently simulate an FDM manufactured part.
\revise{
Furthermore, by distributing the deviation in bead width over several beads near the center of the shape, the outer toolpath is affected less by the deviation, meaning that the dimensional accuracy of the shape of the outline contour is affected less by inaccuracies in the mechanical control system which realizes adaptive bead width.
}{}

The proposed beading scheme greatly improves the process planning for parts with thin contours, which often occur for example in architectural models, prototypes for casings or microstructures.
Meanwhile it leaves most of the toolpaths the same as the uniform width technique in large features, meaning that existing studies which relate process parameters with mechanical properties of the print are still applicable.
\revise{}{Compared to the naive approach of constant width toolpaths our beading scheme is expected to improve the stiffness, dimensional accuracy and visual qualities of the manufactured model.}
\revise{}{It is expected that as distributed beading schemes are implemented in commercial software packages and bead width variation control become commonplace, the practice of design for additive manufacturing can disregards some of the nozzle size considerations.}

\medskip
\revise{}{The presented framework is open source available at \\ \url{github.com/Ultimaker/libArachne}}

\revise{
\subsection{Future work}
The work presented here is evaluated mostly on a computational level, because the effectiveness of a mechanical system to accurately print an adaptive bead width interferes with the physical validation of our techniques.
Future research could be devoted to optimizing the distribution scheme to a specific hardware setup.
An accurate model of manufacturability, print speed and dimensional accuracy would be valuable to fine-tune the beading scheme.
% If we can assign a loss function to several aspects (e.g. manufacturability, print speed, dimensional accuracy) of a toolpath we could deduce the optimal beading scheme.
}{}

\revise{
Our framework uses the skeleton to determine toolpaths based on the local feature size.
Another direction of future research could be devoted to incorporating other information, e.g. properties of the outline such as curvature or visibility, non-local features such as the size of nearby outline features, toolpaths of the previous layer, or voluminal constraints (e.g. for functionally graded materials).
% Furthermore we could also take into account volumetric considerations such as the toolpathing of the previous layer or volume constraint defined by a functionally graded material.
}{}
% Taking a broader perspective we note that the skeletonization decomposes a shape into trapezoids and these trapezoids are decomposed into quads by the toolpath segments which we generate for it.
% It would be interesting to see if similar technique can be used to generate quad dominant meshings for finite element analysis.

%% file: 19_edge_discretization.tex
\section{Edge discretization}\label{edge_discretization}
We calculate the location $l$ of the boundary between a significant and nonsignificant portion of an edge analytically.
For example, the parabolic MAT edge generated from the outline vertex $(0,1)$ and an outline segment aligned with the X-axis follows $y(x) = \nicefrac12 x^2$ and $R(x) = y(x)$.
We can determine the significant portion $[-x_\text{bound}, x_\text{bound}]$ by evaluating $\frac{\partial R}{\partial x} > \cos(\alpha_\text{max} / 2)$, which is $ | x_\text{bound} | = (\tan(\alpha_\text{max} / 2))^{-1}$.
Similarly, a MAT edge generated from two vertices at $(0,0)$ and $(0,1)$ follows $y(x) = \nicefrac12$ and $R(x) = \sqrt{\nicefrac14 + x^2}$.
The boundaries of significance are given by $ | x_\text{bound} | = \nicefrac12 (\tan(\alpha_\text{max} / 2))^{-1}$.
From these we can derive the locations $l = (\pm x_\text{bound}, y(x_\text{bound}))$.
These specific cases can easily be transformed into all possible cases using scaling and rotation operations.

%% file: 20_dataset.tex
\section{Data set}\label{dataset}
The data set we tested on was a custom selected set of open source 3D models found on the internet which was selected to cover a broad range of different types of application and geometry.
Applications range from prototypes, to fixtures and mechanical end-use parts.
The geometry covers a wide range including thin filaments, smooth surfaces, organic shapes, chamfered shapes, small shapes and large shapes.
The models are described in \cref{dataset_description}.

\begin{table}
\caption{3D models used for validation}\label{dataset_description}
\newcommand{\ssize}{\footnotesize}
\begin{tabular}{l l l}
Model Name & Author \\
\hline
\href{https://www.thingiverse.com/thing:26555}{\ssize AirCasting Air Monitor Casing} & \ssize HabitatMap\\
\href{https://www.thingiverse.com/thing:3629434}{\ssize Air hose splitter} & \ssize frizinko\\
\href{https://www.thingiverse.com/thing:1155772}{\ssize Al Hamra Tower} & \ssize TurnerConstructionCompany\\
\href{https://www.thingiverse.com/thing:1498967}{\ssize canon NP-E3 battery cap} & \ssize kosuyoung\\
\href{https://www.thingiverse.com/thing:3567409}{\ssize David} & \ssize Thunk3D\\
\href{https://www.thingiverse.com/thing:3132621}{\ssize Deck Assembly Tool \comment{, Plank Screwing Tool}} & \ssize PSomeone\\
\href{https://www.thingiverse.com/thing:2920060}{\ssize Ender 3 Cable Chain} & \ssize johnniewhiskey\\
\href{https://www.thingiverse.com/thing:2993875}{\ssize Ergonomic Hacksaw Handle} & \ssize mmOne\\
\href{https://www.thingiverse.com/thing:2513922}{\ssize Gap measurement tool} & \ssize ravm84\\
\href{https://www.thingiverse.com/thing:1673030}{\ssize G-Clamp fully printable} & \ssize johann517\\
\href{https://www.youmagine.com/designs/gyroid}{\ssize Gyroid} & \ssize Tim Kuipers\\
\href{https://www.thingiverse.com/thing:1327093}{\ssize 3D Printable Jet Engine} & \ssize CATIAV5FTW\\
\href{https://www.thingiverse.com/thing:2752165}{\ssize Lawn Mower Throttle Replacement} & \ssize Spammington\\
\href{https://www.thingiverse.com/thing:2854328}{\ssize OpenRC F1 Internal gear box mod} & \ssize intoxikated\\
\href{https://www.thingiverse.com/thing:3592328}{\ssize PCB Test Fixture} & \ssize JMadison\\
\href{https://www.thingiverse.com/thing:1078865}{\ssize Pioneer Radio Holder for Ford Focus} & \ssize Perugino\\
\href{https://www.thingiverse.com/thing:34596}{\ssize Replicator Dual Fan Mount} & \ssize aubenc\\
\href{https://www.thingiverse.com/thing:3754728}{\ssize Atuador versão 2 * actuator version 2} & \ssize Caroline Holanda\\
\href{https://www.thingiverse.com/thing:1595179}{\ssize TE Pocket Operator Hard case} & \ssize Salvation76\\
\href{https://www.thingiverse.com/thing:3682303}{\ssize Screw sizer} & \ssize Pierrolalune63\\
\href{http://homepage.tudelft.nl/z0s1z/projects/2017-bone-infill.html}{\ssize Bone-like optimized infill} & \ssize Jun Wu\\
\href{https://www.thingiverse.com/thing:26244}{\ssize Two-Story Spec House} & \ssize pwc-phil\\

\end{tabular}
\end{table}

%% file: 25_accuracy.tex
\section{Accuracy calculation}\label{accuracy_calculation}

\begin{figure}
\centering
\setlength{\figwidth}{.25\columnwidth}
\begin{subfigure}{\figwidth}\centering
\includegraphics[width=\columnwidth]{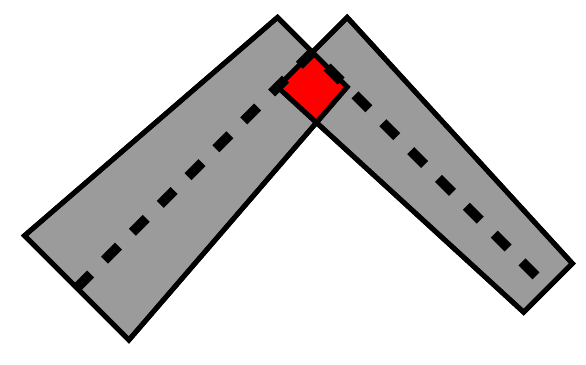}
\caption{Blocky}\label{segment_visualization_blocky}
\end{subfigure}
\begin{subfigure}{\figwidth}\centering
\includegraphics[width=\columnwidth]{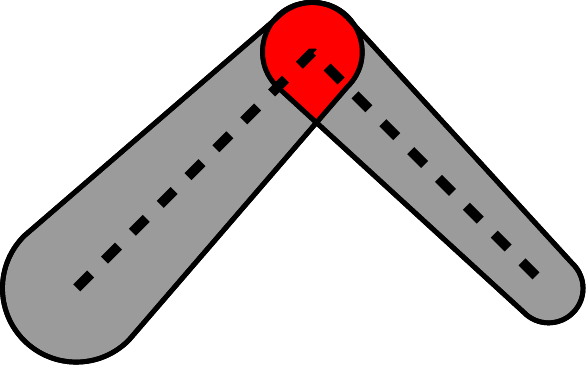}
\caption{Rounded}\label{segment_visualization_rounded}
\end{subfigure}
\begin{subfigure}{\figwidth}\centering
\includegraphics[width=\columnwidth]{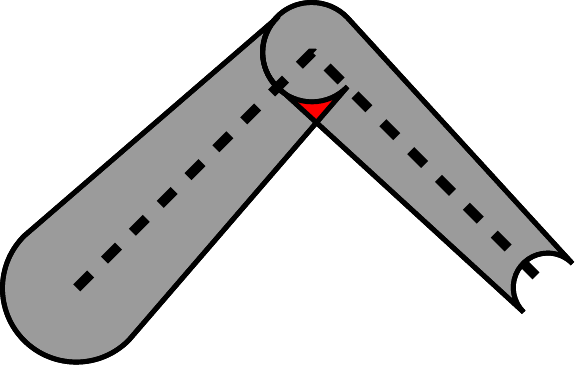}
\caption{Excluded}\label{segment_visualization_excluded}
\end{subfigure}
\caption{
Extruded area of two extrusion segments.
Red areas signify doubly extruded areas.
}
\label{segment_visualization}
\end{figure}

In order to estimate the overfill and underfill, we need to accurately calculate the area covered by a single extrusion path.
If we would simply use an isosceles trapezoidal area, we would get overfill artifacts at corners in the toolpath (\cref{segment_visualization_blocky}).
We therefore use a semi-circle (\cref{segment_visualization_rounded}) with a diameter equal to the starting width in the one end of each segment, and exclude it at the other end, because it will be included in the next segment.
For polyline extrusion paths which are not closed, we also include the semi-circle of the destination location (\cref{segment_visualization_excluded}).

%In order to print such extrusion paths accurately, we can modulate the amount of material flow per millimeter based on this visualization model.
%See \cref{segment_visualization}.

Using boolean operations we can obtain the polygonal regions for overfill and those for underfill.
In order to deal with rounding errors we perform a morphological close of \SI{5}{\micro\meter}, before calculating the total area in \si{\milli\meter\square}.
We also calculate regions which are covered thrice by different extrusion segments and add twice its area to the total overfill area amount.

%% file: 0_main.bbl
\begin{thebibliography}{45}
\providecommand{\natexlab}[1]{#1}
\providecommand{\url}[1]{\texttt{#1}}
\providecommand{\urlprefix}{URL }
\expandafter\ifx\csname urlstyle\endcsname\relax
  \providecommand{\doi}[1]{doi:\discretionary{}{}{}#1}\else
  \providecommand{\doi}[1]{doi:\discretionary{}{}{}\begingroup
  \urlstyle{rm}\url{#1}\endgroup}\fi
\providecommand{\bibinfo}[2]{#2}

\bibitem[{Bates et~al.(2018)Bates, Farrow, and Trask}]{bates2018compressive}
\bibinfo{author}{S.~R. Bates}, \bibinfo{author}{I.~R. Farrow},
  \bibinfo{author}{R.~S. Trask}, \bibinfo{title}{{Compressive behaviour of 3D
  printed thermoplastic polyurethane honeycombs with graded densities}},
  \bibinfo{journal}{Materials {\&} Design} \bibinfo{volume}{162}
  (\bibinfo{year}{2018}) \bibinfo{pages}{130--142},
  \doi{\bibinfo{doi}{10.1016/j.matdes.2018.11.019}}.

\bibitem[{Al-Ketan et~al.(2018)Al-Ketan, Rowshan, and {Abu
  Al-Rub}}]{Al-Ketan2018}
\bibinfo{author}{O.~Al-Ketan}, \bibinfo{author}{R.~Rowshan},
  \bibinfo{author}{R.~K. {Abu Al-Rub}}, \bibinfo{title}{{Topology-mechanical
  property relationship of 3D printed strut, skeletal, and sheet based periodic
  metallic cellular materials}}, \bibinfo{journal}{Additive Manufacturing}
  \bibinfo{volume}{19} (\bibinfo{year}{2018}) \bibinfo{pages}{167--183},
  \doi{\bibinfo{doi}{10.1016/j.addma.2017.12.006}}.

\bibitem[{Maskery et~al.(2018)Maskery, Sturm, Aremu, Panesar, Williams, Tuck,
  Wildman, Ashcroft, and Hague}]{Maskery2018}
\bibinfo{author}{I.~Maskery}, \bibinfo{author}{L.~Sturm},
  \bibinfo{author}{A.~O. Aremu}, \bibinfo{author}{A.~Panesar},
  \bibinfo{author}{C.~B. Williams}, \bibinfo{author}{C.~J. Tuck},
  \bibinfo{author}{R.~D. Wildman}, \bibinfo{author}{I.~A. Ashcroft},
  \bibinfo{author}{R.~J.~M. Hague}, \bibinfo{title}{{Insights into the
  mechanical properties of several triply periodic minimal surface lattice
  structures made by polymer additive manufacturing}},
  \bibinfo{journal}{Polymer} \bibinfo{volume}{152} (\bibinfo{year}{2018})
  \bibinfo{pages}{62--71}, \doi{\bibinfo{doi}{10.1016/j.polymer.2017.11.049}}.

\bibitem[{Zegard and Paulino(2016)}]{Zegard2016SMO}
\bibinfo{author}{T.~Zegard}, \bibinfo{author}{G.~H. Paulino},
  \bibinfo{title}{Bridging topology optimization and additive manufacturing},
  \bibinfo{journal}{Structural and Multidisciplinary Optimization}
  \bibinfo{volume}{53}~(\bibinfo{number}{1}) (\bibinfo{year}{2016})
  \bibinfo{pages}{175--192}, ISSN \bibinfo{issn}{1615-1488},
  \doi{\bibinfo{doi}{10.1007/s00158-015-1274-4}}.

\bibitem[{Wu et~al.(2019)Wu, Wang, and Gao}]{Wu2019a}
\bibinfo{author}{J.~Wu}, \bibinfo{author}{W.~Wang}, \bibinfo{author}{X.~Gao},
  \bibinfo{title}{{Design and Optimization of Conforming Lattice Structures}},
  \bibinfo{journal}{IEEE Transactions on Visualization and Computer Graphics}
  (\bibinfo{year}{2019})
  \bibinfo{pages}{1--14}\urlprefix\url{http://arxiv.org/abs/1905.02902}.

\bibitem[{Cheng et~al.(2019)Cheng, Bai, and To}]{Cheng2019}
\bibinfo{author}{L.~Cheng}, \bibinfo{author}{J.~Bai}, \bibinfo{author}{A.~C.
  To}, \bibinfo{title}{{Functionally graded lattice structure topology
  optimization for the design of additive manufactured components with stress
  constraints}}, \bibinfo{journal}{Computer Methods in Applied Mechanics and
  Engineering} \bibinfo{volume}{344} (\bibinfo{year}{2019})
  \bibinfo{pages}{334--359}, \doi{\bibinfo{doi}{10.1016/j.cma.2018.10.010}}.

\bibitem[{Ding et~al.(2016)Ding, Pan, Cuiuri, Li, and Larkin}]{Ding2016a}
\bibinfo{author}{D.~Ding}, \bibinfo{author}{Z.~Pan},
  \bibinfo{author}{D.~Cuiuri}, \bibinfo{author}{H.~Li},
  \bibinfo{author}{N.~Larkin}, \bibinfo{title}{{Adaptive path planning for
  wire-feed additive manufacturing using medial axis transformation}},
  \bibinfo{journal}{Journal of Cleaner Production} \bibinfo{volume}{133}
  (\bibinfo{year}{2016}) \bibinfo{pages}{942--952},
  \doi{\bibinfo{doi}{10.1016/j.jclepro.2016.06.036}}.

\bibitem[{Xiong et~al.(2019)Xiong, Park, Padmanathan, Dharmawan, Foong, Rosen,
  and Soh}]{Xiong2019}
\bibinfo{author}{Y.~Xiong}, \bibinfo{author}{S.-I. Park},
  \bibinfo{author}{S.~Padmanathan}, \bibinfo{author}{A.~G. Dharmawan},
  \bibinfo{author}{S.~Foong}, \bibinfo{author}{D.~W. Rosen},
  \bibinfo{author}{G.~S. Soh}, \bibinfo{title}{Process planning for adaptive
  contour parallel toolpath in additive manufacturing with variable bead
  width}, \bibinfo{journal}{The International Journal of Advanced Manufacturing
  Technology} \doi{\bibinfo{doi}{10.1007/s00170-019-03954-1}}.

\bibitem[{Jin et~al.(2017{\natexlab{a}})Jin, Du, and He}]{Jin2017JMS}
\bibinfo{author}{Y.~Jin}, \bibinfo{author}{J.~Du}, \bibinfo{author}{Y.~He},
  \bibinfo{title}{{Optimization of process planning for reducing material
  consumption in additive manufacturing}}, \bibinfo{journal}{Journal of
  Manufacturing Systems} \bibinfo{volume}{44}
  (\bibinfo{year}{2017}{\natexlab{a}}) \bibinfo{pages}{65--78},
  \doi{\bibinfo{doi}{10.1016/j.jmsy.2017.05.003}}.

\bibitem[{{Ultimaker}(2019)}]{cura}
\bibinfo{author}{{Ultimaker}}, \bibinfo{title}{Ultimaker Cura 4.2.1 software},
  \urlprefix\url{https://ultimaker.com/software/ultimaker-cura},
  \bibinfo{year}{2019}.

\bibitem[{Livesu et~al.(2017)Livesu, Ellero, Mart{\'{i}}nez, Lefebvre, and
  Attene}]{Livesu2017CGF}
\bibinfo{author}{M.~Livesu}, \bibinfo{author}{S.~Ellero},
  \bibinfo{author}{J.~Mart{\'{i}}nez}, \bibinfo{author}{S.~Lefebvre},
  \bibinfo{author}{M.~Attene}, \bibinfo{title}{{From 3D models to 3D prints: an
  overview of the processing pipeline}}, \bibinfo{journal}{Computer Graphics
  Forum} \bibinfo{volume}{36}~(\bibinfo{number}{2}) (\bibinfo{year}{2017})
  \bibinfo{pages}{537--564}, \doi{\bibinfo{doi}{10.1111/cgf.13147}}.

\bibitem[{{N. Turner} et~al.(2014){N. Turner}, Strong, and {A.
  Gold}}]{N.Turner2014}
\bibinfo{author}{B.~{N. Turner}}, \bibinfo{author}{R.~Strong},
  \bibinfo{author}{S.~{A. Gold}}, \bibinfo{title}{{A review of melt extrusion
  additive manufacturing processes: I. Process design and modeling}},
  \bibinfo{journal}{Rapid Prototyping Journal}
  \bibinfo{volume}{20}~(\bibinfo{number}{3}) (\bibinfo{year}{2014})
  \bibinfo{pages}{192--204}, \doi{\bibinfo{doi}{10.1108/RPJ-01-2013-0012}}.

\bibitem[{Ahn et~al.(2002)Ahn, Montero, Odell, Roundy, and
  Wright}]{ahn2002anisotropic}
\bibinfo{author}{S.~H. Ahn}, \bibinfo{author}{M.~Montero},
  \bibinfo{author}{D.~Odell}, \bibinfo{author}{S.~Roundy},
  \bibinfo{author}{P.~K. Wright}, \bibinfo{title}{{Anisotropic material
  properties of fused deposition modeling ABS}}, \bibinfo{journal}{Rapid
  Prototyping Journal} \bibinfo{volume}{8}~(\bibinfo{number}{4})
  (\bibinfo{year}{2002}) \bibinfo{pages}{248--257},
  \doi{\bibinfo{doi}{10.1108/13552540210441166}}.

\bibitem[{Kuipers et~al.(2019)Kuipers, Wu, and Wang}]{KUIPERS2019CAD}
\bibinfo{author}{T.~Kuipers}, \bibinfo{author}{J.~Wu},
  \bibinfo{author}{C.~C.~L. Wang}, \bibinfo{title}{{CrossFill: Foam Structures
  with Graded Density for Continuous Material Extrusion}},
  \bibinfo{journal}{Computer-Aided Design} \bibinfo{volume}{114}
  (\bibinfo{year}{2019}) \bibinfo{pages}{37--50},
  \doi{\bibinfo{doi}{10.1016/j.cad.2019.05.003}}.

\bibitem[{Han et~al.(2002)Han, Jafari, Danforth, and Safari}]{Han2002JMSE}
\bibinfo{author}{W.~Han}, \bibinfo{author}{M.~A. Jafari},
  \bibinfo{author}{S.~C. Danforth}, \bibinfo{author}{A.~Safari},
  \bibinfo{title}{{Tool Path-Based Deposition Planning in Fused Deposition
  Processes }}, \bibinfo{journal}{Journal of Manufacturing Science and
  Engineering} \bibinfo{volume}{124}~(\bibinfo{number}{2})
  (\bibinfo{year}{2002}) \bibinfo{pages}{462--472},
  \doi{\bibinfo{doi}{10.1115/1.1455026}}.

\bibitem[{Steuben et~al.(2016)Steuben, Iliopoulos, and
  Michopoulos}]{steuben2016implicit}
\bibinfo{author}{J.~C. Steuben}, \bibinfo{author}{A.~P. Iliopoulos},
  \bibinfo{author}{J.~G. Michopoulos}, \bibinfo{title}{{Implicit slicing for
  functionally tailored additive manufacturing}},
  \bibinfo{journal}{Computer-Aided Design} \bibinfo{volume}{77}
  (\bibinfo{year}{2016}) \bibinfo{pages}{107--119},
  \doi{\bibinfo{doi}{10.1016/j.cad.2016.04.003}}.

\bibitem[{Zhao et~al.(2016)Zhao, Chen, Gu, Huang, Garcia, Chen, Tu, Benes,
  Zhang, and Cohen-Or}]{Zhao2016}
\bibinfo{author}{H.~Zhao}, \bibinfo{author}{B.~Chen}, \bibinfo{author}{F.~Gu},
  \bibinfo{author}{Q.-X. Huang}, \bibinfo{author}{J.~Garcia},
  \bibinfo{author}{Y.~Chen}, \bibinfo{author}{C.~Tu},
  \bibinfo{author}{B.~Benes}, \bibinfo{author}{H.~Zhang},
  \bibinfo{author}{D.~Cohen-Or}, \bibinfo{title}{{Connected fermat spirals for
  layered fabrication}}, \bibinfo{journal}{ACM Transactions on Graphics}
  \bibinfo{volume}{35}~(\bibinfo{number}{4}) (\bibinfo{year}{2016})
  \bibinfo{pages}{1--10}, \doi{\bibinfo{doi}{10.1145/2897824.2925958}}.

\bibitem[{Jin et~al.(2017{\natexlab{b}})Jin, He, Fu, Zhang, and
  Du}]{Jin2017RCIM}
\bibinfo{author}{Y.~Jin}, \bibinfo{author}{Y.~He}, \bibinfo{author}{G.~Fu},
  \bibinfo{author}{A.~Zhang}, \bibinfo{author}{J.~Du}, \bibinfo{title}{{A
  non-retraction path planning approach for extrusion-based additive
  manufacturing}}, \bibinfo{journal}{Robotics and Computer-Integrated
  Manufacturing} \bibinfo{volume}{48} (\bibinfo{year}{2017}{\natexlab{b}})
  \bibinfo{pages}{132--144}, \doi{\bibinfo{doi}{10.1016/j.rcim.2017.03.008}}.

\bibitem[{Held and Spielberger(2009)}]{Held2009}
\bibinfo{author}{M.~Held}, \bibinfo{author}{C.~Spielberger}, \bibinfo{title}{{A
  smooth spiral tool path for high speed machining of 2D pockets}},
  \bibinfo{journal}{Computer-Aided Design}
  \bibinfo{volume}{41}~(\bibinfo{number}{7}) (\bibinfo{year}{2009})
  \bibinfo{pages}{539--550}, \doi{\bibinfo{doi}{10.1016/j.cad.2009.04.002}}.

\bibitem[{Huang et~al.(2017)Huang, Lynn, and Kurfess}]{Huang2017}
\bibinfo{author}{N.~Huang}, \bibinfo{author}{R.~Lynn},
  \bibinfo{author}{T.~Kurfess}, \bibinfo{title}{{Aggressive Spiral Toolpaths
  for Pocket Machining Based on Medial Axis Transformation}},
  \bibinfo{journal}{Journal of Manufacturing Science and Engineering}
  \bibinfo{volume}{139}~(\bibinfo{number}{5}),
  \doi{\bibinfo{doi}{10.1115/1.4035720}}.

\bibitem[{McMains et~al.(2000)McMains, Smith, Wang, Sequin, and
  Carlo}]{Mcmains2000DETC}
\bibinfo{author}{S.~McMains}, \bibinfo{author}{J.~Smith},
  \bibinfo{author}{J.~Wang}, \bibinfo{author}{C.~Sequin},
  \bibinfo{author}{S.~Carlo}, \bibinfo{title}{{Layered manufacturing of
  thin-walled parts}}, in: \bibinfo{booktitle}{ASME Design Engineering
  Technical Conference, Baltimore, Maryland}, \bibinfo{organization}{Citeseer},
  \bibinfo{year}{2000}.

\bibitem[{Jin et~al.(2013)Jin, He, and Fu}]{Jin2013adaptive}
\bibinfo{author}{Y.~A. Jin}, \bibinfo{author}{Y.~He}, \bibinfo{author}{J.~Z.
  Fu}, \bibinfo{title}{An Adaptive Tool Path Generation for Fused Deposition
  Modeling}, in: \bibinfo{booktitle}{Advanced Materials Research}, vol.
  \bibinfo{volume}{819}, \bibinfo{pages}{7--12},
  \doi{\bibinfo{doi}{10.4028/www.scientific.net/amr.819.7}},
  \bibinfo{year}{2013}.

\bibitem[{Ding et~al.(2014)Ding, Pan, Cuiuri, and Li}]{Ding2014}
\bibinfo{author}{D.~Ding}, \bibinfo{author}{Z.~Pan},
  \bibinfo{author}{D.~Cuiuri}, \bibinfo{author}{H.~Li}, \bibinfo{title}{{A
  tool-path generation strategy for wire and arc additive manufacturing}},
  \bibinfo{journal}{International Journal of Advanced Manufacturing Technology}
  \bibinfo{volume}{73}~(\bibinfo{number}{1-4}) (\bibinfo{year}{2014})
  \bibinfo{pages}{173--183}, \doi{\bibinfo{doi}{10.1007/s00170-014-5808-5}}.

\bibitem[{Cox et~al.(1994)Cox, Takezaki, Ferguson, Kohkonen, and
  Mulkay}]{Cox1994CAD}
\bibinfo{author}{J.~J. Cox}, \bibinfo{author}{Y.~Takezaki},
  \bibinfo{author}{H.~R.~P. Ferguson}, \bibinfo{author}{K.~E. Kohkonen},
  \bibinfo{author}{E.~L. Mulkay}, \bibinfo{title}{{Space-filling curves in
  tool-path applications}}, \bibinfo{journal}{Computer-Aided Design}
  \bibinfo{volume}{26}~(\bibinfo{number}{3}) (\bibinfo{year}{1994})
  \bibinfo{pages}{215--224},
  \doi{\bibinfo{doi}{https://doi.org/10.1016/0010-4485(94)90044-2}}.

\bibitem[{Griffiths(1994)}]{Griffiths1994}
\bibinfo{author}{J.~G. Griffiths}, \bibinfo{title}{{Toolpath based on Hilbert's
  curve}}, \bibinfo{journal}{Computer-Aided Design}
  \bibinfo{volume}{26}~(\bibinfo{number}{11}) (\bibinfo{year}{1994})
  \bibinfo{pages}{839--844}, \doi{\bibinfo{doi}{10.1016/0010-4485(94)90098-1}}.

\bibitem[{Shaikh et~al.(2016)Shaikh, Kumar, Jain, and Tandon}]{Shaikh2016}
\bibinfo{author}{S.~Shaikh}, \bibinfo{author}{N.~Kumar}, \bibinfo{author}{P.~K.
  Jain}, \bibinfo{author}{P.~Tandon}, \bibinfo{title}{Hilbert curve based
  toolpath for FDM process}, in: \bibinfo{booktitle}{CAD/CAM, Robotics and
  Factories of the Future}, Lecture Notes in Mechanical Engineering,
  \bibinfo{publisher}{Springer}, \bibinfo{pages}{751--759},
  \doi{\bibinfo{doi}{10.1007/978-81-322-2740-3_72}}, \bibinfo{year}{2016}.

\bibitem[{Kao and Prinz(1998)}]{kao1998optimal}
\bibinfo{author}{J.-h. Kao}, \bibinfo{author}{F.~B. Prinz},
  \bibinfo{title}{{Optimal motion planning for deposition in layered
  manufacturing}}, in: \bibinfo{booktitle}{Proceedings of DETC},
  vol.~\bibinfo{volume}{98}, \bibinfo{organization}{Citeseer},
  \bibinfo{pages}{13--16}, \bibinfo{year}{1998}.

\bibitem[{Jin et~al.(2017{\natexlab{c}})Jin, He, and Du}]{Jin2017a}
\bibinfo{author}{Y.~Jin}, \bibinfo{author}{Y.~He}, \bibinfo{author}{J.~Du},
  \bibinfo{title}{{A novel path planning methodology for extrusion-based
  additive manufacturing of thin-walled parts}},
  \bibinfo{journal}{International Journal of Computer Integrated Manufacturing}
  \bibinfo{volume}{30}~(\bibinfo{number}{12})
  (\bibinfo{year}{2017}{\natexlab{c}}) \bibinfo{pages}{1301--1315},
  \doi{\bibinfo{doi}{10.1080/0951192X.2017.1307526}}.

\bibitem[{Moesen et~al.(2011)Moesen, Craeghs, Kruth, and
  Schrooten}]{Moesen2011}
\bibinfo{author}{M.~Moesen}, \bibinfo{author}{T.~Craeghs},
  \bibinfo{author}{J.~P. Kruth}, \bibinfo{author}{J.~Schrooten},
  \bibinfo{title}{{Robust beam compensation for laser-based additive
  manufacturing}}, \bibinfo{journal}{Computer-Aided Design}
  \bibinfo{volume}{43}~(\bibinfo{number}{8}) (\bibinfo{year}{2011})
  \bibinfo{pages}{876--888}, \doi{\bibinfo{doi}{10.1016/j.cad.2011.03.004}}.

\bibitem[{Behandish et~al.(2019)Behandish, Mirzendehdel, and
  Nelaturi}]{Behandish2019a}
\bibinfo{author}{M.~Behandish}, \bibinfo{author}{A.~M. Mirzendehdel},
  \bibinfo{author}{S.~Nelaturi}, \bibinfo{title}{{A Classification of
  Topological Discrepancies in Additive Manufacturing}},
  \bibinfo{journal}{Computer-Aided Design}
  \doi{\bibinfo{doi}{10.1016/j.cad.2019.05.032}}.

\bibitem[{Eiamsa-ard et~al.(2003)Eiamsa-ard, Liou, Landers, and
  Choset}]{eiamsa2003toward}
\bibinfo{author}{K.~Eiamsa-ard}, \bibinfo{author}{F.~W. Liou},
  \bibinfo{author}{R.~G. Landers}, \bibinfo{author}{H.~Choset},
  \bibinfo{title}{Toward automatic process planning of a multi-axis hybrid
  laser aided manufacturing system: skeleton-based offset edge generation}, in:
  \bibinfo{booktitle}{ASME 2003 International Design Engineering Technical
  Conferences and Computers and Information in Engineering Conference},
  \bibinfo{organization}{American Society of Mechanical Engineers Digital
  Collection}, \bibinfo{pages}{227--235},
  \doi{\bibinfo{doi}{10.1115/DETC2003/DAC-48726}}, \bibinfo{year}{2003}.

\bibitem[{Blum et~al.(1967)}]{blum1967transformation}
\bibinfo{author}{H.~Blum}, et~al., \bibinfo{title}{A transformation for
  extracting new descriptors of shape}, \bibinfo{journal}{Models for the
  perception of speech and visual form}
  \bibinfo{volume}{19}~(\bibinfo{number}{5}) (\bibinfo{year}{1967})
  \bibinfo{pages}{362--380}.

\bibitem[{Lee(1982)}]{lee1982medial}
\bibinfo{author}{D.-T.~T. Lee}, \bibinfo{title}{{Medial Axis Transformation of
  a Planar Shape}}, \bibinfo{journal}{IEEE Transactions on Pattern Analysis and
  Machine Intelligence} \bibinfo{volume}{PAMI-4}~(\bibinfo{number}{4})
  (\bibinfo{year}{1982}) \bibinfo{pages}{363--369},
  \doi{\bibinfo{doi}{10.1109/TPAMI.1982.4767267}}.

\bibitem[{Chazelle and Incerpi(1984)}]{chazelle1984}
\bibinfo{author}{B.~Chazelle}, \bibinfo{author}{J.~Incerpi},
  \bibinfo{title}{{Triangulation and shape-complexity}}, \bibinfo{journal}{ACM
  Transactions on Graphics} \bibinfo{volume}{3}~(\bibinfo{number}{2})
  (\bibinfo{year}{1984}) \bibinfo{pages}{135--152},
  \doi{\bibinfo{doi}{10.1145/357337.357340}}.

\bibitem[{Fournier and Montuno(1984)}]{fournier1984}
\bibinfo{author}{A.~Fournier}, \bibinfo{author}{D.~Y. Montuno},
  \bibinfo{title}{{Triangulating Simple Polygons and Equivalent Problems}},
  \bibinfo{journal}{ACM Transactions on Graphics}
  \bibinfo{volume}{3}~(\bibinfo{number}{2}) (\bibinfo{year}{1984})
  \bibinfo{pages}{153--174}, \doi{\bibinfo{doi}{10.1145/357337.357341}}.

\bibitem[{Sch{\"a}ling(2011)}]{schaling2011boost}
\bibinfo{author}{B.~Sch{\"a}ling}, \bibinfo{title}{The boost C++ libraries},
  \bibinfo{publisher}{Boris Sch{\"a}ling}, \bibinfo{year}{2011}.

\bibitem[{Fortune(1987)}]{fortune1986sascg}
\bibinfo{author}{S.~Fortune}, \bibinfo{title}{A sweepline algorithm for Voronoi
  diagrams}, \bibinfo{journal}{Algorithmica}
  \bibinfo{volume}{2}~(\bibinfo{number}{1}) (\bibinfo{year}{1987})
  \bibinfo{pages}{153}, \doi{\bibinfo{doi}{10.1007/BF01840357}}.

\bibitem[{Attali and Montanvert(1996)}]{attali1996modeling}
\bibinfo{author}{D.~Attali}, \bibinfo{author}{A.~Montanvert},
  \bibinfo{title}{{Modeling noise for a better simplification of skeletons}},
  in: \bibinfo{booktitle}{Proceedings of 3rd IEEE International Conference on
  Image Processing}, vol.~\bibinfo{volume}{3}, \bibinfo{organization}{IEEE},
  \bibinfo{pages}{13--16}, \doi{\bibinfo{doi}{10.1109/ICIP.1996.560357}},
  \bibinfo{year}{1996}.

\bibitem[{Sud et~al.(2005)Sud, Foskey, and Manocha}]{Sud2007}
\bibinfo{author}{A.~Sud}, \bibinfo{author}{M.~Foskey},
  \bibinfo{author}{D.~Manocha}, \bibinfo{title}{Homotopy-Preserving Medial Axis
  Simplification}, in: \bibinfo{booktitle}{Proceedings of the 2005 ACM
  Symposium on Solid and Physical Modeling}, \bibinfo{publisher}{Association
  for Computing Machinery}, \bibinfo{pages}{39--50},
  \doi{\bibinfo{doi}{10.1145/1060244.1060250}}, \bibinfo{year}{2005}.

\bibitem[{Kuipers et~al.(2018)Kuipers, Elkhuizen, Verlinden, and
  Doubrovski}]{Kuipers2018}
\bibinfo{author}{T.~Kuipers}, \bibinfo{author}{W.~Elkhuizen},
  \bibinfo{author}{J.~Verlinden}, \bibinfo{author}{E.~Doubrovski},
  \bibinfo{title}{{Hatching for 3D prints: Line-based halftoning for dual
  extrusion fused deposition modeling}}, \bibinfo{journal}{Computers \&
  Graphics} \bibinfo{volume}{74} (\bibinfo{year}{2018})
  \bibinfo{pages}{23--32}, \doi{\bibinfo{doi}{10.1016/j.cag.2018.04.006}}.

\bibitem[{Ertay et~al.(2018)Ertay, Yuen, and Altintas}]{Ertay2018}
\bibinfo{author}{D.~S. Ertay}, \bibinfo{author}{A.~Yuen},
  \bibinfo{author}{Y.~Altintas}, \bibinfo{title}{{Synchronized material
  deposition rate control with path velocity on fused filament fabrication
  machines}}, \bibinfo{journal}{Additive Manufacturing} \bibinfo{volume}{19}
  (\bibinfo{year}{2018}) \bibinfo{pages}{205--213},
  \doi{\bibinfo{doi}{10.1016/j.addma.2017.05.011}}.

\bibitem[{Tronvoll et~al.(2019)Tronvoll, Popp, Elverum, and
  Welo}]{tronvoll2019investigating}
\bibinfo{author}{S.~A. Tronvoll}, \bibinfo{author}{S.~Popp},
  \bibinfo{author}{C.~W. Elverum}, \bibinfo{author}{T.~Welo},
  \bibinfo{title}{Investigating pressure advance algorithms for filament-based
  melt extrusion additive manufacturing: theory, practice and simulations},
  \bibinfo{journal}{Rapid Prototyping Journal}
  \bibinfo{volume}{25}~(\bibinfo{number}{5}) (\bibinfo{year}{2019})
  \bibinfo{pages}{830--839}, \doi{\bibinfo{doi}{10.1108/RPJ-10-2018-0275}}.

\bibitem[{Johnson(2017)}]{johnson2014clipper}
\bibinfo{author}{A.~Johnson}, \bibinfo{title}{{Clipper 6.4.2 - an open source
  freeware library for clipping and offsetting lines and polygons}},
  \bibinfo{year}{2017}.

\bibitem[{Liu and Shapiro(2016)}]{Liu2016CAD}
\bibinfo{author}{X.~Liu}, \bibinfo{author}{V.~Shapiro},
  \bibinfo{title}{Homogenization of material properties in additively
  manufactured structures}, \bibinfo{journal}{Computer-Aided Design}
  \bibinfo{volume}{78} (\bibinfo{year}{2016}) \bibinfo{pages}{71 -- 82},
  \doi{\bibinfo{doi}{https://doi.org/10.1016/j.cad.2016.05.017}}.

\bibitem[{{Wu} et~al.(2018){Wu}, {Aage}, {Westermann}, and
  {Sigmund}}]{wu2017infill}
\bibinfo{author}{J.~{Wu}}, \bibinfo{author}{N.~{Aage}},
  \bibinfo{author}{R.~{Westermann}}, \bibinfo{author}{O.~{Sigmund}},
  \bibinfo{title}{Infill Optimization for Additive Manufacturing -- Approaching
  Bone-Like Porous Structures}, \bibinfo{journal}{IEEE Transactions on
  Visualization and Computer Graphics}
  \bibinfo{volume}{24}~(\bibinfo{number}{2}) (\bibinfo{year}{2018})
  \bibinfo{pages}{1127--1140}, \doi{\bibinfo{doi}{10.1109/TVCG.2017.2655523}}.

\end{thebibliography}
